\documentclass[%
notitlepage,
twocolumn,
superscriptaddress,
amsmath,amssymb,
prb,
floatfix,
]{revtex4-2}

\usepackage[strict]{changepage}
\usepackage{graphicx}
\usepackage{dcolumn}
\usepackage{bm}
\usepackage{units}
\usepackage{cancel}

\usepackage{subfigure}
\usepackage{xcolor}
\usepackage{mathtools}
\usepackage{bbold}

\usepackage{mleftright}

\usepackage{framed}
\colorlet{shadecolor}{gray!20}

\usepackage{empheq}

\usepackage[most]{tcolorbox}

\usepackage{rotating}

\tcbset{colback=yellow!10!white, colframe=black!50!black, 
        highlight math style= {enhanced, 
            colframe=black,colback=red!10!white,boxsep=0pt}
        }

\newcommand*{\Poloptwo}[2]{{{}\hat{P}^{\vphantom{\dagger}}_{}}{\vphantom{P}}_{#1}^{#2}}

\newcommand*{\Poldagtwo}[2]{{{}\hat{P}^{\dagger}}{\vphantom{P}}_{#1}^{#2}}

\newcommand*{\Pndagtwo}[2]{{{}\hat{P}^{\vphantom{\dagger}}_{}}{\vphantom{P}}_{#1}^{#2}}

\newcommand*{\Pdagtwo}[2]{{{}\hat{P}^{\dagger}}{\vphantom{P}}_{#1}^{#2}}

\newcommand*{\Nex}[1]{N_{#1}}
\newcommand*{\Nextwo}[2]{N_{#1}^{#2}}

\newcommand*{\adagtwo}[2]{{\hat{a}^{\dagger}_{}}{\vphantom{a}}_{#1}^{#2}}
\newcommand*{\andagtwo}[2]{{\hat{a}^{\vphantom{\dagger}}_{}}{\vphantom{a}}_{#1}^{#2}}

\newcommand*{\vdagtwo}[2]{{\hat v^{\dagger}_{}}{\vphantom{\hat v}}_{#1}^{#2}}
\newcommand*{\vndagtwo}[2]{{\hat v_{}^{\vphantom{\dagger}}}{\vphantom{\hat v}}_{#1}^{#2}}
\newcommand*{\cdagtwo}[2]{{\hat c_{}^{\dagger}}{\vphantom{\hat c}}_{#1}^{#2}}
\newcommand*{\cndagtwo}[2]{{\hat c_{}^{\vphantom{\dagger}}}{\vphantom{\hat c}}_{#1}^{#2}}

\newcommand{\ExWFtwo}[2]{\varphi_{#1}^{#2}}
\newcommand{\ExWFstartwo}[2]{\varphi^*{\vphantom{\varphi}}_{#1}^{#2}}

\newcommand{\custompm}{\mathbin{\rotatebox[origin=c]{80}{$-$}\kern-0.6ex{\raisebox{-0.5ex}{$+$}}}}

\usepackage{placeins}
\usepackage{appendix}

\definecolor{custom}{RGB}{15,45,150}
\usepackage[urlcolor=blue]{hyperref}
\hypersetup{colorlinks=true,linktoc=page,citecolor = blue,linkcolor=blue}

\setcitestyle{numbers,square}

\begin{document}
\makeatletter
\renewcommand\@biblabel[1]{[#1]}
\makeatletter

\preprint{APS/123-QED}
\title{Many-Body Rashba Spin-Orbit Interaction and Exciton Spin Relaxation in Atomically Thin Semiconductor Structures}
\author{Henry Mittenzwey}
\email{henry.mittenzwey@uni-giessen.de}
\affiliation{Institut für Theoretische Physik und Zentrum für Materialforschung, Justus-Liebig-Universität Gie{\ss}en, 35392 Gie{\ss}en, Germany}
\affiliation{Nichtlineare Optik und Quantenelektronik, Institut für Physik und Astronomie (IFPA), Technische Universität Berlin, 10623 Berlin, Germany}
\author{Andreas Knorr}
\affiliation{Nichtlineare Optik und Quantenelektronik, Institut für Physik und Astronomie (IFPA), Technische Universität Berlin, 10623 Berlin, Germany}
\begin{abstract}
    We propose a pair spin-orbit interaction (PSOI) mechanism by establishing a mesoscopic many-particle Rashba Hamiltonian. In lowest order, this Hamiltonian self-consistently describes exciton spin relaxation in monolayer transition metal dichalcogenides (TMDC) due to local electric fields caused by spatial asymmetries in the dielectric environment. 
    For a monolayer MoSe$_2$ on a SiO$_2$ substrate above 77\,K showing a bright-dark splitting in the meV range, the local electric field causes fast intravalley spin relaxation on a sub-picosecond timescale, whereas it is negligible for other TMDCs with larger bright-dark splitting. 
\end{abstract}
\date{\today}
\maketitle
\section{Introduction}

Atomically thin semiconductors \cite{wang2018colloquium} host a broad variety of incoherent exciton scattering mechanisms. Based on the valley-selective optical selection rules, next to spin-conserving exciton-phonon scattering \cite{selig2018dark,brem2020phonon,wallauer2021momentum,katzer2023exciton}, Coulomb scattering mechanisms such as 
spin-conserving Dexter interaction \cite{berghauser2018inverted} and spin-nonconserving intra- and intervalley exchange \cite{guo2019exchange,maialle1993exciton,yu2014valley,selig2019ultrafast,selig2020suppression,kwong2021effect} via double spin flips are relevant. 
Moreover, single spin flips can occur via, e.g., spin-Zeeman interaction induced by externally applied in-plane magnetic fields \cite{snoeken2025theory,robert2020measurement,vasconcelos2018dark,scharf2017magnetic,zhang2017magnetic} or Rashba interaction \cite{article:THEORY_DispersionBychkovRashba1984,book:Spin_orbit_coupling_Winkler2003}. 
Rashba spin-orbit interaction (SOI) can already affect linear optical spectra of a spin-forbidden mixture of $s$-$p$ orbital excitons in monolayers \cite{cao2024emergent,cao2025tunable}. However, these signatures are very weak and have not been measured until now. For $s$-orbital excitonic states, which are usually measured in absorption or photoluminescence due to their large optical oscillator strength, no brightening of spin-dark excitonic states in out-of-plane electric fields can occur, since Rashba spin coupling vanishes at zero center-of-mass (COM) momentum within the light cone \cite{mittenzwey2025ultrafast}. In contrast, in-plane magnetic fields already brighten spin-forbidden $s$-orbital excitons at zero COM momentum \cite{snoeken2025theory,feierabend2020brightening,robert2020measurement}. 
Thus, Rashba spin coupling is only relevant for incoherent $s$-orbital excitons occupying COM momenta well beyond the light cone. In contrast to Coulomb-exchange-induced double spin flips or Zeeman-interaction-induced single spin flips, single spin flip scattering mechanisms induced by electric fields are less understood, as they are often included via external parameters \cite{ochoa2013spin,dery2015polarization,song2013transport,yang2020exciton} or resulting from a complex band structure \cite{junior2022first,deilmann2020ab,wang2018intravalley,molina2017ab}. 
Recently, chiral phonons have been identified as an important single spin-flip mechanism \cite{lagarde2024efficient,chan2025exciton,zhang2022ab}. 
Besides the one-body SOI, it has been shown, that many-body-, Breit- or pair spin-orbit interaction (PSOI) due to self-induced Coulomb fields also occurs \cite{breit1929effect,yi2000breit,gindikin2018spin,gindikin2022spin,liu2024ordered,rajagopal1998spin}, which completes the electron-electron interaction picture. 
In particular, such process can induce a superconducting phase in thin films under suitable conditions \cite{gindikin2025electron}.

In this work, we derive a many-body Rashba PSOI mechanism \cite{article:THEORY_DispersionBychkovRashba1984,article:THEORY_Bychkov_Rashba_Coupling_Kormanyos2014,gindikin2022spin,chang1993many},
which results from self-consistently described out-of-plane electric fields $\mathbf E(\mathbf r,z)$ induced by the dielectric environment and examine its consequences for excitons in atomically thin semiconductors.

The nonrelativistic limit of the Dirac equation results in the Pauli equation, where the spin-orbit interaction appears. The corresponding Hamiltonian reads \cite{book:Spin_orbit_coupling_Winkler2003}:
\begin{align}
    H_{\text{BR}} = -\frac{\alpha_{\text{BR}}}{\hbar}\boldsymbol{\sigma}\cdot\left(\mathbf E(\mathbf r)\times\mathbf p\right),
    \label{eq:Hamiltonian_Rashba_General}
\end{align}
where
\mbox{$\boldsymbol{\sigma}=\begin{pmatrix}
    \sigma_x & \sigma_y & \sigma_z
\end{pmatrix}^{\top}$} is the vector of the Pauli matrices $\sigma_i$, $\mathbf E(\mathbf r)$ is an electric field, \mbox{$\mathbf p=\frac{\hbar}{\mathrm i}\nabla$} is the momentum operator and $\alpha_{\text{BR}}$ is the Rashba coupling constant. Note that, in solids, the Rashba coupling constant $\alpha_{\text{BR}}$ is greatly enhanced compared to the atomic constant in vacuum \cite{article:THEORY_Bychkov_Rashba_Coupling_Kormanyos2014,book:Spin_orbit_coupling_Winkler2003}. 
In Eq.~\eqref{eq:Hamiltonian_Rashba_General}, the electric field \mbox{$\mathbf E(\mathbf r)$} acts on the electrons and couples the spin $\boldsymbol{\sigma}$ to the orbital angular momentum $\mathbf L = \mathbf r\times\mathbf p$. In contrast to atoms, where the electric field occurs due to the nucleus and exhibits radial symmetry, in mesoscopically structured semiconductors, the electric field $\mathbf E(\mathbf r)$ acting on Bloch electrons is not necessarily radial \cite{combescot2019spin}, and, hence, induces distinct physical effects depending on the geometric arrangement of atoms: E.g., in-plane microscopic fields induced by the ionic background, which vary on the scale of the unit cell, are responsible for the SOI-induced spin-splitting of valence- and conduction bands \cite{kormanyos2015k,kosmider2013large,article:THEORY_spin_valley_selective_excitation_Yao2012}, while out-of-plane fields, external or intrinsic, induce spin flips \cite{article:THEORY_Bychkov_Rashba_Coupling_Kormanyos2014}. 
Recently, it has been shown, that PSOI, which is often not considered in \textit{ab initio} calculations for solids, is non-negligible in determining the valence- and conduction band splitting in WSe$_2$ monolayers \cite{boccuni2024unveiling}.

Starting from Eq.~\eqref{eq:Hamiltonian_Rashba_General}, we derive a second-quantized description of the Rashba interaction caused by the optically induced charge density resulting in an electric field $\mathbf E(\mathbf r,z)$ due to spatial asymmetries in the dielectric environment of an atomically thin semiconductor (Sec.~\ref{sec:Electron-Hole Many-Particle Rashba Hamiltonian}). Using an excitonic description (Sec.~\ref{sec:Excitonic Rashba Hamiltonian}) in combination with exciton-phonon-interaction-generating non-zero COM-momentum excitons (Sec.~\ref{sec:Equations of Motion for Exciton-Phonon Scattering}), the described PSOI leads to a phonon-assisted exciton spin relaxation. We illustrate the new mechanism for a monolayer MoSe$_2$ and quantify its impact on the incoherent exciton dynamics via microscopic calculations (Sec.~\ref{sec:Spin Relaxation in Monolayer MoSe$_2$}).

\section{Electron-Hole Many-Particle Rashba Hamiltonian due to Self-Consistent Fields in an Inhomogeneous Dielectric Environment}
\label{sec:Electron-Hole Many-Particle Rashba Hamiltonian}
\begin{figure}
    \centering
    \includegraphics[width=0.7\linewidth]{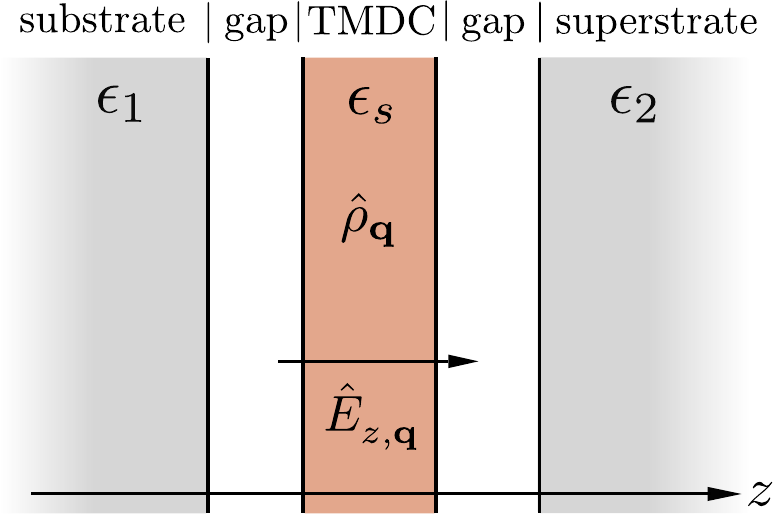}
    \caption{Sketch of the sample geometry. $\epsilon_{1/2}$ is the static dielectric constant of substrate/superstrate and $\epsilon_s$ is the dielectric constant of the thin semiconductor, which are separated by a vacuum gap of width $l$. $\hat \rho_{\mathbf q}$, cf.\ Eq.~\eqref{eq:ChargeDensityOperator}, is the charge density and $\hat E_{z,\mathbf q}$ is the (operator-valued) electric field induced by surface charges at the boundaries, cf.\ Eq.~\eqref{eq:ElectricFieldConfined}.}
    \label{fig:structure}
\end{figure}

We consider the following geometry, cf.~Fig.~\ref{fig:structure}: A monolayer TMDC ($\epsilon_s$) is sandwiched between a substrate ($\epsilon_1$) and superstrate ($\epsilon_2$) dielectric material with a small vacuum gap in between due to van~der~Waals interaction \cite{florian2018dielectric,druppel2017diversity}. An induced charge density $\hat \rho_{\mathbf q}$ via, e.g., optical excitation, accumulates surface charges at dielectric boundaries, which induce a net out-of-plane electric field $\hat E_{z,\mathbf q}$ in an inhomogeneous dielectric environment ($\epsilon_1\neq\epsilon_2$).

To investigate the consequences thereof, we employ the method of second quantization for Bloch electrons \cite{book:graphene_carbon_nanotubes_malic2013}. The Rashba Hamiltonian from Eq.~\eqref{eq:Hamiltonian_Rashba_General} reads:
\begin{align}
\label{eq:RashbaHamiltonian}
    \hat H_{\text{BR}} = &\, \sum_{\substack{\mathbf k,\mathbf q,\\\lambda_1,\lambda_2,\\\xi,\xi^{\prime},s}}\hat E_{z,\mathbf q+\mathbf K^{\xi^{\prime}}-\mathbf K^{\xi}}
    S_{\mathbf k+\mathbf q,\mathbf k}^{\lambda_1,\lambda_2,\xi^{\prime},\xi,\bar s,s}
    \adagtwo{\lambda_1,\mathbf k+\mathbf q}{\xi^{\prime},\bar s}\andagtwo{\lambda_2,\mathbf k}{\xi,s}.
\end{align}
A detailed derivation of Eq.~\ref{eq:RashbaHamiltonian} is given in App.~\ref{app:RashbaHamiltonianDerivation}. Here, $\hat E_{z,\mathbf q}$ is
the spatially inhomogeneous, quantum-confined out-of-plane ($z$-direction) oriented electric field, cf.\ Fig.~\ref{fig:structure}, defined as:
\begin{align}
    \hat E_{z,\mathbf q} = \frac{1}{\mathcal A}\int\mathrm dz\,\left|\zeta(z)\right|^2\hat E_{z,\mathbf q}(z),
    \label{eq:local_field_quantum_confined}
\end{align}
where $\mathcal A$ is the in-plane area of the atomically thin semiconductor, $\zeta(z)$ are the electronic confinement wave functions and $\hat E_{z,\mathbf q}(z)$ is the in-plane wave vector $\mathbf q$-dependent Fourier transform
of the real-space field $\hat E_z(\mathbf r,z)$, $\mathbf r$ being the in-plane coordinate. We note, that we neglect Umklapp processes, i.e., we neglect spatial variations of the electric field on the scale of a unit cell. Also, we note, that the quantum-confined electric field $\hat E_{z,\mathbf q}$ in Eq.~\eqref{eq:local_field_quantum_confined} carries equal units of V\,nm$^{-1}$ as the electric field in real space $E_z(\mathbf r,z)$ (in contrast to the true Fourier transform $\hat E_{z,\mathbf q}(z)$), since it is defined over the semiconductor area $\mathcal A$. In Eq.~\eqref{eq:RashbaHamiltonian}, $\mathbf K^{\xi}$ is a valley momentum at valley $\xi$, with reference point $\mathbf K^{\Gamma} = \mathbf 0$ at the $\Gamma$ valley, $\hat a^{(\dagger)}\vphantom{\hat a}_{\lambda,\mathbf k}^{\xi,s}$ is the annihilation (creation) operator of an electron in band $\lambda$ with momentum $\mathbf k$ relative to valley $\xi$ and spin $s$. Note that, if \mbox{$s=\,\uparrow$}, then \mbox{$\bar s=\,\downarrow$} and vice versa.
The Rashba matrix element $S_{\mathbf k+\mathbf q,\mathbf k}^{\lambda_1,\lambda_2,\xi,\xi,\bar s,s}$ results from the expansion of the Rashba coupling in single-particle Bloch wave functions:
\begin{align}
\Psi_{\mathbf k,\lambda}^{\xi,s}(\mathbf r,z) = \frac{1}{\sqrt{\mathcal{A}}}\mathrm e^{\mathrm i\mathbf k\cdot\mathbf r}u_{\lambda,\mathbf k}^{\xi,s}(\mathbf r)\zeta(z)\chi_s,
\end{align}
with Bloch factor $u_{\lambda,\mathbf k}^{\xi,s}(\mathbf r)$ and spin wave function $\chi_s$ and is defined as:
\begin{align}
\begin{split}
\label{eq:RashbaMatrixElementGeneral}
    &S_{\mathbf k+\mathbf q,\mathbf k}^{\lambda_1,\lambda_2,\xi^{\prime},\xi,\bar s,s}=-\frac{\alpha_{\text{BR}}}{\hbar}\left(
    \vphantom{\left(\mathbf p_{\mathbf k+\mathbf q,\mathbf k}^{\lambda_1,\lambda_2,\xi,\xi,\bar s,s}\right)_x - \left(\mathbf p_{\mathbf k+\mathbf q,\mathbf k}^{\lambda_1,\lambda_2,\xi,\xi,\bar s,s}\right)_y}
    \mathrm i\left(\delta_{s,\uparrow}-\delta_{s,\downarrow}\right)\right.\\
    &\quad\quad\quad\left.\times\left(\mathbf p_{\mathbf k+\mathbf q,\mathbf k}^{\lambda_1,\lambda_2,\xi^{\prime},\xi,\bar s,s}\right)_x - \left(\mathbf p_{\mathbf k+\mathbf q,\mathbf k}^{\lambda_1,\lambda_2,\xi^{\prime},\xi,\bar s,s}\right)_y\right),
    \end{split}
\end{align}
where $\mathbf p_{\mathbf k+\mathbf q,\mathbf k}^{\lambda,\lambda^{\prime},\xi^{\prime},\xi,\bar s,s}$ is the in-plane momentum matrix element, cf.~Eq.~\eqref{eq:momentum_matrix_element} in App.~\ref{app:RashbaHamiltonianDerivation}. 
At first glance, Eq.~\eqref{eq:RashbaMatrixElementGeneral}, and therefore Eq.~\eqref{eq:RashbaHamiltonian}, seem to break the rotational symmetry in the $x$-$y$-plane. Actually, this is not the case, as a rotation in the $x$-$y$-plane simultaneously acts on both the spin vector $\boldsymbol{\sigma}$ and the cross product of field and momentum $\mathbf E(\mathbf r,z)\times\mathbf p$, which always ensures rotational invariance of the Rashba Hamiltonian in Eq.~\eqref{eq:RashbaHamiltonian}. 
In Eq.~\eqref{eq:RashbaHamiltonian}, an external electric field $\hat E_{z,\mathbf q}(z)$ can be applied in $z$-direction with vanishing in-plane variation, so that \mbox{$\hat E_{z,\mathbf q}(z)\equiv E_{z}^{\text{ext}}\mathcal{A}\delta_{\mathbf q,\mathbf 0}$}, where $E_z^{\text{ext}}$ can be generated by, e.g., static electrical gating \cite{kovalchuk2025revealing}, or it can be an in-plane propagating optical field \cite{wang2017plane} with an almost vanishing in-plane photon momentum. 
Static fields couple mainly via the intraband Rashba interaction with $\lambda_1=\lambda_2$, whereas optical fields couple mainly via the interband Rashba interaction with \mbox{$\lambda_1\neq\lambda_2$}.

In the case considered here, i.e., without electrical gating or in-plane propagating optical fields, the electric field is induced by a self-arranging charge-density fluctuation of electrons and holes $\hat\rho_{\mathbf q}^{}(z)$ within the atomically thin semiconductor itself. 
The occurring charge distribution $\hat\rho_{\mathbf q}^{}(z)$ results from the dielectric boundary conditions for the longitudinal fields $\mathbf E_{}^{\parallel}$ and $\mathbf D_{}^{\parallel}$ implied by the macroscopic Maxwell equations. The boundary conditions for the normal vector $\mathbf n$ in $z$-direction read: \mbox{$\mathbf n\times\mathbf E_i^{\parallel}=\mathbf n\times\mathbf E_j^{\parallel}$} and \mbox{$\mathbf n\cdot\mathbf D_i^{\parallel}=\mathbf n\cdot\mathbf D_j^{\parallel}$} at the interface, cf.~Fig.~\ref{fig:structure}, between layer $i$ and layer $j$ inducing electronic charges $\hat \rho_{\mathbf q}$ in the heterostructure, that have to be adjusted self-consistently.
To account for this self-consistency in the charge $\hat \rho_{\mathbf q}$, we solve the inhomogeneous Poisson equation for a layered dielectric environment with the operator-valued charge density $\hat\rho_{\mathbf q}^{}(z)$ as a source.
The solution can be obtained by the Green's function $G_{\mathbf q}(z,z^{\prime})$ of Poisson's equation. The local, quantum-confined electric field can be expressed as, cf.~App.~\ref{app:ElectricField}:
\begin{align}
\begin{split}
\label{eq:ElectricFieldConfined}
    &\hat E_{z,\mathbf q} =   -\frac{1}{\mathcal A}\int\mathrm dz\,\left|\zeta(z)\right|^2\partial_z\int\mathrm dz^{\prime}G_{\mathbf q}(z,z^{\prime})
    \hat\rho_{\mathbf q}^{}(z^{\prime}).
    \end{split}
\end{align}
The occurring source charge density reads:
\begin{align}
\label{eq:ChargeDensityOperator}
    \hat\rho_{\mathbf q}^{}(z) = &\,  -e\left|\zeta(z)\right|^2\sum_{\substack{\mathbf k,\xi,\xi^{\prime},s,\\\lambda,\lambda^{\prime}}}\overline \Upsilon_{\lambda,\lambda^{\prime},\mathbf k-\mathbf q,\mathbf k}^{\xi,\xi^{\prime},s,s}\adagtwo{\lambda,\mathbf k-\mathbf q}{\xi,s}\andagtwo{\lambda^{\prime},\mathbf k}{\xi^{\prime},s}
    ,
\end{align}
where $e$ is the elementary charge and $\overline \Upsilon_{\lambda,\lambda^{\prime},\mathbf k-\mathbf q,\mathbf k}^{\xi,\xi^{\prime},s,s}$ is the (reduced) form factor, given by:
\begin{align}
    \overline \Upsilon_{\lambda,\lambda^{\prime},\mathbf k-\mathbf q,\mathbf k}^{\xi,\xi^{\prime},s,s} = \frac{1}{\mathcal A_{\text{UC}}}\int_{\mathcal A_{\text{UC}}}\mathrm d^2r\,\,u^*_{}\vphantom{u}_{\lambda,\mathbf k-\mathbf q}^{\xi,s}(\mathbf r)u_{\lambda^{\prime},\mathbf k}^{\xi^{\prime},s}(\mathbf r),
    \label{eq:FormFactor}
\end{align}
with unit cell area $\mathcal A_{\text{UC}}$ and integration over the lattice-periodic Bloch factor $u_{\lambda,\mathbf k}^{\xi,s}(\mathbf r)$.
In contrast to Ref.~\cite{mittenzwey2025ultrafast}, where the charge density is treated in a classical mean-field limit, the crucial point of the present work is to preserve the operator-valued character and thus quantum fluctuations in Eq.~\eqref{eq:ChargeDensityOperator}. This way, not only Hartree contributions but also Fock- or higher correlations can be considered and improve the treatment of many-body interactions.
Using Eqs.~\eqref{eq:ElectricFieldConfined}--\eqref{eq:FormFactor}, we arrive at the following form of the Rashba Hamiltonian from Eq.~\eqref{eq:RashbaHamiltonian}:
\begin{multline}
\label{eq:RashbaHamiltonianManyBody}
    \hat H_{\text{BR}} = \sum_{\substack{\mathbf k,\mathbf k^{\prime},\mathbf q,\\\lambda_1,\lambda_2,\lambda_3,\lambda_4,\\\xi,\xi^{\prime},\xi^{\prime\prime},s,s^{\prime}}}E_{z,\mathbf q+\mathbf K^{\xi^{\prime}}-\mathbf K^{\xi}}^{\text{loc}}\overline \Upsilon_{\lambda_2,\lambda_3,\mathbf k^{\prime}-\mathbf q,\mathbf k^{\prime}}^{\xi^{\prime\prime}+\xi-\xi^{\prime},\xi^{\prime\prime},s^{\prime},s^{\prime}}\\
    \times S_{\mathbf k+\mathbf q,\mathbf k}^{\lambda_1,\lambda_4,\xi^{\prime},\xi,\bar s,s}
    \left(\adagtwo{\lambda_1,\mathbf k+\mathbf q}{\xi^{\prime},\bar s}
    \adagtwo{\lambda_2,\mathbf k^{\prime}-\mathbf q}{\xi^{\prime\prime}+\xi-\xi^{\prime},s^{\prime}}\andagtwo{\lambda_3,\mathbf k^{\prime}}{\xi^{\prime\prime},s^{\prime}}
    \andagtwo{\lambda_4,\mathbf k}{\xi,s}\right.\\
    \left.\quad\quad + \delta_{\mathbf k^{\prime},\mathbf k+\mathbf q}^{\lambda_3,\lambda_1}\delta_{s^{\prime},\bar s}^{\xi^{\prime},\xi^{\prime\prime}} \adagtwo{\lambda_2,\mathbf k}{\xi,\bar s}\andagtwo{\lambda_4,\mathbf k}{\xi,s}\right).
\end{multline}
Eq.~\eqref{eq:RashbaHamiltonianManyBody} is the full many-body Rashba Hamiltonian, which, at this level, accounts for various single spin flip scattering $s\rightarrow \bar s$ analog to the many-body Coulomb Hamiltonian \cite{mittenzwey2026coulombinteraction}, e.g.~intraband (\mbox{$\lambda_1=\lambda_4$} and \mbox{$\lambda_2=\lambda_3$}), interband (\mbox{$\lambda_1\neq\lambda_4$} and \mbox{$\lambda_2\neq\lambda_3$}) and Auger-like (\mbox{$\lambda_1=\lambda_4$} and \mbox{$\lambda_2\neq\lambda_3$} or vice versa) electron-electron, hole-hole and electron-hole intravalley (\mbox{$\xi=\xi^{\prime}$}) and intervalley (\mbox{$\xi^{\prime}\neq\xi$}) scattering. Note, that in contrast to the spin-conserving many-body Coulomb interaction, the one-particle contribution ($\mathbf r=\mathbf 0$-self-energy, second term in Eq.~\eqref{eq:RashbaHamiltonianManyBody}) arising from normal ordering \cite{haug2009quantum} \textit{does} contribute to the dynamics, since -- in low-wavenumber approximation -- it is not simply proportional to the particle number operator $\hat N = \adagtwo{\lambda,\mathbf k}{\xi,s}\andagtwo{\lambda,\mathbf k}{\xi,s}$ but describes a spin-flip process $s \rightarrow \bar s$ via $\adagtwo{\lambda,\mathbf k}{\xi,\bar s}\andagtwo{\lambda,\mathbf k}{\xi,s}$.
 
As mediating quantity, in Eq.~\eqref{eq:RashbaHamiltonianManyBody}, the (classical) local electric field $E_{z,\mathbf q}^{\text{loc}}$ can be identified:
\begin{align}
\begin{split}
\label{eq:ElectricFieldLocalExplicit}
     E_{z,\mathbf q}^{\text{loc}} = \frac{e}{\mathcal A2\epsilon_0\epsilon_{s,\bot}}\frac{g_{\mathbf q}}{f_{\mathbf q}},
    \end{split}
\end{align}
where
$\epsilon_0$ is the vacuum permittivity and $\epsilon_{s,\bot}$ is the relative out-of-plane dielectric constant of the thin semiconductor. The unitless functions $g_{\mathbf q}$ and $f_{\mathbf q}$, which encode the momentum dependence determined by the dielectric structure, are given in Eq.~\eqref{eq:g_function} and Eq.~\eqref{eq:f_function} in App.~\ref{app:ElectricField}, respectively.
The electric field in Eq.~\eqref{eq:ElectricFieldLocalExplicit} stems solely from the homogeneous solution of the Poisson equation, which arises due to surface charges determined by the electrostatic boundary conditions. Out-of-plane fields induced by the corresponding real charge, which corresponds to the inhomogeneous solution of the Poisson equation, vanish under $z$-integration within our assumptions
of a symmetric out-of-plane charge distribution $|\zeta(z)|^2$ in Eq.~\eqref{eq:ElectricFieldConfined} or Eq.~\eqref{eq:local_field_quantum_confined}. We note, that this is not always the case, since, e.g., chalcogen vacancies can induce spatial asymmetries in $z$-direction in the charge distribution \cite{ozcan2024point,min2024chalcogen}.

To evaluate Eq.~\eqref{eq:RashbaHamiltonianManyBody}, we assume a perfectly flat TMDC monolayer with a $z$-dependent dielectric environment $\epsilon(z)$ without any vacancies \cite{latini2015excitons}. 
Consequently, if we set \mbox{$\epsilon_1=\epsilon_2$}, i.e., equal substrate and superstrate material in Fig.~\ref{fig:structure}, the field in Eq.~\eqref{eq:ElectricFieldLocalExplicit} vanishes:
\begin{align}
\label{eq:LocalEletcricFieldSymmZero}
    E_{z,\mathbf q}^{\text{loc}}\Big|_{\epsilon_1=\epsilon_2} = 0.
\end{align}
On the other hand, if substrate and superstrate differ in their respective dielectric constants \mbox{$\epsilon_1\neq\epsilon_2$}, the local electric field is nonzero:
\begin{align}
\label{eq:LocalEletcricFieldSymmNonzero}
    E_{z,\mathbf q}^{\text{loc}}\Big|_{\epsilon_1\neq\epsilon_2} \neq 0.
\end{align}
This is the situation depicted in Fig.~\ref{fig:structure}.

\begin{figure}[h!]
\includegraphics[width=1\linewidth]{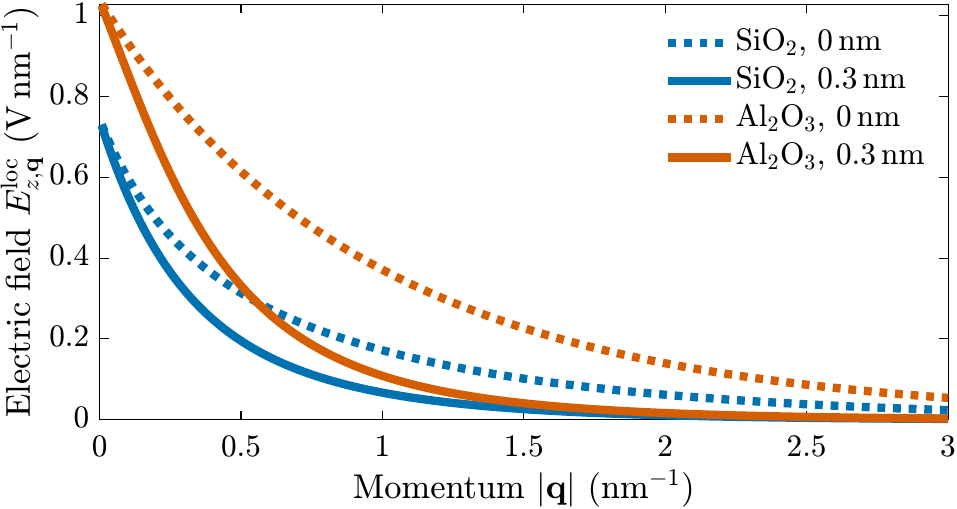}
\caption{Local electric field $E_{z,\mathbf q}^{\text{loc}}$ from Eq.~\eqref{eq:ElectricFieldLocalExplicit} for an example MoSe$_2$ monolayer (\mbox{$\epsilon_{s,\perp} = 7.2$} \cite{laturia2018dielectric}) on a SiO$_2$ substrate (\mbox{$\epsilon_1=3.9$} \cite{xue2011scanning}, \mbox{$\epsilon_2=1$}) and on a sapphire substrate (\mbox{$\epsilon_1=\sqrt{\epsilon_{1,\parallel}\epsilon_{1,\perp}}=\sqrt{11.6\cdot 9.4}$} \cite{fontanella1974low}, \mbox{$\epsilon_2=1$}) for a substrate distance $l$ of 0\,nm and 0.3\,nm.}
\label{fig:LocalFieldMoS2}
\end{figure}

In Fig.~\ref{fig:LocalFieldMoS2}, we depict the local electric field in Eq.~\eqref{eq:ElectricFieldLocalExplicit} for an example MoSe$_2$ monolayer on a SiO$_2$ and sapphire substrate for different substrate distances. Here, the local electric field is well-localized in momentum space and reaches values larger than 1\,V\,nm$^{-1}$ at $\mathbf q=\mathbf 0$ on a sapphire substrate. 
At first glance, the calculated fields are in the range of V\,nm$^{-1}$ and appear to be rather strong. However, the strength of the local electric field cannot be directly compared to the total field acting in the sample, since it is weighted within the momentum sum in Eq.~\eqref{eq:RashbaHamiltonianManyBody}, so that the overall impact of the local electric field is provided by the momentum integral of the curves in Fig.~\ref{fig:LocalFieldMoS2} times the semiconductor area $\mathcal A$. Typical weighted internal fields are discussed versus applied external fields in Sec.~\ref{sec:Excitonic Rashba Hamiltonian}. Moreover, a distance $l$ between the atomically thin semiconductor and the substrate, which always occurs to some degree in van~der~Waals-bound TMDC samples on substrates on the order of \mbox{$l=0.3$\,nm} \cite{florian2018dielectric,druppel2017diversity}, affects the localization. The larger the distance $l$, the stronger the electric field is localized around zero momentum resulting in a decreased overall impact of the local electric field on the electrons and holes.
Due to the relatively strong localization around small in-plane momenta $\mathbf q$, intervalley scattering occurring due to the derived Rashba interaction, cf.~Eq.~\eqref{eq:RashbaHamiltonianManyBody}, at large valley momenta \mbox{$|\mathbf q|= K = \frac{4\pi}{3a_0} = 12.6\,$nm$^{-1}$} ($K$ valley-momentum transfer with lattice constant $a_0$ \cite{kormanyos2015k}) is strongly suppressed.

\section{Excitonic Rashba Hamiltonian}
\label{sec:Excitonic Rashba Hamiltonian}
To illustrate the basic impact of the spin-flip processes introduced by the charge- and field fluctuations of the dielectric environment, we restrict the analysis to intraband electron-electron, hole-hole and electron-hole scattering mediated by the Rashba Hamiltonian in Eq.~\eqref{eq:RashbaHamiltonianManyBody} in full analogy to the four-particle Coulomb interaction \cite{rohlfing2000electron,qiu2015nonanalyticity}. For optical excitation at the $K$ points, we apply the low-wavenumber approximation to the scattering matrix elements in Eq.~\eqref{eq:RashbaHamiltonianManyBody}, which is valid for small momentum transfers $\mathbf q$, and obtain:
\begin{multline}
    \hat H_{\text{BR}} \\
    = \sum_{\substack{\mathbf k,\mathbf k^{\prime},\mathbf q,\lambda,\lambda^{\prime}\\\xi,\xi^{\prime},s,s^{\prime}}}E_{z,\mathbf q}^{\text{loc}}S_{\mathbf k}^{\lambda,\xi,\bar s,s}
    \adagtwo{\lambda,\mathbf k+\mathbf q}{\xi,\bar s}
    \adagtwo{\lambda^{\prime},\mathbf k^{\prime}-\mathbf q}{\xi^{\prime},s^{\prime}}\andagtwo{\lambda^{\prime},\mathbf k^{\prime}}{\xi^{\prime},s^{\prime}}
    \andagtwo{\lambda,\mathbf k}{\xi,s},
    \label{eq:RashbaHamiltonianManyBodySmallQ}
\end{multline}
where $S_{\mathbf k}^{\lambda,\xi,\bar s,s}\equiv S_{\mathbf k,\mathbf k}^{\lambda,\lambda,\xi,\xi,\bar s,s}$ from Eq.~\eqref{eq:RashbaMatrixElementGeneral}.
In the excitonic picture, using the approach and definitions in Ref.~\cite{katsch2018theory}, the Hamiltonian in Eq.~\eqref{eq:RashbaHamiltonianManyBodySmallQ} 
transforms as follows, cf.~App.~\ref{app:excitonic_hamiltonian}:
\begin{align}
\begin{split}
\label{eq:RashbaHamiltonianIntrinsicExciton}
    &\hat H_{\text{BR-X}} = \sum_{\substack{\mu,\nu,\mathbf Q,\\\xi,\xi^{\prime},s,s^{\prime}}}\left(S_{\mu,\nu,\mathbf Q}^{h,\xi,\xi^{\prime},\bar s,s^{\prime}}\Pdagtwo{\mu,\mathbf Q}{\xi,\xi^{\prime},s,s^{\prime}}\Pndagtwo{\nu,\mathbf Q}{\xi,\xi^{\prime},\bar s,s^{\prime}}\right.\\
     &\,\left.\quad\quad\quad\quad\quad- S_{\mu,\nu,\mathbf Q}^{e,\xi,\xi^{\prime},s,\bar s^{\prime}}\Pdagtwo{\mu,\mathbf Q}{\xi,\xi^{\prime},s,s^{\prime}}\Pndagtwo{\nu,\mathbf Q}{\xi,\xi^{\prime},s,\bar s^{\prime}}\right),
     \end{split}
\end{align}
which closely resembles the Hamiltonian describing the Rashba interaction on interlayer excitons derived in Ref.~\cite{mittenzwey2025ultrafast}. In contrast to our many-body density-independent description in Eq.~\eqref{eq:RashbaHamiltonianIntrinsicExciton}, however, the Rashba interaction in Ref.~\cite{mittenzwey2025ultrafast} originates from a density-dependent Hartree-limit. The latter should naturally emerge from Eq.~\eqref{eq:RashbaHamiltonianManyBody}, if we also included higher-order contributions in the expansion in electron-hole pairs in App.~\ref{app:excitonic_hamiltonian}. In Eq.~\eqref{eq:RashbaHamiltonianIntrinsicExciton}, $\hat P^{(\dagger)}\vphantom{P}_{\mu,\mathbf Q}^{\xi,\xi^{\prime},s,s^{\prime}}$ are the excitonic annihilation (creation) operators, $\mu$ is the excitonic quantum number and $\mathbf Q$ is the center-of-mass momentum. Regarding the index pairs of valley $\xi,\xi^{\prime}$ and spin $s,s^{\prime}$, the first index always belongs to the hole and the second to the electron.

\begin{figure}
    \centering
    \subfigure[]{\includegraphics[width=0.21\linewidth]{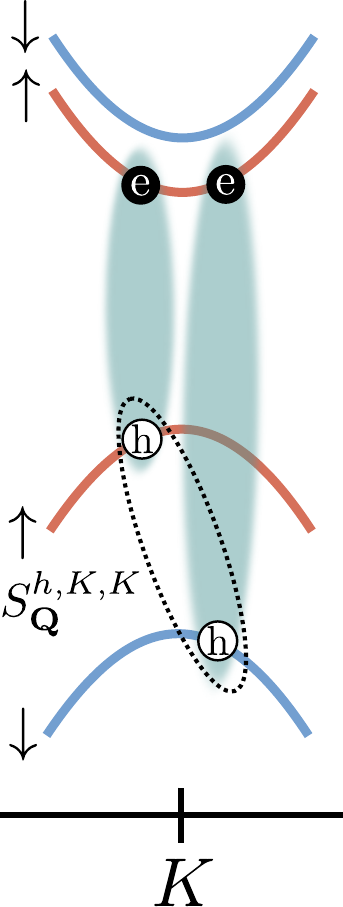}
    \label{fig:ElectronHoleSpinCoupling1}}
    \hspace{0.1cm}
    \subfigure[]{\includegraphics[width=0.21\linewidth]{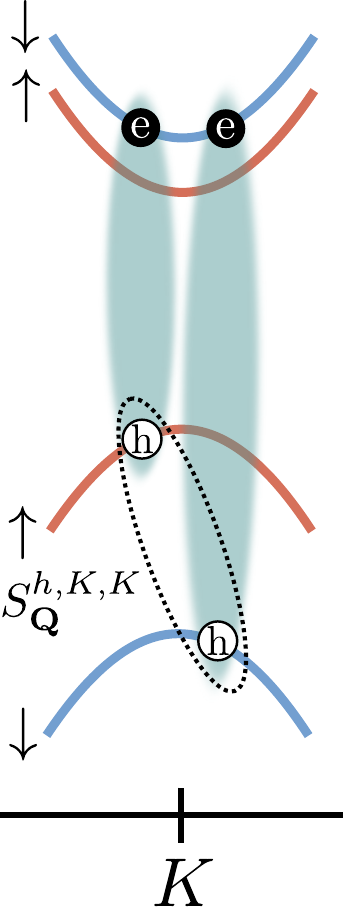}
    \label{fig:ElectronHoleSpinCoupling2}}
    \hspace{0.1cm}
    \subfigure[]{\includegraphics[width=0.21\linewidth]{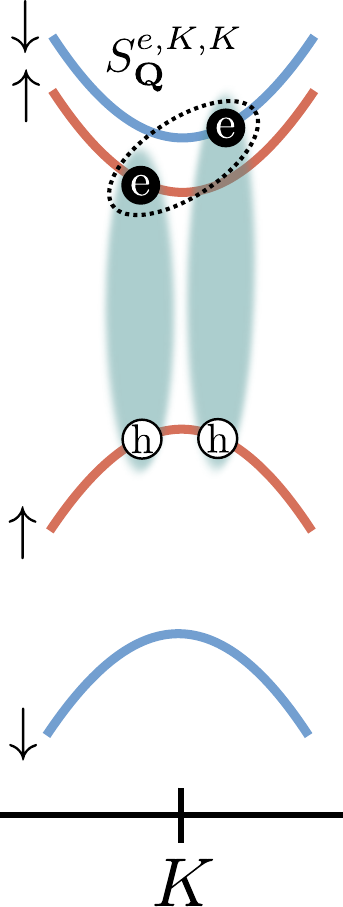}
    \label{fig:ElectronHoleSpinCoupling3}}
    \hspace{0.1cm}
    \subfigure[]{\includegraphics[width=0.21\linewidth]{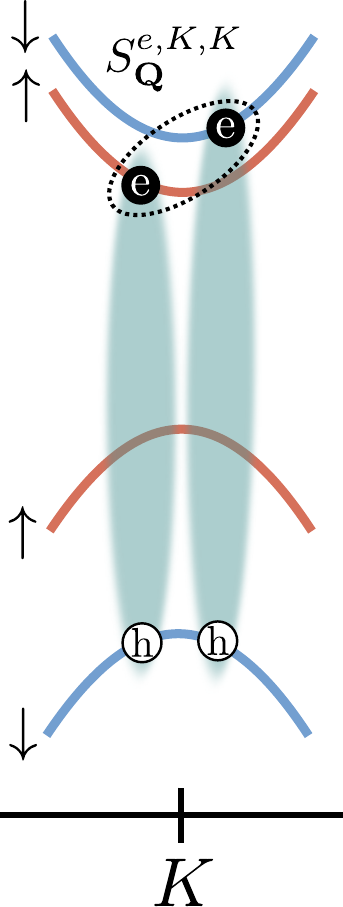}
    \label{fig:ElectronHoleSpinCoupling4}}
    \caption{Scheme of hole spin coupling between spin-bright A and spin-dark B excitons (a), hole spin coupling between spin-dark A and spin-bright B excitons (b), electron spin coupling between spin-bright and spin-dark A excitons (c) and electron spin coupling between spin-dark B and spin-bright B excitons (d) at the $K$ valley induced by the corresponding Rashba coupling matrix elements $S_{\mathbf Q}^{h/e,K,K}$ in Eq.~\eqref{eq:RashbaMatrixElementIntrinsicHole} and Eq.~\eqref{eq:RashbaMatrixElementIntrinsicElectron}.}
    \label{fig:ElectronHoleSpinCoupling}
\end{figure}

The many-particle Rashba Hamiltonian in Eq.~\eqref{eq:RashbaHamiltonianIntrinsicExciton} encodes single hole spin-flip processes between spin-bright A and spin-dark B excitons, cf.\ Fig.~\ref{fig:ElectronHoleSpinCoupling1} and between spin-dark A and spin-bright B excitons, cf.\ Fig.~\ref{fig:ElectronHoleSpinCoupling2}, as well as single electron spin flip processes between spin-bright A and spin-dark A excitons, cf.\ Fig.~\ref{fig:ElectronHoleSpinCoupling3}, and between spin-bright B and spin-dark B excitons, cf.\ Fig.~\ref{fig:ElectronHoleSpinCoupling4}. Note, that, in Fig.~\ref{fig:ElectronHoleSpinCoupling}, we only depict the Rashba intravalley spin-flip processes for intravalley excitons at the $K$ valley with \mbox{$\xi=\xi^{\prime} = K$}, while intravalley spin-flip processes are equally possible for intravalley excitons at the $K^{\prime}$ valley with \mbox{$\xi=\xi^{\prime}=K^{\prime}$} or for any intervalley configuration with \mbox{$\xi\neq\xi^{\prime}$}, e.g., for \mbox{$\xi=K$} and \mbox{$\xi^{\prime}=K^{\prime}$}.

The corresponding excitonic Rashba matrix elements occurring in Eq.~\eqref{eq:RashbaHamiltonianIntrinsicExciton} for single hole spin flips, cf.\ Fig.~\ref{fig:ElectronHoleSpinCoupling1} and Fig.~\ref{fig:ElectronHoleSpinCoupling2}, read:
\begin{multline}
    S_{\mu,\nu,\mathbf Q}^{h,\xi,\xi^{\prime},s,s^{\prime}}
    =  \sum_{\mathbf q,\mathbf q^{\prime},\mathbf q^{\prime\prime}}\alpha_{\text{BR},v}^{\xi}\left( \mathrm i\left(\delta_{\bar s,\uparrow} - \delta_{\bar s,\downarrow}\right)q_x - q_y \right)\\
    \times \left(
    E_{z,\mathbf q-\mathbf q^{\prime}}^{\text{loc},h} \delta_{\mathbf q^{\prime\prime},\mathbf q^{\prime}}
    +
    E_{z,\mathbf q-\mathbf q^{\prime}}^{\text{loc},e} \delta_{\mathbf q^{\prime\prime},\mathbf q}
    +
    E_z^{\text{ext}}\delta_{\mathbf q^{\prime\prime},\mathbf q^{\prime}}^{\mathbf q^{\prime},\mathbf q}
    \right)\\
    \times \ExWFstartwo{\mu,\mathbf q^{\prime\prime}+\beta_{\xi,\xi^{\prime}}^{\bar s,s^{\prime}}\mathbf Q}{\xi,\xi^{\prime},\bar s,s^{\prime}}\ExWFtwo{\nu,\mathbf q^{\prime}+\beta_{\xi,\xi^{\prime}}^{s,s^{\prime}}\mathbf Q}{\xi,\xi^{\prime},s,s^{\prime}},
    \label{eq:RashbaMatrixElementIntrinsicHole}
\end{multline}
the excitonic Rashba matrix elements for single electron spin flips, cf.\ Fig.~\ref{fig:ElectronHoleSpinCoupling3} and Fig.~\ref{fig:ElectronHoleSpinCoupling4}, read:
\begin{multline}
    S_{\mu,\nu,\mathbf Q}^{e,\xi,\xi^{\prime},s,s^{\prime}}
     =  \sum_{\mathbf q,\mathbf q^{\prime},\mathbf q^{\prime\prime}}\alpha_{\text{BR},c}^{\xi^{\prime}}\left( \mathrm i\left(\delta_{s^{\prime},\uparrow} - \delta_{s^{\prime},\downarrow}\right)q_x - q_y \right)\\
     \times \left(E_{z,\mathbf q-\mathbf q^{\prime}}^{\text{loc},h}\delta_{\mathbf q^{\prime\prime},\mathbf q^{\prime}} + E_{z,\mathbf q-\mathbf q^{\prime}}^{\text{loc},e}\delta_{\mathbf q^{\prime\prime},\mathbf q} + E_z^{\text{ext}}\delta_{\mathbf q^{\prime\prime},\mathbf q^{\prime}}^{\mathbf q^{\prime},\mathbf q}\right)\\
     \times \ExWFstartwo{\mu,\mathbf q^{\prime\prime}-\alpha_{\xi,\xi^{\prime}}^{s,\bar s^{\prime}}\mathbf Q}{\xi,\xi^{\prime},s,\bar s^{\prime}}
     \ExWFtwo{\nu,\mathbf q-\alpha_{\xi,\xi^{\prime}}^{s,s^{\prime}}\mathbf Q}{\xi,\xi^{\prime},s,s^{\prime}}.
     \label{eq:RashbaMatrixElementIntrinsicElectron}
\end{multline}
Here, $\alpha_{\text{BR},v/c}^{\xi}$ is the Rashba coupling constant for holes/electrons at high-symmetry point $\xi$ \cite{article:THEORY_Bychkov_Rashba_Coupling_Kormanyos2014}, $\ExWFtwo{\mu,\mathbf q}{\xi,\xi^{\prime}, s,s^{\prime}}$ is the excitonic wave function as a solution of the Wannier equation, cf.~Eq.~\eqref{eq:WannierEquation}, where the first (second) index always belongs to the hole (electron), $\mathbf q$ is the relative momentum and $\alpha_{\xi,\xi^{\prime}}^{s,s^{\prime}}$, $\beta_{\xi,\xi^{\prime}}^{s,s^{\prime}}$ are the effective-mass ratios of the corresponding exciton configuration \cite{katsch2018theory}:
\begin{align}
    \alpha_{\xi,\xi^{\prime}}^{s,s^{\prime}} = \frac{m_e^{\xi^{\prime},s^{\prime}}}{m_h^{\xi,s}+m_e^{\xi^{\prime},s^{\prime}}}, \quad \beta_{\xi,\xi^{\prime}}^{s,s^{\prime}} = \frac{m_h^{\xi,s}}{m_h^{\xi,s}+m_e^{\xi^{\prime},s^{\prime}}},
\end{align}
with effective mass $m_{e/h}^{\xi,s}$ of the involved electron/hole \cite{kormanyos2015k}. 
The use of $\alpha$ and $\beta$ in the transformation to $\mathbf q$ and $\mathbf Q$ avoids mixed terms in the kinetic energy. 
Notably, the local fields $E_{z,\mathbf q}^{\text{loc},i}$ for $i=e,h$ in Eq.~\eqref{eq:RashbaMatrixElementIntrinsicHole} and Eq.~\eqref{eq:RashbaMatrixElementIntrinsicElectron} are defined as:
\begin{align}
    E_{z,\mathbf q}^{\text{loc},i} = (\delta_{i,e}-\delta_{i,h}) E_{z,\mathbf q}^{\text{loc}},
    \label{eq:LocalElectricField_i}
\end{align}
with $E_{z,\mathbf q}^{\text{loc}}$ from Eq.~\eqref{eq:ElectricFieldLocalExplicit}. In Eq.~\eqref{eq:LocalElectricField_i}, $i$ denotes the (quasi-) particle, which \textit{induces} the field, which is not necessarily equivalent to the (quasi-) particle the field \textit{acts} upon. 
We rewrote Eq.~\eqref{eq:RashbaMatrixElementIntrinsicHole} and Eq.~\eqref{eq:RashbaMatrixElementIntrinsicElectron} in a way that they can address a more general situation than the dielectric-environment-induced field fluctuations by including a spatially homogeneous external electric field $E_z^{\text{ext}}$ arising from electrical gating. This way, the excitonic Rashba Hamiltonian in Eq.~\eqref{eq:RashbaHamiltonianIntrinsicExciton} combines two physically distinct fields, a local electric field $E_{z,\mathbf q}^{\text{loc},i}$ due to many-particle interactions inducing PSOI and an arbitrary in-plane spatially homogeneous external electric field $E_z^{\text{ext}}$ inducing SOI, in one consistent description. 
The external electric field $E_z^{\text{ext}}$ and the local electric fields $E_{z,\mathbf q}^{\text{loc},i}$ acting on the corresponding electron/hole may amplify or counteract each other, depending on the dielectric environment and the direction of $E_z^{\text{ext}}$.

\begin{figure}
    \centering
    \subfigure[]{\includegraphics[width=0.48\linewidth]{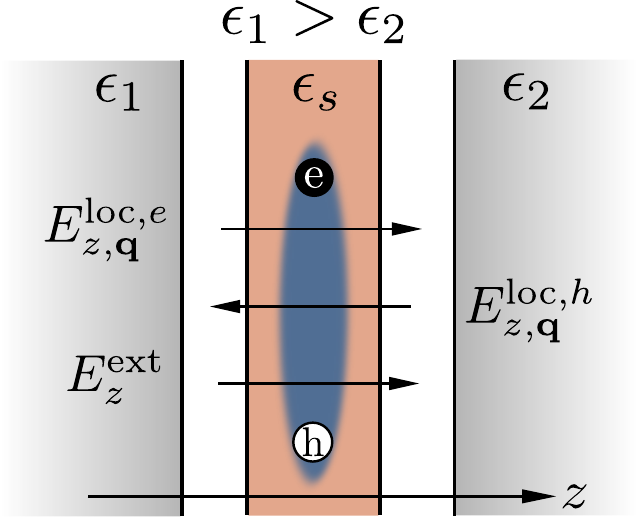}
    \label{fig:LocalFieldInfluenceExciton1}}
    \subfigure[]{\includegraphics[width=0.48\linewidth]{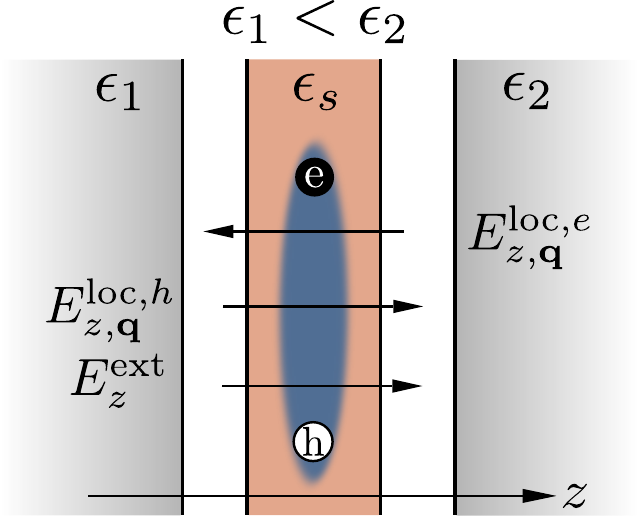}
    \label{fig:LocalFieldInfluenceExciton2}}
    \caption{Alignment of the respective local electric field $E_{z,\mathbf q}^{\text{loc},i}$ induced by the individual hole ($i=h$) and electron ($i=e$) of an exciton (shaded ellipse) in a thin semiconductor $\epsilon_s$ for different dielectric environments $\epsilon_1$ and $\epsilon_2$. $E_z^{\text{ext}}$ is an external electric field chosen as polarized in positive $z$-direction.}
    \label{fig:LocalFieldInfluenceExciton}
\end{figure}

To examine this aspect in more detail, in Fig.~\ref{fig:LocalFieldInfluenceExciton}, we depict the alignment of the local electric fields $E_z^{\text{loc},i}$ for $i=e,h$ induced by the corresponding electron ($e$) or hole ($h$) of the exciton, cf.~Eq.~\eqref{eq:LocalElectricField_i}, in comparison to an arbitrary external electric field $E_z^{\text{ext}}$ polarized in positive $z$-direction. We observe, that the individual local fields induced by the negatively charged electrons $E_z^{\text{loc},e}$ and positively charged holes $E_z^{\text{loc},h}$ point in opposite directions. Even though these local fields are of fully quantum-mechanical origin, as they represent electron-hole interactions and self-energy corrections arising from Fock contributions, cf.~App.~\ref{app:excitonic_hamiltonian}, they are in line with a classical interpretation: Within a classical picture, negative (positive) charges located in the TMDC-monolayer region $\epsilon_s$ induce net electric fields pointing towards a positive (negative) $z$-direction if $\epsilon_1 > \epsilon_2$ (Fig.~\ref{fig:LocalFieldInfluenceExciton}(a)) and vice versa if $\epsilon_1<\epsilon_2$ (Fig.~\ref{fig:LocalFieldInfluenceExciton}(b)) as a usual behavior caused by the electrostatic boundary conditions, where high-$\epsilon$ regions accumulate surface charges at the boundaries. Hence, an external electric field can amplify or suppress the total field acting on the electron or hole depending on its direction. However, care must be taken regarding specific values, since the interaction mechanisms between electrons and holes regarding specific local fields $E_z^{\text{loc},i}$ differ: Electron- (hole-) induced local electric fields $E_z^{\text{loc},e/h}$ interact with electrons (holes) via self-interaction and with holes (electrons) via electron-hole interaction, which are not the same.

\begin{figure}[h!]
\includegraphics[width=1\linewidth]{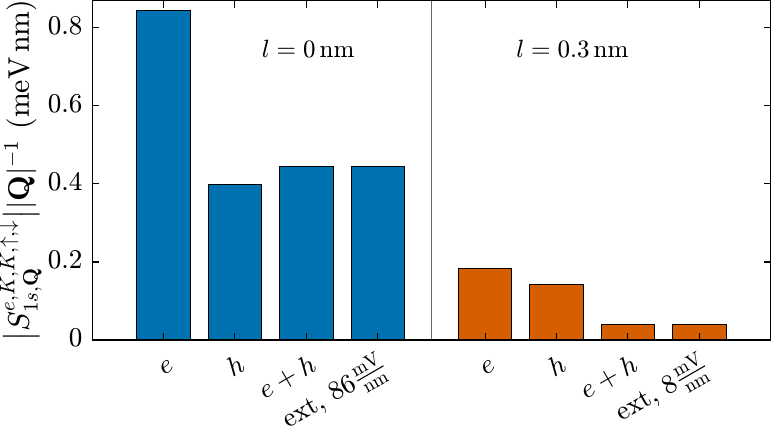}
\caption{Excitonic Rashba matrix element due to local electron-induced fields $E_{z,\mathbf q}^{\text{loc},e}$ (``$e$''), local hole-induced fields $E_{z,\mathbf q}^{\text{loc},h}$ (``$h$'') and total local fields (``$e+h$''), cf.~Eq.~\eqref{eq:LocalElectricField_i}, from Eq.~\eqref{eq:RashbaMatrixElementIntrinsicElectron} for interlayer distances $l$ of 0\,nm and 0.3\,nm (blue and red coloring) in a MoSe$_2$ monolayer on a sapphire substrate. ``ext'' denotes the external contribution from Eq.~\eqref{eq:RashbaMatrixElementIntrinsicElectron}, where the magnitude of the external field $E_z^{\text{ext}}$ is chosen to match the total local-field interaction strength (``$e+h$'').}
\label{fig:SXeLocalFieldContributions}
\end{figure}

In Fig.~\ref{fig:SXeLocalFieldContributions}, we depict the individual contributions via self-interaction (``$e$'') with local fields induced by the electrons themselves and electron-hole interaction (``$h$'') with local fields induced by the holes and their combined contribution (``$e+h$'') for electron spin flips in terms of the Rashba energy length:
\begin{align}
    \text{Energy length} = \big|S_{1s,\mathbf Q}^{e,K,K,\uparrow,\downarrow}\big| |\mathbf Q|^{-1},
    \label{eq:rashba_energy_length}
\end{align}
for two different interlayer distances $l$ of 0\,nm and 0.3\,nm. ``ext'' denotes the Rashba energy length due to an externally applied electric field tuned to match the total local-field contribution (``$e+h$''). Since the absolute value of the Rashba matrix element (energy) $\big|S_{1s,\mathbf Q}^{e,K,K,\uparrow,\downarrow}\big|$ is well linear over a wide range of center-of-mass momenta $\mathbf Q$, we can safely use the energy length, i.e., the Rashba matrix element divided by the center-of-mass momentum, cf.~Eq.~\eqref{eq:rashba_energy_length}, 
for comparison. 
We stress that comparing the individual field strengths of the local field $E_{z,\mathbf q}^{\text{loc},i}$ and an externally applied electric field $E_z^{\text{ext}}$ is meaningless, as their net influence on the spin hybridization of the involved excitons for a given field strength is quite different. 
Hence, the only meaningful comparison is obtained via the corresponding excitonic Rashba matrix elements directly. 
In Fig.~\ref{fig:SXeLocalFieldContributions}, the electron-electron self-energy contribution (``$e$'') outcompetes the hole-electron contribution (``$h$''), so that the overall Rashba energy length (``$e+h$'') is even slightly larger than the individual hole-electron contribution at $l=0\,$nm and is equivalent to an external electric field of $E_z^{\text{ext}} = 86\,$mV\,nm$^{-1}$. 
This value, while already substantial, is much lower than the \mbox{$\mathbf q=\mathbf 0$}-values of the local electric field $E_{z,\mathbf q}^{\text{loc}}$ at approximately 1\,V\,nm$^{-1}$ in Fig.~\ref{fig:LocalFieldMoS2}. 
At larger van~der~Waals-distances of $l=0.3\,$nm \cite{florian2018dielectric,druppel2017diversity}, not only both individual contributions decrease, but also their difference shrinks, so that the overall Rashba energy length is greatly reduced and corresponds to an equivalent external electric field of $E_z^{\text{ext}} = 8\,$mV\,nm$^{-1}$. 
Note that, for the conditions in Fig.~\ref{fig:SXeLocalFieldContributions}, to counteract (amplify) the local-field interaction, the external field has to be applied in the negative (positive) $z$-direction if $\epsilon_1 > \epsilon_2$, as the self-energy contribution caused by $E^{\text{loc},e}_z$, which points in positive $z$-direction, cf.~Fig.~\ref{fig:LocalFieldInfluenceExciton}, always dominates. For $\epsilon_1 < \epsilon_2$, vice versa.

\begin{figure}[h!]
\includegraphics[width=1\linewidth]{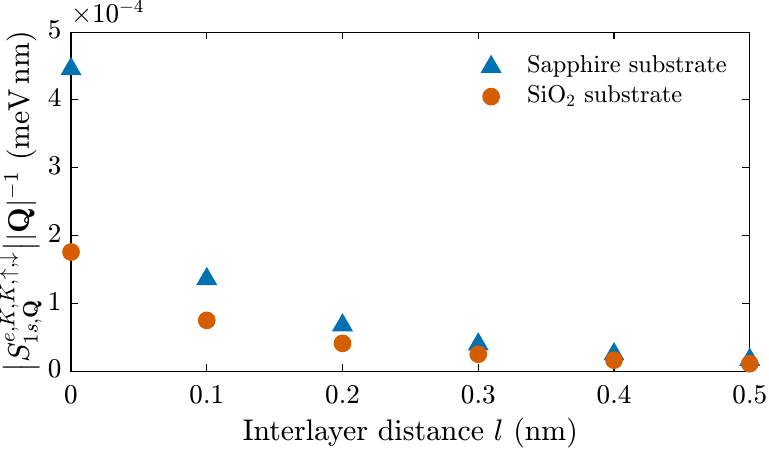}
\caption{Excitonic Rashba matrix element with a local electric field $E_{z,\mathbf q}^{\text{loc}}$ from Eq.~\eqref{eq:RashbaMatrixElementIntrinsicElectron} for different interlayer distances $l$ in a MoSe$_2$ monolayer on a sapphire substrate (blue triangles) and on a SiO$_2$ substrate (red circles).}
\label{fig:SXeLocalFieldDistance}
\end{figure}

\begin{figure}[h!]
\begin{tabular}{cc}
    \begin{tabular}{c}
    \hspace{-0.2cm}
        \includegraphics[width=0.487\linewidth]{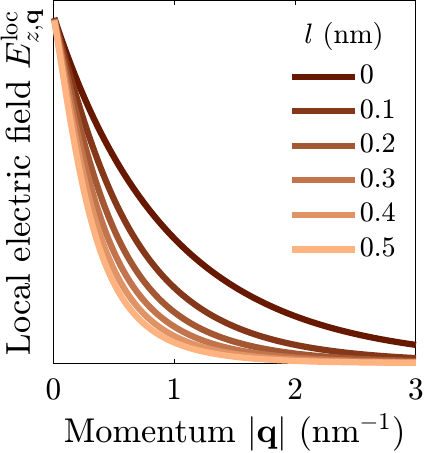}
    \end{tabular}
    \begin{tabular}{c}
    \vspace{0.23cm}
    \hspace{-0.25cm}
        \includegraphics[width=0.495\linewidth]{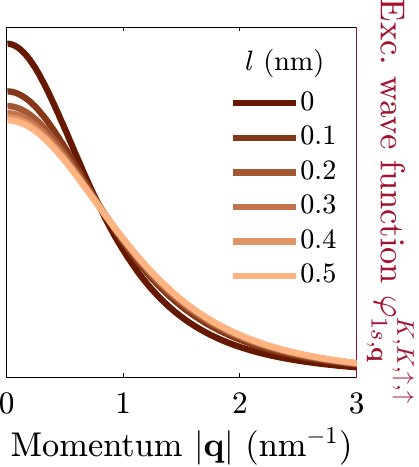}
    \end{tabular}
\end{tabular}
\caption{Local electric field from Eq.~\eqref{eq:ElectricFieldLocalExplicit} (left) and excitonic wave functions obtained by solving Eq.~\eqref{eq:WannierEquation} (right) for a MoSe$_2$ monolayer on a sapphire substrate for different interlayer distances $l$.}
\label{fig:LocalFieldWaveFunctionDistances}
\end{figure}

In Fig.~\ref{fig:SXeLocalFieldDistance}, we depict the (total) Rashba energy length, cf.~Eq.~\eqref{eq:rashba_energy_length}, of a local electric field from Eq.~\eqref{eq:RashbaMatrixElementIntrinsicElectron} depending on the interlayer distance $l$ for a MoSe$_2$ monolayer on a sapphire substrate (blue triangles) and on a SiO$_2$ substrate (red circles), cf.~also ``$e+h$''-case in Fig.~\ref{fig:SXeLocalFieldContributions}. 
Overall, as already indicated by Fig.~\ref{fig:SXeLocalFieldContributions}, the magnitude of the excitonic Rashba matrix element for a local electric field decreases by increasing interlayer separation, cf.~Fig.~\ref{fig:SXeLocalFieldDistance}. This behavior stems from the fact, that a larger interlayer distance $l$ leads to a stronger localization of the local electric field $E_{z,\mathbf q}^{\text{loc}}$ in momentum space, cf.~Fig.~\ref{fig:LocalFieldWaveFunctionDistances}(left), which entails a weaker localization in real space, i.e., the increasing interlayer distance gradually causes the local electric field being less spatially inhomogeneous. This, in turn, reduces the overlap with the excitonic wave functions $\ExWFtwo{\mu,\mathbf q}{K,K,s,s^{\prime}}$ on the one hand, cf.\ Eq.~\eqref{eq:RashbaMatrixElementIntrinsicElectron}, which exhibit an opposite behavior, i.e., a weaker localization in momentum space, cf.~Fig.~\ref{fig:LocalFieldWaveFunctionDistances}(right), with respect to an increasing interlayer separation due to decreased substrate screening, and, on the other hand, it causes the individual competing contributions $E_{z,\mathbf q}^{\text{loc},e}$ and $E_{z,\mathbf q}^{\text{loc},h}$ in Eq.~\eqref{eq:RashbaMatrixElementIntrinsicElectron} (and similarly regarding hole spin flips in Eq.~\eqref{eq:RashbaMatrixElementIntrinsicHole}) to approach each other, since they exactly compensate each other in the limit of a spatially homogeneous local electric field $E_{z,\mathbf q}^{\text{loc}}\rightarrow E_{z}^{\text{loc}}\delta_{\mathbf q,\mathbf 0}$, where it holds:
\begin{align}
    S_{\mu,\nu,\mathbf Q}^{e/h,\xi,\xi^{\prime},s,s^{\prime}}\Big|_{E_{z,\mathbf q}^{\text{loc}}\rightarrow E_{z}^{\text{loc}}\delta_{\mathbf q,\mathbf 0}} \rightarrow 0.
\end{align}

Hence, the most relevant tuning knobs of the local-field driven spin-orbit interaction are the dielectric mismatch between superstrate $\epsilon_1$ and substrate $\epsilon_2$ as well as the van~der~Waals-distance $l$. While the former can be relatively easy adjusted by exfoliating or growing the monolayer TMDC on the desired substrate, tuning the latter involves, e.g., applying pressure in the out-of-plane direction \cite{steeger2023pressure}.

\section{Equations of Motion for Exciton-Phonon Scattering}
\label{sec:Equations of Motion for Exciton-Phonon Scattering}
To activate the Rashba spin-flip process, the occurrence of excitonic occupations $\Nex{\mu,\mathbf Q}$ at \mbox{$\mathbf Q\neq\mathbf 0$} is necessary, cf.\ Eq.~\eqref{eq:RashbaMatrixElementIntrinsicHole} and Eq.~\eqref{eq:RashbaMatrixElementIntrinsicElectron}. 
For this, we choose exciton-phonon scattering as the most relevant scattering mechanism to generate
incoherent excitonic occupations $\Nex{\mu,\mathbf Q}$ \cite{selig2018dark,brem2020phonon,thranhardt2000quantum,katzer2023exciton}:
\begin{align}
    \Nextwo{\mu,\mathbf Q}{\xi,\xi^{\prime},s,s^{\prime}} = \left\langle \Pdagtwo{\mu,\mathbf Q}{\xi,\xi^{\prime},s,s^{\prime}}\Pndagtwo{\mu,\mathbf Q}{\xi,\xi^{\prime},s,s^{\prime}}\right\rangle_{\text{c}},
\end{align}
at above-cryogenic temperatures after optical excitation. 
To formulate the theory in an appropriate basis for exciton-phonon scattering, we perform a diagonalization of the excitonic Rashba Hamiltonian from Eq.~\eqref{eq:RashbaHamiltonianIntrinsicExciton} into the spin-diagonal basis \mbox{$s,s^{\prime}\rightarrow \mathcal S$} analog to Ref.~\cite{mittenzwey2025ultrafast}. Here, $\mathcal S$ is the excitonic spin-diagonal quantum number, which emerges by diagonalizing the excitonic spin eigenvalue equation. For details, we refer to App.~\ref{app:SpinEigenvalueEquation}. 
By adding the exciton-phonon Hamiltonian, the Boltzmann-type equations \cite{katzer2023exciton,selig2018dark} in the spin-hybrid basis for \mbox{$\mu=1s$} excitonic states can be derived \cite{mittenzwey2025ultrafast}:

\begin{align}
\begin{split}
    \partial_t\Nextwo{\mathbf Q}{\xi,\xi^{\prime},\mathcal S} = &\, \sum_{\substack{\mathbf K,\mathcal S^{\prime},\\\xi^{\prime\prime},\xi^{\prime\prime\prime}}}\left(\Gamma^{\text{in},\xi^{\prime\prime},\xi^{\prime\prime\prime},\mathcal S^{\prime},\xi,\xi^{\prime},\mathcal S}_{\mathbf K,\mathbf Q}\Nextwo{\mathbf K}{\xi^{\prime\prime},\xi^{\prime\prime\prime},\mathcal S^{\prime}}\right.\\
    &\,\quad\quad\quad -\left. \Gamma^{\text{out},\xi^{\prime\prime},\xi^{\prime\prime\prime},\mathcal S^{\prime},\xi,\xi^{\prime},\mathcal S}_{\mathbf K,\mathbf Q}\Nextwo{\mathbf Q}{\xi,\xi^{\prime},\mathcal S}\right).
    \end{split}
    \label{eq:Boltzmann_Equation_SpinHybridBasis}
\end{align}
Here, $\Gamma^{\text{in}/\text{out},j,i}_{\mathbf K,\mathbf Q}$ denote the phonon-assisted scattering rates between states $\mathbf Q,i$ and $\mathbf K,j$, which encode intra-
(\mbox{$\xi^{\prime\prime}=\xi$} and \mbox{$\xi^{\prime\prime\prime}=\xi^{\prime}$})
as well as intervalley
(\mbox{$\xi^{\prime\prime}\neq\xi$} or \mbox{$\xi^{\prime\prime\prime}\neq\xi^{\prime}$})
scattering of the corresponding holes and electrons, respectively, forming the exciton. All phonon-related parameters such as phonon modes and electron-phonon coupling elements in deformation potential approximation for the corresponding materials are taken from DFT calculations \cite{li2013intrinsic,jin2014intrinsic}. 
Due to Rashba coupling-induced mixing of electron/hole spins in the conduction/valence bands, reflected by the excitonic spin-diagonal quantum number $\mathcal S$, phonon-assisted scattering in Eq.~\eqref{eq:Boltzmann_Equation_SpinHybridBasis} does not conserve the spin anymore and phonon-assisted spin relaxation can occur.
We note, that the excitonic spin-diagonal quantum number $\mathcal S$ corresponds to the old spins $s,s^{\prime}$ at zero center-of-mass momentum \mbox{$\mathbf Q\rightarrow \mathbf 0$} or in case of a vanishing out-of-plane electric field \mbox{$E_{z,\mathbf q}^{\text{tot}} \rightarrow 0$}: \mbox{$\mathcal S\rightarrow s,s^{\prime}$}.

From the spin-hybrid excitonic occupations $\Nextwo{\mathbf Q}{\xi,\xi^{\prime},\mathcal S}$ obtained by solving Eq.~\eqref{eq:Boltzmann_Equation_SpinHybridBasis}, the total momentum-integrated excitonic occupations in the old spin basis $\Nextwo{}{\xi,\xi^{\prime},s,s^{\prime}}$ can be retrieved:
\begin{align}
    \Nextwo{}{\xi,\xi^{\prime},s,s^{\prime}} = \frac{1}{\mathcal{A}}\sum_{\mathbf Q,\mathcal S}\left|\mathcal C_{\mathbf Q,\mathcal S}^{\xi,\xi^{\prime},s,s^{\prime}}\right|^2\Nextwo{\mathbf Q}{\xi,\xi^{\prime},\mathcal S}.
    \label{eq:NTotalOldBasis}
\end{align}
Here, $\mathcal C_{\mathbf Q,\mathcal S}^{\xi,\xi^{\prime},s,s^{\prime}}$ are the excitonic spin eigenfunctions solving Eq.~\eqref{eq:SpinEigenvalueEquation}, cf.\ App.~\ref{app:SpinEigenvalueEquation}.

The aim of this work is to present and quantify this new many-body Rashba spin-flip mechanism, which arises as a further effect on top of all other known spin-flip effects. Hence, we leave the analysis of the interplay of Rashba single spin-flip mechanisms with other Coulomb-scattering mechanisms such as double spin-flip intervalley exchange \cite{selig2020suppression,selig2019ultrafast} or spin-conserving intervalley Coulomb scattering such as Dexter coupling \cite{berghauser2018inverted,dogadov2026diss} for future work. 
However, the theory can be straightforwardly extended to account for these mechanisms as well, similar to the exciton-phonon interaction described above.
Also, for a full picture, density-dependent mechanisms such as Auger scattering \cite{erkensten2021dark,steinhoff2021microscopic} need to be considered. In this contribution however, we stay in linear order of the exciton density $N$.

In the following, we examine the local-electric-field-induced phonon-assisted spin relaxation in one example atomically thin semiconductor, a monolayer MoSe$_2$.

\section{Spin Relaxation in Monolayer \protect{M\lowercase{o}S\lowercase{e}$_2$}}
\label{sec:Spin Relaxation in Monolayer MoSe$_2$}

\begin{figure}
    \centering
    \includegraphics[width=1\linewidth]{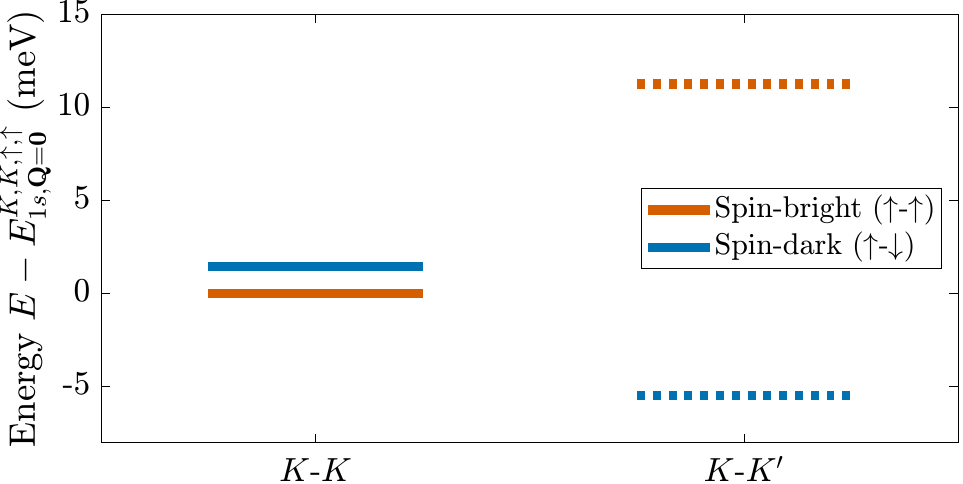}
    \caption{Energies at zero COM $\mathbf Q=\mathbf 0$ for the four considered exciton configurations of a MoSe$_2$ monolayer on a sapphire substrate.}
    \label{fig:excitonic_energies}
\end{figure}

We start the analysis with a monolayer MoSe$_2$ on a sapphire substrate, where the momentum-direct spin-dark exciton is energetically located at approximately $1.45\,$meV \cite{lu2019magnetic,robert2020measurement} above the optically addressable spin-bright exciton. This small bright-dark splitting is a direct consequence of excitonic effects, as the bare conduction band splitting is on the order of 20\,meV \cite{kormanyos2015k}. Within the used effective-mass approach, this implies a short-range intravalley exchange blue shift \cite{qiu2015nonanalyticity,yu2014valley} of the spin-bright momentum-direct, i.e., intravalley $K$-$K$ A$_{1s}$ exciton of 5.5\,meV, from which we can obtain an upper estimate of the short-range intravalley exchange blue shift of the spin-bright momentum-indirect, i.e., intervalley $K$-$K^{\prime}$ exciton of 9.8\,meV \cite{li2022intervalley,yang2022relaxation,he2020valley} following the procedure in Ref.~\cite{dogadov2026diss}. Note, that in our simulations we assume a fixed bright-dark splitting, which is experimentally known from h-BN-encapsulated MoSe$_2$ monolayer samples \cite{lu2019magnetic,robert2020measurement}, irrespective of the chosen substrate material, which might not always be valid. Here, \textit{ab initio} calculations \cite{echeverry2016splitting}, which properly take into account substrate effects with meV precision, are needed. 
However, the average environmental dielectric constants $\frac{1}{2}(\epsilon_1+\epsilon_2)$ for h-BN encapsulation ($\epsilon_1=\epsilon_2=4.8$ \cite{latini2015excitons}) and sapphire substrates ($\epsilon_1 = \sqrt{11.6\cdot 9.4}$ \cite{fontanella1974low} and $\epsilon_2=1$) are almost equal, so that we expect a similar net screening and, hence, a similar Coulomb exchange interaction in both cases. Note that, on a SiO$_2$ substrate, monolayer MoSe$_2$ preserves its bright ground state \cite{molas2017brightening}. In Fig.~\ref{fig:excitonic_energies}, we depict the energies at zero center-of-mass momentum of the four considered exciton species.

Due to the small bright-dark splitting of the $K$-$K$ excitons allowing efficient spin hybridization, a monolayer MoSe$_2$ is a promising candidate for efficient excitonic Rashba spin relaxation induced by out-of-plane electric fields. In our case, the electric field is a local field $\mathbf E_{z,\mathbf q}^{\text{loc}}$ self-consistently induced by the spatial asymmetries of the dielectric environment of the MoSe$_2$ monolayer sandwiched between vacuum and a SiO$_2$ substrate, cf.\ Eq.~\eqref{eq:ElectricFieldLocalExplicit}.

\begin{figure}[h!]
\includegraphics[width=1\linewidth]{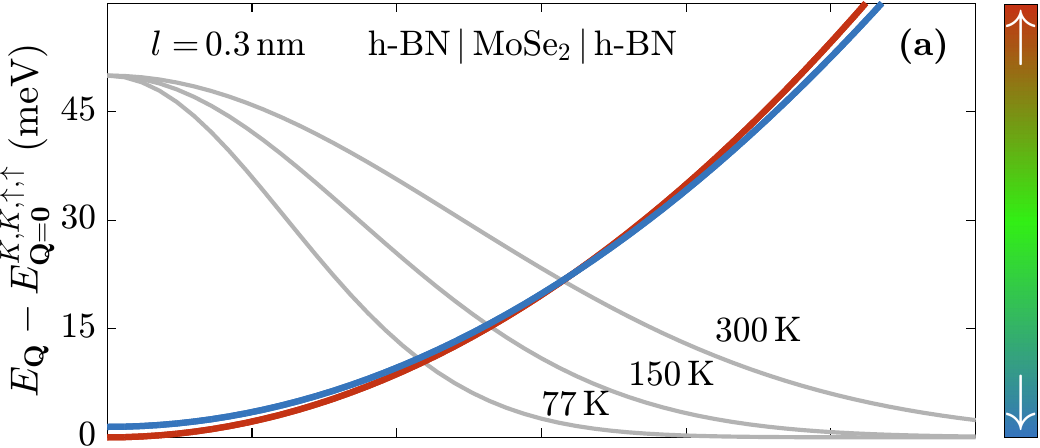}
    \includegraphics[width=1\linewidth]{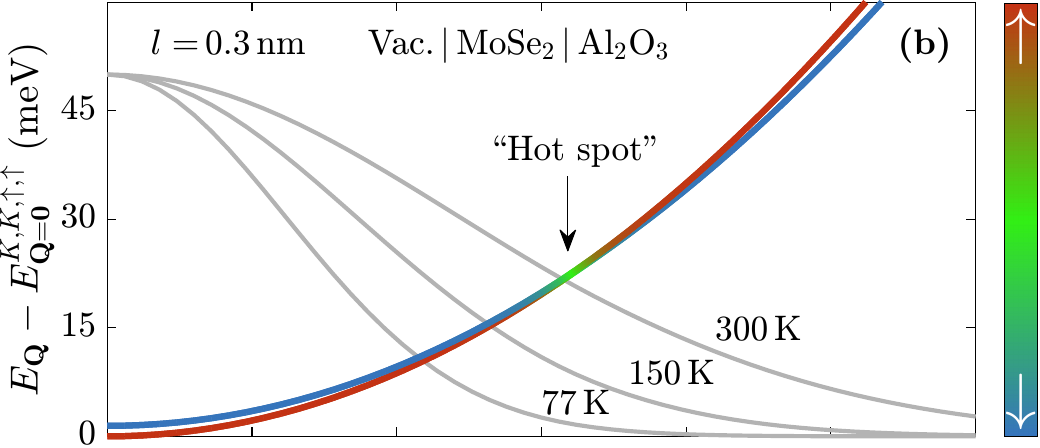}
    \includegraphics[width=1\linewidth]{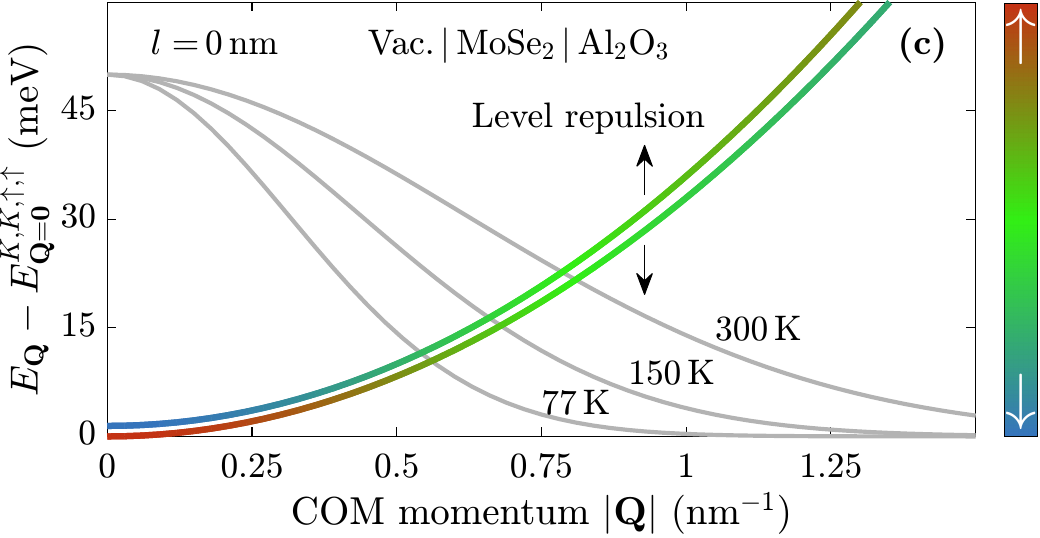}
\caption{Dispersions of the spin-bright $E_{\mathbf Q}^{K,K,\uparrow,\uparrow}$ (red solid line) and spin-dark $E_{\mathbf Q}^{K,K,\uparrow,\downarrow}$ (blue solid line) excitons at the $K$ valley in a MoSe$_2$ monolayer encapsulated in h-BN with $l=0.3\,$nm (a), and on a sapphire substrate for $l=0.3\,$nm (b) and $l=0\,$nm (c), cf.\ also Fig.~\ref{fig:MoSe2onSiO2ScatteringSketch}. The green color gradient denotes the degree of the electron spin hybridization via a local electric field. Grey solid lines denote Boltzmann distributions of the spin-bright excitonic occupation at various temperatures.}
\label{fig:MoSe2onSiO2Dispersion}
\end{figure}

In Fig.~\ref{fig:MoSe2onSiO2Dispersion}, we depict the full momentum-dependent exciton dispersions of the lowest spin-bright and spin-dark excitons at the $K$ valley, where the green color gradient denotes the degree of the spin hybridization of the electronic part of the excitons. Due to the slightly smaller exciton mass of the spin-bright exciton and the small bright-dark splitting, a hybridization ``hot spot'' \cite{yang2020exciton} around a COM momentum of 0.8\,nm$^{-1}$ occurs. At this region, the initially non-hybridized bands, cf.~Fig.~\ref{fig:MoSe2onSiO2Dispersion}(a) (h-BN-encapsulated MoSe$_2$ monolayer, symmetric dielectric environment) cross, which translates into an anti-crossing, i.e., level-repulsion behavior, as soon as a spin-hybridizing electric field is present, cf.~Fig.~\ref{fig:MoSe2onSiO2Dispersion}(b) (MoSe$_2$ monolayer on a sapphire substrate with $l=0.3\,$nm, asymmetric dielectric environment), which increases with increasing local electric field, cf.~Fig.~\ref{fig:MoSe2onSiO2Dispersion}(c) (MoSe$_2$ monolayer on a sapphire substrate with $l=0\,$nm). The grey solid lines denote Boltzmann distributions at various temperatures, which are important for the interpretation of the exciton dynamics later on. 

\begin{figure}[h!]
\includegraphics[width=0.32\linewidth]{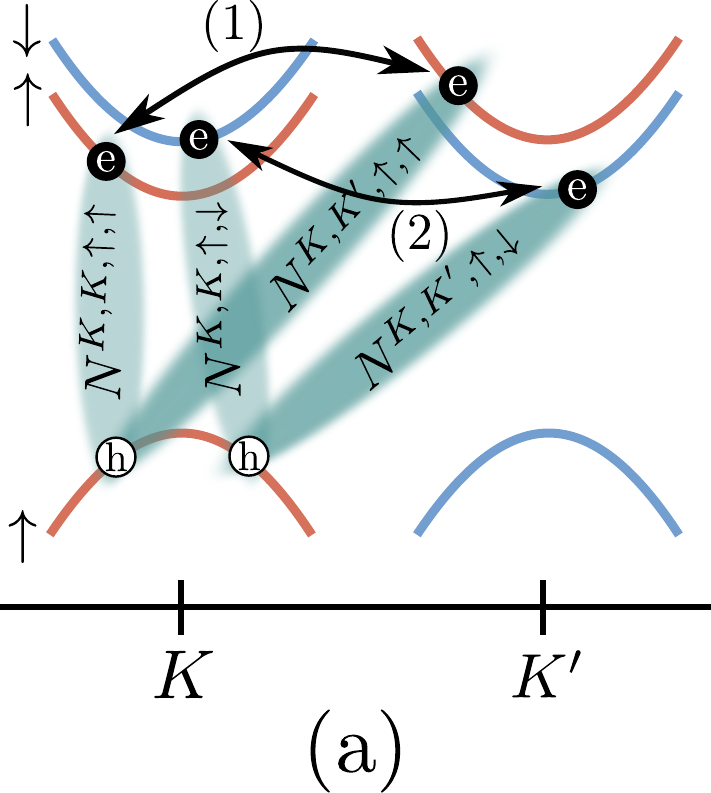}
\includegraphics[width=0.32\linewidth]{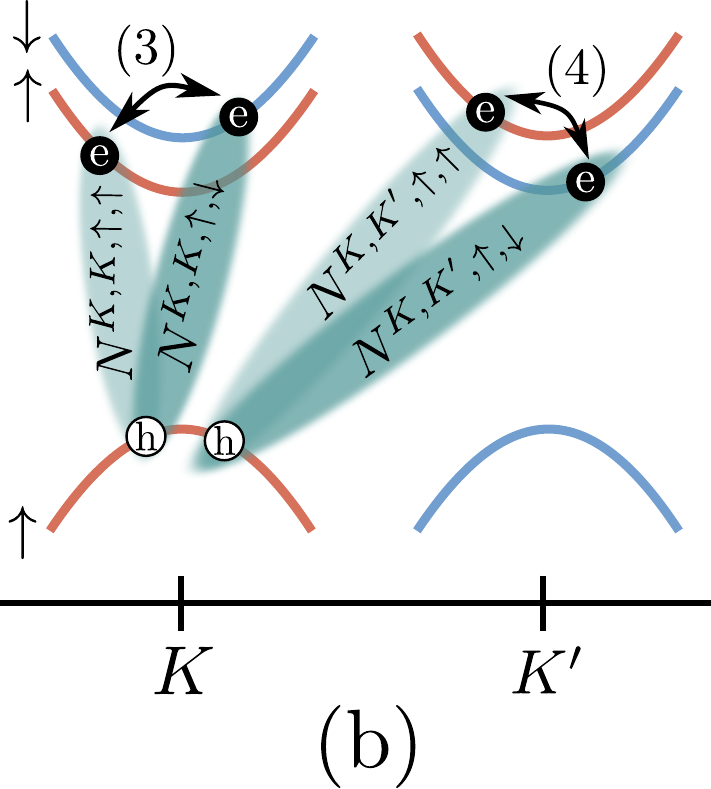}
\includegraphics[width=0.32\linewidth]{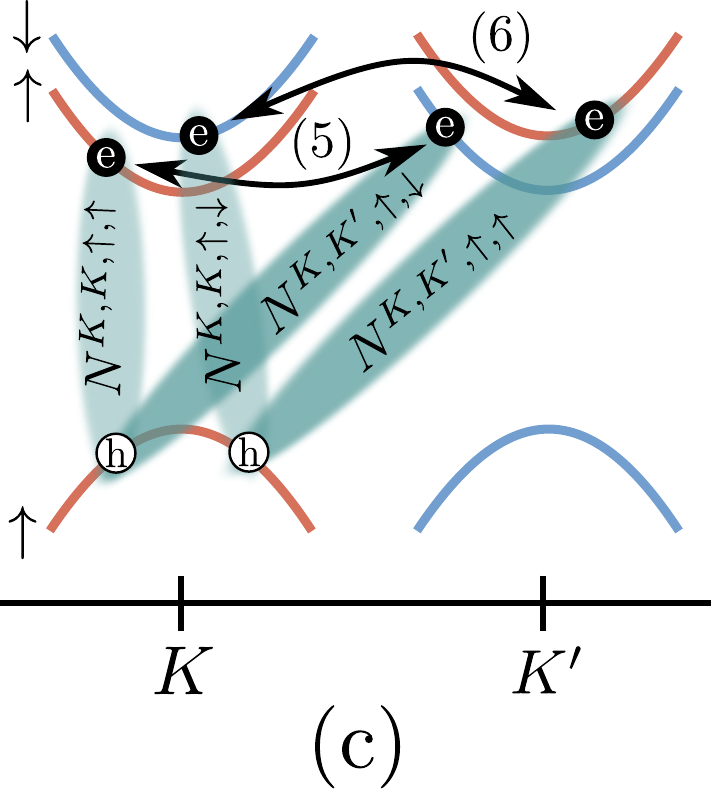}
\caption{Phonon-assisted spin-conserving intervalley scattering (a), spin-nonconserving intravalley scattering (b) and spin-nonconserving intervalley scattering (c) in a MoSe$_2$ monolayer.}
\label{fig:MoSe2onSiO2ScatteringSketch}
\end{figure}

In the numerical simulations, we take into account spin-bright and spin-dark intra- ($\Nextwo{\mathbf Q}{K,K,\uparrow,\uparrow}$, $\Nextwo{\mathbf Q}{K,K,\uparrow,\downarrow}$) and intervalley ($\Nextwo{\mathbf Q}{K,K^{\prime},\uparrow,\uparrow}$, $\Nextwo{\mathbf Q}{K,K^{\prime},\uparrow,\downarrow}$) excitonic occupations, cf.\ also Fig.~\ref{fig:excitonic_energies} and Fig.~\ref{fig:MoSe2onSiO2ScatteringSketch}, in a spin-hybrid description, cf.\ Eq.~\eqref{eq:Boltzmann_Equation_SpinHybridBasis}. We neglect any hole spin flips, as they are suppressed due to the large A-B splitting. We note, that Rashba coupling between distinct valleys $K$ and $K^{\prime}$ (intervalley spin hybridization) is also strongly suppressed, cf.~the discussion to Fig.~\ref{fig:LocalFieldMoS2}, and therefore neglected. However, intervalley phonon-assisted spin flips are still possible due to the softening of the spin selection rules of the exciton-phonon scattering via intravalley spin hybridization.

As the initial condition for our numerical simulations, we assume a normalized spin-bright excitonic occupation at the $K$ valley within the light cone:
\begin{align}
    \Nextwo{\mathbf Q}{K,K,\uparrow,\uparrow}(t=0) = \delta_{\mathbf Q,\mathbf 0},
    \label{eq:initial_conditions}
\end{align}
which roughly approximates the situation after optically exciting the A$_{1s}$ excitonic transition at the $K$ valley with $\sigma_+$-light.

\begin{figure}[h!]
\begin{tabular}{cc}
    \begin{tabular}{c}
      \begin{turn}{90} Exciton density\end{turn}
      \end{tabular}
\begin{tabular}{c}
\hspace{-2mm}\includegraphics[width=0.9\linewidth]{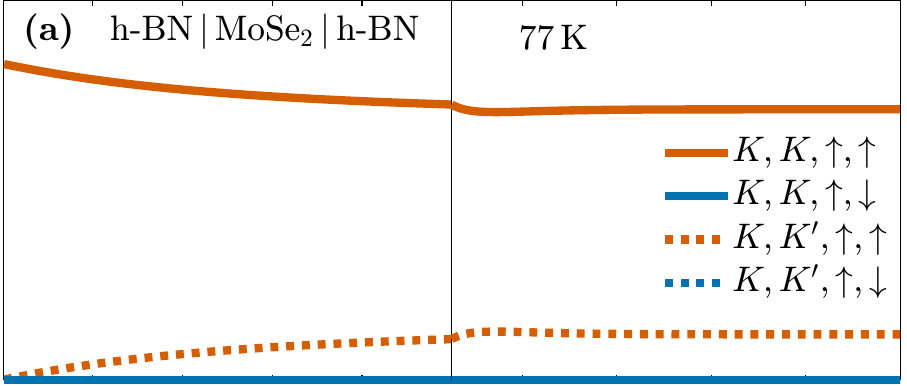}\\[-1mm]
\hspace{-2mm}\includegraphics[width=0.9\linewidth]{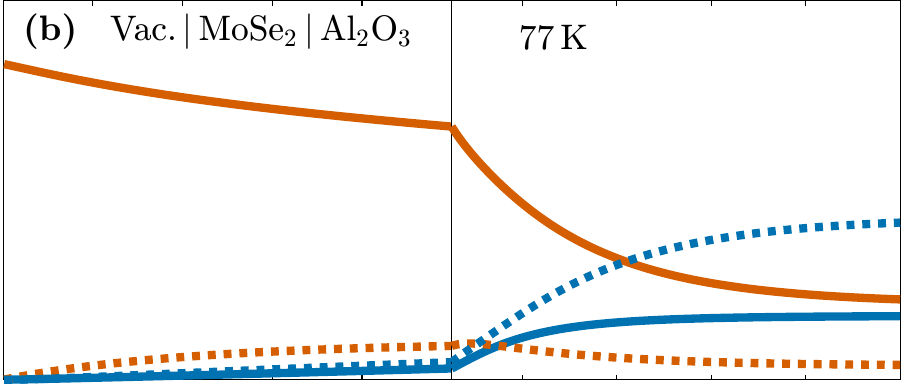}\\[-1mm]
\hspace{-2mm}\includegraphics[width=0.9\linewidth]{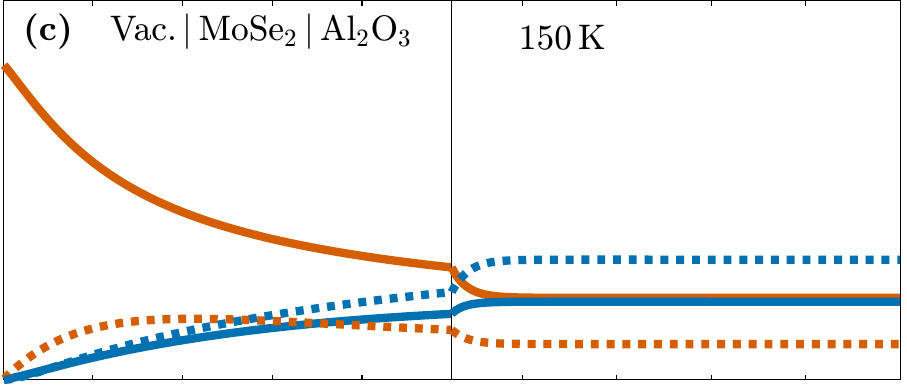}\\[-1mm]
\hspace{-1.5mm}\includegraphics[width=0.913\linewidth]{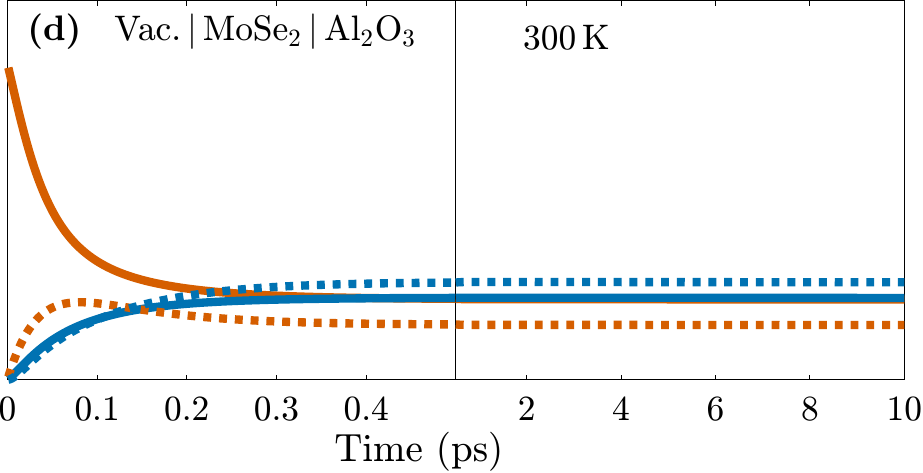}
\end{tabular}
\end{tabular}
\caption{Dynamics of the total excitonic occupations $\Nextwo{}{\xi,\xi^{\prime},s,s^{\prime}}$ from Eq.~\eqref{eq:NTotalOldBasis} for a MoSe$_2$ monolayer encapsulated in h-BN (a) with $\epsilon_1=\epsilon_2=4.8$ \cite{latini2015excitons} and on a sapphire substrate (b,c,d) with $\epsilon_1 = \sqrt{11.6\cdot 9.4}$ \cite{fontanella1974low} and $\epsilon_2=1$ with $l=0.3\,$nm at various temperatures.}
\label{fig:MoSe2onSiO2Dynamics}
\end{figure}

In Fig.~\ref{fig:MoSe2onSiO2Dynamics}, we display the momentum-integrated excitonic occupations $\Nextwo{}{\xi,\xi^{\prime},s,s^{\prime}}$ from Eq.~\eqref{eq:NTotalOldBasis}, which are obtained by projecting the spin-hybrid solutions of Eq.~\eqref{eq:Boltzmann_Equation_SpinHybridBasis} into the old spin basis, for a MoSe$_2$ monolayer encapsulated in h-BN yielding a zero local field \mbox{$E_{z,\mathbf q}^{\text{loc}}=0$} at \mbox{$T=77\,$K}, Fig.~\ref{fig:MoSe2onSiO2Dynamics}(a), and for a MoSe$_2$ monolayer on a sapphire substrate yielding a nonzero local field \mbox{$E_{z,\mathbf q}^{\text{loc}}\neq 0$} at increasing temperatures, Fig.~\ref{fig:MoSe2onSiO2Dynamics}(b--d). We do not apply any external fields, i.e., \mbox{$E_{z}^{\text{ext}}=0$} always holds. We note, that the total exciton density $N = \sum_{\xi,\xi^{\prime},s,s^{\prime}} \Nextwo{}{\xi,\xi^{\prime},s,s^{\prime}}$ is always conserved. 
At initial time, the optically addressable, momentum-direct spin-bright excitonic occupation is populated (red solid line), which reflects our initial conditions in Eq.~\eqref{eq:initial_conditions}.

First, we consider the zero-electric-field case in a symmetric dielectric structure (\mbox{$E_{z,\mathbf q}^{\text{loc}}=0$}, Eq.~\eqref{eq:LocalElectricField_i}) depicted in Fig.~\ref{fig:MoSe2onSiO2Dynamics}(a). Here, the spin-bright excitonic intravalley occupation $\Nextwo{}{K,K,\uparrow,\uparrow}$ scatters into the spin-bright intervalley occupation $\Nextwo{}{K,K^{\prime},\uparrow,\uparrow}$ (red dotted line) via spin-conserving phonon-assisted intervalley scattering $\Nextwo{}{K,K,\uparrow,\uparrow}\rightarrow\Nextwo{}{K,K^{\prime},\uparrow,\uparrow}$, cf.\ process (1) in Fig.~\ref{fig:MoSe2onSiO2ScatteringSketch}(a) within 6\,ps at 77\,K. Spin-dark excitonic occupations remain zero.

Next, in Fig.~\ref{fig:MoSe2onSiO2Dynamics}(b), we allow for a nonzero local electric field $E_{z,\mathbf q}^{\text{loc}}\neq 0$ in Eq.~\eqref{eq:LocalElectricField_i}, which emerges due to the asymmetric dielectric structure. This way, next to spin-conserving intervalley scattering, phonon-assisted single spin-nonconserving intravalley scattering into $\Nextwo{}{K,K,\uparrow,\downarrow}$ (blue solid line) within 6\,ps is activated, cf.\ process (3) in Fig.~\ref{fig:MoSe2onSiO2ScatteringSketch}(b). 
Additionally, spin-dark intervalley occupations $\Nextwo{}{K,K^{\prime},\uparrow,\downarrow}$ (blue dotted line) emerge via two paths: On the one hand, it involves two two-step processes. The first two-step process occurs via intervalley spin-conserving scattering into $\Nextwo{}{K,K^{\prime},\uparrow,\uparrow}$, cf.\ process (1) in Fig.~\ref{fig:MoSe2onSiO2ScatteringSketch}(a), and a subsequent intravalley spin-nonconserving scattering into $\Nextwo{}{K,K^{\prime},\uparrow,\downarrow}$, cf.\ process (4) in Fig.~\ref{fig:MoSe2onSiO2ScatteringSketch}(b). The second occurring two-step process involves intravalley spin-nonconserving scattering first, cf.\ process (3) in Fig.~\ref{fig:MoSe2onSiO2ScatteringSketch}(b), and intervalley spin-conserving scattering afterwards, cf.\ process (2) in Fig.~\ref{fig:MoSe2onSiO2ScatteringSketch}(a). On the other hand, spin-nonconserving intervalley scattering from $\Nextwo{}{K,K,\uparrow,\uparrow}$ into $\Nextwo{}{K,K^{\prime},\uparrow,\downarrow}$ also occurs as a one-step process due to the combined action of valley-momentum phonons and spin hybridization, cf.\ process (5) in Fig.~\ref{fig:MoSe2onSiO2ScatteringSketch}(c). However, even though both scattering paths contribute, the first path is significantly more efficient. Similarly, process (6) in Fig.~\ref{fig:MoSe2onSiO2ScatteringSketch}(c) is also of minor importance.

Overall, the spin-dark excitonic occupations $\Nextwo{}{K,K,\uparrow,\downarrow}$ and $\Nextwo{}{K,K^{\prime},\uparrow,\downarrow}$ rise on a similar timescale as the spin-conserving intervalley scattering, cf.\ Fig.~\ref{fig:MoSe2onSiO2Dynamics}(a), until an overall equilibrium after 8\,ps is reached. The reason for this comparably fast equilibration is the small excitonic bright-dark splitting of 1.45\,meV, which renders spin hybridization and therefore spin relaxation very efficient. Note that, comparing the zero-field case in Fig.~\ref{fig:MoSe2onSiO2Dynamics}(a) to the nonzero-field case in Fig.~\ref{fig:MoSe2onSiO2Dynamics}(b) at an equal temperature of 77\,K, the efficiency of the spin-conserving intervalley scattering channel, cf.~process (1) in Fig.~\ref{fig:MoSe2onSiO2ScatteringSketch}(a), does not change in the presence of a spin-hybridizing local field, as the ratio between $\Nextwo{}{K,K,\uparrow,\uparrow}$ and $\Nextwo{}{K,K^{\prime},\uparrow,\uparrow}$ is approximately preserved.

At higher temperatures such as 150\,K in Fig.~\ref{fig:MoSe2onSiO2Dynamics}(c) and 300\,K in Fig.~\ref{fig:MoSe2onSiO2Dynamics}(d), the overall spin-conserving intervalley and spin-nonconserving intra- as well as intervalley scattering dramatically increases: At 150\,K an overall equilibrium is reached after 1\,ps, which speeds up to 0.2\,ps at 300\,K. This overall speed-up occurs due to thermal activation of phonons and a thermalization of excitonic occupations. The former increases the induced absorption and emission from and into the phonon bath at increasing temperature and the latter results in a temperature-dependent broadening of the (Boltzmann) distribution in momentum space $\mathbf Q$, cf.\ grey solid lines in Fig.~\ref{fig:MoSe2onSiO2Dispersion}. The latter phenomenon is especially responsible for a speed-up of the spin relaxation, as excitonic states at larger center-of-mass momenta $\mathbf Q$ become increasingly populated, where the spin relaxation is more efficient 
due to the momentum-dependent spin coupling and, most importantly, due to an increasing probability of scattering via the ``hot spot'', cf.\ Fig.~\ref{fig:MoSe2onSiO2Dispersion}. In Tab.~\ref{tab:spin_relaxation_rates}, we depict the explicit spin relaxation times, which are obtained by extracting the risetime of the spin-dark intravalley occupations at the $K$ valley from the numerical simulations in Fig.~\ref{fig:MoSe2onSiO2Dynamics}.

\begin{table}[]
    \centering
    \begin{tabular}{l|lll}
      & 77\,K & 150\,K & 300\,K\\
     \hline
     MoSe$_2$ on Al$_2$O$_3$ ($l=0.3\,$nm) & 1.9\,ps &  0.28\,ps & 76\,fs \\
     MoSe$_2$ on Al$_2$O$_3$ ($l=0\,$nm) & 0.37\,ps &  62\,fs & 14\,fs \\
     MoSe$_2$ on SiO$_2$ ($l=0.3\,$nm) & 2.6\,ps &  0.35\,ps  &  99\,fs \\
     MoSe$_2$ on SiO$_2$ ($l=0\,$nm) & 0.57\,ps &  77\,fs  &  16\,fs \\
     MoS$_2$ on Al$_2$O$_3$ ($l=0.3\,$nm) & 14\,ns &  4.5\,ns & 0.58\,ns \\
     MoS$_2$ on Al$_2$O$_3$ ($l=0\,$nm) & 0.17\,ns &  50\,ps & 10\,ps
    \end{tabular}
    \caption{Spin relaxation times for spin flips of intravalley excitons for MoSe$_2$ and MoS$_2$ monolayers on a sapphire and SiO$_2$ substrate in units of ps$^{-1}$. The spin relaxation times are obtained by extracting the risetime of spin-dark intravalley occupations at the $K$ valley from numerical simulations.}
    \label{tab:spin_relaxation_rates}
\end{table}

However, care should be taken in comparing specific numbers to distinct experiments, as a specific dielectric environment has been chosen for the calculation. Also, the bright-dark splitting and van~der~Waals distance plays a critical role. To illustrate this, we have also performed numerical simulations for vanishing van~der~Waals distances and a MoS$_2$ monolayer for the same and distinct selected dielectric environments. The extracted spin relaxation times are displayed in Tab.~\ref{tab:spin_relaxation_rates} and show a dramatic decrease by decreasing van~der~Waals distance, which is expected, cf.~Fig.~\ref{fig:SXeLocalFieldContributions}. For a MoS$_2$ monolayer within the conditions in Fig.~\ref{fig:MoSe2onSiO2Dynamics}, the spin relaxation times are 14\,ns at 77\,K, which speed up to 0.58\,ns at 300\,K. Compared to a MoSe$_2$ monolayer, they are orders of magnitude slower, which is a direct consequence of the larger bright-dark splitting of 14\,meV \cite{robert2020measurement} compared to 1.45\,meV in MoSe$_2$ rendering the spin hybridization much less efficient. Other monolayer materials such as WSe$_2$ or WSe$_2$ behave similar, as their bright-dark splitting is even larger (40--55\,meV \cite{molas2019probing,mapara2022bright,molas2017brightening,zhou2017probing}). In these materials, other spin relaxation mechanisms such as spin-nonconserving scattering via chiral phonons \cite{lagarde2024efficient,chan2025exciton,zhang2022ab} are expected to dominate over a Rashba PSOI induced by spatial asymmetries of the dielectric environment. To clarify the role of chiral phonons \cite{lagarde2024efficient,chan2025exciton,zhang2022ab} compared Rashba PSOI in monolayer MoSe$_2$, future work may shed more light on this issue. However, in TMDC heterostructures such as MoSe$_2$/MoS$_2$ bilayers, intra-exciton fields due to charge-separated interlayer exciton densities can induce a significant many-body Rashba effect \cite{mittenzwey2025ultrafast}. Moreover, for a quantitative comparison in real-world experiments, double spin flips \cite{maialle1993exciton,yu2014valley,selig2019ultrafast}, which depend on the specific material \cite{selig2020suppression,dogadov2026diss}, have to be additionally considered in the analysis. 
Thus, the exact choice of the material and dielectric environment conditions determine, which spin relaxation mechanism is dominant.

\section{Conclusion}
In this work, we established a Rashba single spin-flip mechanism due to local out-of-plane electric fields self-consistently induced by spatial asymmetries in the dielectric environment in atomically thin semiconductors. We developed an excitonic many-body description, explicitly calculated the local electric fields for a given dielectric environment and quantified the Rashba coupling strength. For a monolayer MoSe$_2$, we demonstrated very efficient spin relaxation on a sub-picosecond (150\,K) to a hundred-femtosecond (300\,K) timescale via a phonon-assisted single electron spin-flip channel.

\acknowledgments

Funded by the Deutsche Forschungsgemeinschaft (DFG, German Research Foundation) -- Project No.\ 420760124 (H.M., A.K.); 556436549 (A.K.). H.M.\ acknowledges funding by project 21209528 (``proof of trust'').

We thank Michiel Snoeken (Technische Universität Berlin) 
for discussions and Malte Selig for inspiration.

\section*{Data Availability Statement}
The data supporting the findings of this work are available from the authors within reasonable request.

\FloatBarrier
\clearpage
\onecolumngrid

\appendix

\setcounter{table}{0}
\renewcommand{\thetable}{A\Roman{table}}

\section{Derivation of the Many-Particle Rashba Hamiltonian}
\label{app:RashbaHamiltonianDerivation}
Spin-orbit Hamiltonian $H_{\text{BR}}$ in "first quantization", cf.~Eq.~\eqref{eq:Hamiltonian_Rashba_General} from the main text \cite{book:Spin_orbit_coupling_Winkler2003}:
\begin{align}
    H_{\text{BR}}^{} = -\frac{\alpha_{\text{BR}}}{\hbar}\boldsymbol{\sigma}\cdot\left(\mathbf E(\mathbf r,z)\times\mathbf p\right).
    \label{eq:Hamiltonian_Rashba_GeneralApp}
\end{align}
The second-quantized Hamiltonian $\hat H_{\text{BR}}$ is obtained via an expansion in Bloch wave functions:
\begin{align}
\label{eq:WaveFunctionElectronic}
\psi_{\mathbf k,\lambda}^{s}(\mathbf r,z) = \frac{1}{\sqrt{\mathcal{A}}}\mathrm e^{\mathrm i\mathbf k\cdot\mathbf r}u_{\lambda,\mathbf k}^{}(\mathbf r)\zeta(z)\chi_s,
\end{align}
and reads:
\begin{align}
    \hat H_{\text{BR}} = \sum_{\substack{\mathbf k_1,\mathbf k_2,\lambda_1,\lambda_2,\\s_1,s_2}}
    S_{\mathbf k_1,\mathbf k_2,\lambda_1,\lambda_2}^{s_1,s_2}
    \adagtwo{\mathbf k_1,\lambda_1}{s_1}\andagtwo{\mathbf k_2,\lambda_2}{s_2},
\end{align}
where the Rashba matrix element is given by:
\begin{align}
    S_{\mathbf k_1,\mathbf k_2,\lambda_1,\lambda_2}^{s_1,s_2}=\int\mathrm d^3r\,{\psi}^{*}_{}{\vphantom{\psi}}_{\mathbf k_1,\lambda_1}^{s_1}(\mathbf r)H_{\text{BR}}^{}\psi{\vphantom{\psi}}_{\mathbf k_2,\lambda_2}^{s_2}(\mathbf r).
    \label{eq:RashbaMatrixElementStart}
\end{align}
With the wave functions from Eq.~\eqref{eq:WaveFunctionElectronic}, it follows:
\begin{align}
\begin{split}
    S_{\mathbf k_1,\mathbf k_2,\lambda_1,\lambda_2}^{s_1,s_2}
    =&\,-\frac{\alpha_{\text{BR}}}{\hbar}\frac{1}{\mathcal A}\int\mathrm d^2r\,\mathrm dz\,\mathrm{e}^{-\mathrm i\mathbf k_1\cdot\mathbf r}u^*_{}\vphantom{u}_{\mathbf k_1,\lambda_1}^{s_1}(\mathbf r)\zeta_{}^{*}(z)\chi_{s_1}^{*}\hat E_z^{}(\mathbf r,z)\left(\sigma_yp_x-\sigma_xp_y\right)
    \mathrm{e}^{\mathrm i\mathbf k_2\cdot\mathbf r}u_{\mathbf k_2,\lambda_2}^{s_2}(\mathbf r)\zeta_{}^{}(z)\chi_{s_2}^{},
\end{split}
\end{align}
with the Rashba coupling constant $\alpha_{\text{BR}}$ in units of $[\alpha_{\text{BR}}]=e\,\mathrm{nm}^2$.
Writing the electric field in Fourier representation $\hat E_z^{}(\mathbf r,z)=\frac{1}{\mathcal A}\sum_{\mathbf q,\mathbf G}\mathrm e^{\mathrm i(\mathbf q+\mathbf G)\cdot\mathbf r}\hat E_{z,\mathbf q+\mathbf G}^{}(z)$ \cite{wiser1963dielectric}, where $\mathbf q$ is a crystal momentum within the first Brillouin zone and $\mathbf G$ is a reciprocal lattice vector, and identifying the quantum-confined electric field as:
\begin{align}
\hat E_{z,\mathbf q}=\frac{1}{\mathcal A}\int\,\mathrm dz\,\hat E_{z,\mathbf q}(z)|\zeta(z)|^2,
\label{eq:EzQuantumConfined}
\end{align}
yields:
\begin{align}
\begin{split}
S_{\mathbf k_1,\mathbf k_2,\lambda_1,\lambda_2}^{s_1,s_2}=
    -\frac{\alpha_{\text{BR}}}{\hbar}\sum_{\mathbf q,\mathbf G}\frac{1}{\mathcal A}\hat E_{z,\mathbf q+\mathbf G}
    \int\mathrm d^2r\,\mathrm{e}^{\mathrm i(\mathbf q+\mathbf G)\cdot\mathbf r}\mathrm{e}^{-\mathrm i\mathbf k_1\cdot\mathbf r}u^*_{}\vphantom{u}_{\mathbf k_1,\lambda_1}^{s_1}(\mathbf r) 
    \chi_{s_1}^{*}\left(\sigma_yp_x-\sigma_xp_y\right)\chi_{s_2}^{}\mathrm{e}^{\mathrm i\mathbf k_2\cdot\mathbf r}
    u_{\mathbf k_2,\lambda_2}^{s_2}(\mathbf r).
    \end{split}
\end{align}
Letting the momentum operator act on the Bloch wave function yields:
\begin{multline}
S_{\mathbf k_1,\mathbf k_2,\lambda_1,\lambda_2}^{s_1,s_2}\\
= -\frac{\alpha_{\text{BR}}}{\hbar}\sum_{\mathbf q,\mathbf G}\frac{1}{\mathcal A}\hat E_{z,\mathbf q+\mathbf G}
    \int\mathrm d^2r\,\mathrm{e}^{-\mathrm i(\mathbf k_1-\mathbf k_2-\mathbf q+\mathbf G)\cdot\mathbf r}u^*_{}\vphantom{u}_{\mathbf k_1,\lambda_1}^{s_1}(\mathbf r) 
    \chi_{s_1}^{*}\left(\sigma_y(\hbar k_{2,x}+p_x)-\sigma_x(\hbar k_{2,y}+p_y)\right)\chi_{s_2}^{}
    u_{\mathbf k_2,\lambda_2}^{s_2}(\mathbf r).
\end{multline}
Expanding the integral over the unit cells yields:
\begin{multline}
S_{\mathbf k_1,\mathbf k_2,\lambda_1,\lambda_2}^{s_1,s_2}=
    -\frac{\alpha_{\text{BR}}}{\hbar}\sum_{\mathbf q,\mathbf G}\frac{1}{\mathcal A}\hat E_{z,\mathbf q+\mathbf G}
    \sum_{\mathbf R}\mathrm{e}^{-\mathrm i(\mathbf k_1-\mathbf k_2-\mathbf q-\mathbf G)\cdot\mathbf R}\\
    \times\int_{\mathcal A_{\text{UC}}}\mathrm d^2r\,\mathrm{e}^{-\mathrm i(\mathbf k_1-\mathbf k_2-\mathbf q-\mathbf G)\cdot\mathbf r}u^*_{}\vphantom{u}_{\mathbf k_1,\lambda_1}^{s_1}(\mathbf r) 
    \chi_{s_1}^{*}\left(\sigma_y(\hbar k_{2,x}+p_x)-\sigma_x(\hbar k_{2,y}+p_y)\right)\chi_{s_2}^{}
    u_{\mathbf k_2,\lambda_2}^{s_2}(\mathbf r).
\end{multline}
The lattice sum yields:
\begin{align}
    \sum_{\mathbf R}\mathrm{e}^{-\mathrm i(\mathbf k_1-\mathbf k_2-\mathbf q)\cdot\mathbf R} = N\delta_{\mathbf k_1,\mathbf k_2+\mathbf q},
\end{align}
where 
$N$ is the total number of unit cells. Note, that $\mathrm e^{\mathrm i\mathbf G\cdot\mathbf R}=1$. Considering further the action of the Pauli matrices on the spin eigenfunctions:
\begin{align}
    \sigma_x\chi_s^{} = \chi_{\bar s}^{},\quad \sigma_y\chi_s^{} = \mathrm i\left(\delta_{s,\uparrow}-\delta_{s,\downarrow}\right)\chi_{\bar s}^{},
\end{align}
yields:
\begin{multline}
S_{\mathbf k_1,\mathbf k_2,\lambda_1,\lambda_2}^{s_1,s_2}=
    -\frac{\alpha_{\text{BR}}}{\hbar}\sum_{\mathbf q,\mathbf G}\hat E_{z,\mathbf q+\mathbf G}
    \delta_{\mathbf k_1,\mathbf k_2+\mathbf q}\delta_{s_1,\bar s_2}\\
    \times\frac{1}{\mathcal A_{\text{UC}}}\int_{\mathcal A_{\text{UC}}}\mathrm d^2r\,\mathrm{e}^{\mathrm i\mathbf G\cdot\mathbf r}u^*_{}\vphantom{u}_{\mathbf k_2+\mathbf q,\lambda_1}^{\bar s_2}(\mathbf r) 
    \left(\mathrm i\left(\delta_{s,\uparrow}-\delta_{s,\downarrow}\right)(\hbar k_{2,x}+p_x)-(\hbar k_{2,y}+p_y)\right)
    u_{\mathbf k_2,\lambda_2}^{s_2}(\mathbf r).
    \label{eq:rashba_matrix_element_all_umklapp}
\end{multline}
Identifying the momentum matrix element as:
\begin{align}
    \mathbf p_{\mathbf k_1,\mathbf k_2,\lambda_1,\lambda_2}^{s_1,s_2}(\mathbf G) = \frac{1}{\mathcal A_{\text{UC}}}\int_{\mathcal A_{\text{UC}}}\, u^*_{}\vphantom{u}_{\lambda_1,\mathbf k_1}^{s_1}(\mathbf r)\mathrm{e}^{\mathrm i\mathbf G\cdot\mathbf r}\left(\hbar\mathbf k_2+\mathbf p\right)u_{\lambda_2,\mathbf k_2}^{s_2}(\mathbf r),
    \label{eq:momentum_matrix_element}
\end{align}
expanding over the band extrema with $\mathbf k\rightarrow \mathbf k+\mathbf K^{\xi}$ and $\mathbf q\rightarrow \mathbf q+\mathbf K^{\xi^{\prime}}$ and neglecting 
Umklapp processes with $\mathbf G\neq\mathbf 0$ yields Eq.~\eqref{eq:RashbaHamiltonian} in the main text.

\FloatBarrier
\section{Calculation of the Local Electric Field}
\label{app:ElectricField}
Starting point is the generalized Poisson equation for the Coulomb potential $\hat \phi(\mathbf r)$ \cite{pasenow2005excitonic,zimmermann2016poisson}:
\begin{align}
	\nabla\cdot\left(\epsilon(\mathbf r)\nabla\hat \phi(\mathbf r)\right)=-\hat{\rho}(\mathbf r),
    \label{eq:poisson_equation_generalized}
\end{align}
where $\epsilon(\mathbf r)$ is the spatially dependent dielectric function of the background and $\hat{\rho}(\mathbf r)$ is the active charge density. Redefining $\mathbf r\equiv \left(\mathbf r,z\right)$ and applying a Fourier transformation for the in-plane coordinates leads to:
\begin{align}
	-|\mathbf q|^2\hat{\phi}_{\mathbf q}(z)+\frac{1}{\epsilon_{\parallel}(z)}\partial_z\left(\epsilon_{\bot}(z)\partial_z\hat{\phi}_{\mathbf q}(z)\right)=-\frac{1}{\epsilon_{\parallel}(z)}\hat{\rho}_{\mathbf q}(z),
\end{align}
where we explicitly take into account distinct in-plane ($\epsilon_{\parallel}$) and out-of-plane ($\epsilon_{\perp}$) dielectric functions. In Eq.~\eqref{eq:poisson_equation_generalized} the inhomogeneity is caused by the operator-valued charge density from Eq.~\eqref{eq:ChargeDensityOperator} in the main text.
We reformulate the Poisson equation for the Green's function $G_{\mathbf q}(z,z^{\prime})$, defined by $\hat{\phi}_{\mathbf q}(z)=\int_{-\infty}^{\infty}\mathrm dz^{\prime}\,G_{\mathbf q}(z,z^{\prime})\hat{\rho}_{\mathbf q}(z^{\prime})$:
\begin{align}
	-|\mathbf q|^2G_{\mathbf q}(z,z^{\prime})+\frac{1}{\epsilon_{\parallel}(z)}\partial_z\left(\epsilon_{\bot}(z)\partial_zG_{\mathbf q}(z,z^{\prime})\right)=-\frac{1}{\epsilon_{\parallel}(z)}\delta(z-z^{\prime}).
\end{align}
By defining the Poisson equation piecewise in a geometry of five dielectric slabs (we assume a small interlayer gap $l$ (van~der~Waals-distance) between the TMDC monolayer and substrate and superstrate \cite{florian2018dielectric,druppel2017diversity}, cf.~Fig.~\ref{fig:structure} in the main text), where the inhomogeneity is located in the middle layer at $i=s$:
\begin{align}
	\label{eq:greens_function_equation_volumes_inhom}
    -|\mathbf q|^2G_{i,j,\mathbf q}(z,z^{\prime}) + \partial_z^2G_{i,j,\mathbf q}(z,z^{\prime})=-\frac{1}{\epsilon_{i}}\delta(z-z^{\prime})\delta_{i,j}^{i,s},
\end{align}
where $i,j$ denote the corresponding slab, and employing the electrostatic boundary conditions: \mbox{$\mathbf n\times\mathbf E_i^{\parallel}=\mathbf n\times\mathbf E_j^{\parallel}$} and \mbox{$\mathbf n\cdot\mathbf D_i^{\parallel}=\mathbf n\cdot\mathbf D_j^{\parallel}$} \cite{jackson1999classical}, which translate into:
\begin{align}
\begin{split}
G_{1,j,\mathbf q}(-\infty,z^{\prime})= {}& G_{n,j,\mathbf q}(\infty,z^{\prime})=0,\\
	G_{i,j,\mathbf q}(z_i,z^{\prime})={}&G_{i+1,j,\mathbf q}(z_i,z^{\prime}),\\
    \epsilon_{i}\partial_zG_{i,j,\mathbf q}(z,z^{\prime})\Big|_{z=z_i}={}&\epsilon_{i+1}\partial_zG_{i+1,j,\mathbf q}(z,z^{\prime})\Big|_{z=z_{i}},\\
\end{split}
\label{eq:boundary_conditions}
\end{align}
the Green's function $G_{\mathbf q}(z,z^{\prime}) \equiv G_{s,s,\mathbf q}(z,z^{\prime})$ can be solved for:
\begin{align}
\begin{split}
    &G_{\mathbf q}(z,z^{\prime}) =  \frac{1}{2\epsilon_0\epsilon_s|\mathbf q|}\mathrm e^{-\bar \epsilon_s|\mathbf q||z-z^{\prime}|}\\
    &\, +\frac{1}{2\epsilon_0\epsilon_s|\mathbf q|f_{\mathbf q}}\left(\left(\epsilon_{s,-}\epsilon_{2,+}\mathrm e^{l|\mathbf q|}-\epsilon_{s,+}\epsilon_{2,-}\mathrm e^{-l|\mathbf q|}\right)\left(\epsilon_{1,+}\epsilon_{s,-}\mathrm e^{l|\mathbf q|}-\epsilon_{1,-}\epsilon_{s,+}\mathrm e^{-l|\mathbf q|}\right)\mathrm e^{-\bar \epsilon_s d|\mathbf q|}\left(\mathrm e^{\bar \epsilon_s|\mathbf q|(z-z^{\prime})}+\mathrm e^{-\bar \epsilon_s|\mathbf q|(z-z^{\prime})}\right)\right.\\
    &\,\left.
    +\left(\epsilon_{s,-}\epsilon_{2,+}\mathrm e^{l|\mathbf q|}-\epsilon_{s,+}\epsilon_{2,-}\mathrm e^{-l|\mathbf q|}\right)\left(\epsilon_{1,+}\epsilon_{s,+}\mathrm e^{l|\mathbf q|}-\epsilon_{1,-}\epsilon_{s,-}\mathrm e^{-l|\mathbf q|}\right)\mathrm e^{\bar \epsilon_s|\mathbf q|(z+z^{\prime})}\right.\\
    &\,\left.
    +\left(\epsilon_{s,+}\epsilon_{2,+}\mathrm e^{l|\mathbf q|}-\epsilon_{s,-}\epsilon_{2,-}\mathrm e^{-l|\mathbf q|}\right)\left(\epsilon_{1,+}\epsilon_{s,-}\mathrm e^{l|\mathbf q|}-\epsilon_{1,-}\epsilon_{s,+}\mathrm e^{-l|\mathbf q|}\right)\mathrm e^{-\bar \epsilon_s|\mathbf q|(z+z^{\prime})}
    \right),
    \end{split}
    \label{eq:GreensFunction}
\end{align}
where we used the abbreviations:
\begin{align}
    \epsilon_{i,\pm}=\epsilon_i\pm 1, \quad \epsilon_i = \sqrt{\epsilon_{\parallel,i}\epsilon_{\perp,i}}, \quad \bar \epsilon_i = \sqrt{\frac{\epsilon_{\parallel,i}}{\epsilon_{\perp,i}}}.
\end{align}
The electric field $\hat E_{z,\mathbf q}(z)$, which is generated by the Coulomb potential $\hat{\phi}_{\mathbf q}(z)$, is then obtained by:
\begin{align}
    \hat E_{z,\mathbf q}(z) = -\partial_z\hat{\phi}_{\mathbf q}(z) = -\partial_z\int\mathrm dz^{\prime}\,G_{\mathbf q}(z,z^{\prime})\hat{\rho}_{\mathbf q}(z^{\prime}),
    \label{eq:local_field_no_coulomblocalfields}
\end{align}
which yields the quantum-confined electric field $\hat E_{z,\mathbf q}$:
\begin{align}
    \hat E_{z,\mathbf q} = \frac{1}{\mathcal A}\int\mathrm dz\,|\zeta(z)|^2 \hat E_{z,\mathbf q}(z)
    = -\frac{1}{\mathcal A}\int\mathrm dz\,|\zeta(z)|^2\partial_z\int\mathrm dz^{\prime}\,G_{\mathbf q}(z,z^{\prime})\hat{\rho}_{\mathbf q}(z^{\prime}),
\end{align}
cf.\ also Eq.~\eqref{eq:ElectricFieldConfined} in the main text. By separating the $z$-dependence of the charge density $\hat{\rho}_{\mathbf q}(z) \equiv -e\left|\zeta(z)\right|^2 \hat \rho_{\mathbf q}$, we can identify a classical quantum-confined electric field $E_{z,\mathbf q}^{\text{loc}}$ via $\hat E_{z,\mathbf q} \equiv E_{z,\mathbf q}^{\text{loc}}\hat \rho_{\mathbf q}$:
\begin{align}
    E_{z,\mathbf q}^{\text{loc}} = \frac{e}{\mathcal A}\int\mathrm dz\,|\zeta(z)|^2 \partial_z\int\mathrm dz^{\prime}\,G_{\mathbf q}(z,z^{\prime})|\zeta(z^{\prime})|^2.
    \label{eq:app_quantum_confined_field}
\end{align}
By performing the $z$-derivative and employing confinement wave functions $\zeta(z)$ of the form:
\begin{align}
    \zeta(z) = \sqrt{\frac{2}{d}}\cos\left(\frac{\pi}{d}z\right),\quad-\frac{d}{2}\leq z \leq\frac{d}{2},
\end{align}
the $z$- and $z^{\prime}$-integrals in Eq.~\eqref{eq:app_quantum_confined_field} can be performed, which yields:
\begin{align}
    E_{z,\mathbf q}^{\text{loc}} = \frac{e}{\mathcal A2\epsilon_0\epsilon_{s,\bot}}\frac{g_{\mathbf q}}{f_{\mathbf q}},
\end{align}
cf.~Eq.~\eqref{eq:ElectricFieldLocalExplicit} in the main text, where:
\begin{align}
\begin{split}
    g_{\mathbf q} = &\, \left(\left(\epsilon_{s,-}\epsilon_{2,+}\mathrm e^{l|\mathbf q|} - \epsilon_{s,+}\epsilon_{2,-}\mathrm e^{-l|\mathbf q|}\right)
    \left(\epsilon_{1,+}\epsilon_{s,+}\mathrm e^{l|\mathbf q|} - \epsilon_{1,-}\epsilon_{s,-}\mathrm e^{-l|\mathbf q|}\right)
    \right.\\
    &\,\left.-
    \left(\epsilon_{s,+}\epsilon_{2,+}\mathrm e^{l|\mathbf q|} - \epsilon_{s,-}\epsilon_{2,-}\mathrm e^{-l|\mathbf q|}\right)
    \left(\epsilon_{1,+}\epsilon_{s,-}\mathrm e^{l|\mathbf q|} - \epsilon_{1,-}\epsilon_{s,+}\mathrm e^{-l|\mathbf q|}\right)
    \right)\left(\frac{8\pi^2\sinh\left(\frac{\bar\epsilon_s|\mathbf q|d}{2}\right)}{\bar\epsilon_s|\mathbf q|d\left(4\pi^2+\left(\bar\epsilon_s|\mathbf q|d\right)^2\right)}\right)^2,
    \end{split}
    \label{eq:g_function}
\end{align}
and:
\begin{align}
\begin{split}
    f_{\mathbf q} = &\, -\epsilon_{s,-}\epsilon_{s,-}\left(\epsilon_{1,+}\epsilon_{2,+}\mathrm e^{-\bar \epsilon_sd|\mathbf q|}\mathrm e^{2l|\mathbf q|} - \epsilon_{1,-}\epsilon_{2,-}\mathrm e^{\bar \epsilon_sd|\mathbf q|}\mathrm e^{-2l|\mathbf q|}\right)
    -2\epsilon_{s,+}\epsilon_{s,-}\left(\epsilon_1\epsilon_2-1\right)\left(\mathrm e^{\bar \epsilon_sd|\mathbf q|}-\mathrm e^{-\bar \epsilon_sd|\mathbf q|}\right)\\
    &\, + 
    \epsilon_{s,+}\epsilon_{s,+}\left(\epsilon_{1,+}\epsilon_{2,+}\mathrm e^{\bar \epsilon_sd|\mathbf q|}\mathrm e^{2l|\mathbf q|} - \epsilon_{1,-}\epsilon_{2,-}\mathrm e^{-\bar \epsilon_sd|\mathbf q|}\mathrm e^{-2l|\mathbf q|}\right)
    .
    \end{split}
    \label{eq:f_function}
\end{align}

\section{Development of the Excitonic PSOI Hamiltonian}
\label{app:excitonic_hamiltonian}
The many-body Rashba Hamiltonian in low-wavenumber approximation restricted to intraband processes reads, cf.\ Eq.~\eqref{eq:RashbaHamiltonianManyBodySmallQ} in the main text:
\begin{align}
\begin{split}
\label{eq:RashbaHamiltonianManyBodySmallQ_app}
    \hat H_{\text{BR}} = &\, \sum_{\substack{\mathbf k,\mathbf k^{\prime},\mathbf q,\\\xi,\xi^{\prime},s,s^{\prime}}}E_{z,\mathbf q}^{\text{loc}}
    \left(\,S_{\mathbf k}^{v,\xi,\bar s,s}
    \vdagtwo{\mathbf k+\mathbf q}{\xi,\bar s}
    \vdagtwo{\mathbf k^{\prime}-\mathbf q}{\xi^{\prime},s^{\prime}}
    \vndagtwo{\mathbf k^{\prime}}{\xi^{\prime},s^{\prime}}
    \vndagtwo{\mathbf k}{\xi,s}
    +
    S_{\mathbf k}^{c,\xi,\bar s,s}
    \cdagtwo{\mathbf k+\mathbf q}{\xi,\bar s}
    \cdagtwo{\mathbf k^{\prime}-\mathbf q}{\xi^{\prime},s^{\prime}}
    \cndagtwo{\mathbf k^{\prime}}{\xi^{\prime},s^{\prime}}
    \cndagtwo{\mathbf k}{\xi,s}\right.\\
    &\,\quad\quad\quad\quad +\left.
    S_{\mathbf k}^{v,\xi,\bar s,s}
    \vdagtwo{\mathbf k+\mathbf q}{\xi,\bar s}
    \cdagtwo{\mathbf k^{\prime}-\mathbf q}{\xi^{\prime},s^{\prime}}
    \cndagtwo{\mathbf k^{\prime}}{\xi^{\prime},s^{\prime}}
    \vndagtwo{\mathbf k}{\xi,s}
    +
    S_{\mathbf k}^{c,\xi,\bar s,s}
    \cdagtwo{\mathbf k+\mathbf q}{\xi,\bar s}
    \vdagtwo{\mathbf k^{\prime}-\mathbf q}{\xi^{\prime},s^{\prime}}
    \vndagtwo{\mathbf k^{\prime}}{\xi^{\prime},s^{\prime}}
    \cndagtwo{\mathbf k}{\xi,s}
    \right)\\
    &\, + \sum_{\substack{\mathbf k,\mathbf q,\xi,s}}E_{z,\mathbf q}^{\text{loc}}\left( S_{\mathbf k}^{v,\xi,\bar s,s} \vdagtwo{\mathbf k}{\xi,\bar s}\vndagtwo{\mathbf k}{\xi,s} + S_{\mathbf k}^{c,\xi,\bar s,s} \cdagtwo{\mathbf k}{\xi,\bar s}\cndagtwo{\mathbf k}{\xi,s} \right),
    \end{split}
\end{align}
In excitonic normal ordering, we obtain:
\begin{align}
\begin{split}
    \hat H_{\text{BR}} = &\,\sum_{\substack{\mathbf k,\mathbf k^{\prime},\mathbf q,\\\xi,\xi^{\prime},s,s^{\prime}}}E_{z,\mathbf q}^{\text{loc}}
    \left(
    S_{\mathbf k}^{v,\xi,\bar s,s}
    \vndagtwo{\mathbf k^{\prime}}{\xi^{\prime},s^{\prime}}
    \vndagtwo{\mathbf k}{\xi,s}
    \vdagtwo{\mathbf k+\mathbf q}{\xi,\bar s}
    \vdagtwo{\mathbf k^{\prime}-\mathbf q}{\xi^{\prime},s^{\prime}}
    +
    S_{\mathbf k}^{c,\xi,\bar s,s}
    \cdagtwo{\mathbf k+\mathbf q}{\xi,\bar s}
    \cdagtwo{\mathbf k^{\prime}-\mathbf q}{\xi^{\prime},s^{\prime}}
    \cndagtwo{\mathbf k^{\prime}}{\xi^{\prime},s^{\prime}}
    \cndagtwo{\mathbf k}{\xi,s}\right)\\
    &\, - \sum_{\substack{\mathbf k,\mathbf k^{\prime},\mathbf q,\\\xi,\xi^{\prime},s,s^{\prime}}}E_{z,\mathbf q}^{\text{loc}} \left(S_{\mathbf k}^{v,\xi,\bar s,s} \cdagtwo{\mathbf k^{\prime}-\mathbf q}{\xi^{\prime},s^{\prime}}\cndagtwo{\mathbf k^{\prime}}{\xi^{\prime},s^{\prime}}\vndagtwo{\mathbf k}{\xi,s}\vdagtwo{\mathbf k+\mathbf q}{\xi,\bar s}
    + S_{\mathbf k}^{c,\xi,\bar s,s} \cdagtwo{\mathbf k+\mathbf q}{\xi,\bar s}\cndagtwo{\mathbf k}{\xi,s}\vndagtwo{\mathbf k^{\prime}}{\xi^{\prime},s^{\prime}}\vdagtwo{\mathbf k^{\prime}-\mathbf q}{\xi^{\prime},s^{\prime}}\right)\\
    &\, + \sum_{\substack{\mathbf k,\mathbf k^{\prime},\\\xi,\xi^{\prime},s,s^{\prime}}}E_{z,\mathbf 0}^{\text{loc}}\left(S_{\mathbf k}^{c,\xi,\bar s,s} \cdagtwo{\mathbf k}{\xi,\bar s}\cndagtwo{\mathbf k}{\xi,s} - S_{\mathbf k}^{v,\xi,\bar s,s} \vndagtwo{\mathbf k}{\xi,s}\vdagtwo{\mathbf k}{\xi,\bar s} \right)\\
    &\, +\sum_{\substack{\mathbf k,\mathbf q,\xi,s}}E_{z,\mathbf q}^{\text{loc}}S_{\mathbf k}^{v,\xi,\bar s,s} \left(\vndagtwo{\mathbf k+\mathbf q}{\xi,s}\vdagtwo{\mathbf k+\mathbf q}{\xi,\bar s} + \vndagtwo{\mathbf k}{\xi,s}\vdagtwo{\mathbf k}{\xi,\bar s} \right)\\
    &\, + \sum_{\substack{\mathbf k,\mathbf q,\xi,s}}E_{z,\mathbf q}^{\text{loc}}\left( S_{\mathbf k}^{c,\xi,\bar s,s} \cdagtwo{\mathbf k}{\xi,\bar s}\cndagtwo{\mathbf k}{\xi,s} -  S_{\mathbf k}^{v,\xi,\bar s,s} \vndagtwo{\mathbf k}{\xi,s}\vdagtwo{\mathbf k}{\xi,\bar s} \right).
    \end{split}
    \label{eq:RashbaHamiltonianManyBodySmallQ_app2}
\end{align}
In the following, we neglect the first line in Eq.~\eqref{eq:RashbaHamiltonianManyBodySmallQ_app2}, since we aim for a density-independent description. 
In the third line of Eq.~\eqref{eq:RashbaHamiltonianManyBodySmallQ_app2}, the $E_{z,\mathbf 0}$-contributions, i.e., the spatially independent Hartree contributions due to a filled valence band, do not contribute, since they cancel out with the $\mathbf q=\mathbf 0$-contributions of the electron-ion interaction (not shown here). 
Moreover, the second terms in the brackets in the fourth (valence-band self-energy from the many-body term $\vdagtwo{}{}\vdagtwo{}{}\vndagtwo{}{}\vndagtwo{}{}$) and fifth (valence-band self-energy from the one-body term $\vdagtwo{}{}\vndagtwo{}{}$) line in Eq.~\eqref{eq:RashbaHamiltonianManyBodySmallQ_app2} cancel. Note that, in contrast to the many-body Coulomb interaction, the one-body contributions due to self-interacting fields in the last line in Eq.~\eqref{eq:RashbaHamiltonianManyBodySmallQ_app2} do not vanish in the calculation of the quantum kinetics, since they encode spin-flip processes and are not just proportional to the particle-density operator $\hat N = \adagtwo{\lambda,\mathbf k}{\xi,s}\andagtwo{\lambda,\mathbf k}{\xi,s}$. 
Applying the unit-operator method \cite{katsch2018theory,ivanov1993self} in the third, fourth and fifth lines of Eq.~\eqref{eq:RashbaHamiltonianManyBodySmallQ_app2} in lowest order, $\mathbb{1} = \sum_{\mathbf k,\xi,s}\cdagtwo{\mathbf k}{\xi,s}\cndagtwo{\mathbf k}{\xi,s} + \sum_{\mathbf k,\xi,s}\vndagtwo{\mathbf k}{\xi,s}\vdagtwo{\mathbf k}{\xi,s} + \mathcal O(Na_B^2)^2$, we arrive at:
\begin{align}
\begin{split}
    \hat H_{\text{BR}} =  - \sum_{\substack{\mathbf k,\mathbf k^{\prime},\mathbf q,\\\xi,\xi^{\prime},s,s^{\prime}}}E_{z,\mathbf q}^{\text{loc}} 
    &\left(
     S_{\mathbf k}^{v,\xi,\bar s,s} \Pdagtwo{\mathbf k,\mathbf k^{\prime}+\mathbf q}{\xi,\xi^{\prime},s,s^{\prime}}\Pndagtwo{\mathbf k-\mathbf q,\mathbf k^{\prime}}{\xi,\xi^{\prime},\bar s,s^{\prime}}
     +
     S_{\mathbf k^{\prime}}^{c,\xi^{\prime},\bar s^{\prime},s^{\prime}} \Pdagtwo{\mathbf k,\mathbf k^{\prime}+\mathbf q}{\xi,\xi^{\prime},s,\bar s^{\prime}}\Pndagtwo{\mathbf k-\mathbf q,\mathbf k^{\prime}}{\xi,\xi^{\prime},s,s^{\prime}}\right.\\
     &\left.
    - S_{\mathbf k}^{v,\xi,\bar s,s}
    \Pdagtwo{\mathbf k+\mathbf q,\mathbf k^{\prime}}{\xi,\xi^{\prime},s,s^{\prime}}\Pndagtwo{\mathbf k+\mathbf q,\mathbf k^{\prime}}{\xi,\xi^{\prime},\bar s,s^{\prime}}
    - S_{\mathbf k^{\prime}}^{c,\xi^{\prime},\bar s^{\prime},s^{\prime}}
    \Pdagtwo{\mathbf k,\mathbf k^{\prime}}{\xi,\xi^{\prime},s,\bar s^{\prime}}\Pndagtwo{\mathbf k,\mathbf k^{\prime}}{\xi,\xi^{\prime}, s,s^{\prime}}
    \right)
    \end{split}
\end{align}
After an expansion in excitonic wave functions \cite{katsch2018theory}:
\begin{align}
    \Pndagtwo{\mathbf k,\mathbf k^{\prime}}{\xi,\xi^{\prime},s,s^{\prime}} = \sum_{\mu}\ExWFtwo{\mu,\alpha_{\xi,\xi^{\prime}}^{s,s^{\prime}}\mathbf k+\beta_{\xi,\xi^{\prime}}^{s,s^{\prime}}\mathbf k^{\prime}}{\xi,\xi^{\prime},s,s^{\prime}}\Pndagtwo{\mu,\mathbf k^{\prime}-\mathbf k}{\xi,\xi^{\prime},s,s^{\prime}},
\end{align}
where $\ExWFtwo{\mu,\mathbf q}{\xi,\xi^{\prime},s,s^{\prime}}$ solves the Wannier equation:
\begin{align}
    \left(\tilde{E}_{\text{gap}}^{\xi,\xi^{\prime},s,s^{\prime}} + \frac{\hbar^2\mathbf q^2}{2m_{\text{r},\xi,\xi^{\prime}}^{s,s^{\prime}}}\right)\ExWFtwo{\mu,\mathbf q}{\xi,\xi^{\prime},s,s^{\prime}} - \sum_{\mathbf q^{\prime}}V_{\mathbf q-\mathbf q^{\prime}}\ExWFtwo{\mu,\mathbf q^{\prime}}{\xi,\xi^{\prime},s,s^{\prime}} 
    = E_{\mu}^{\xi,\xi^{\prime},s,s^{\prime}}\ExWFtwo{\mu,\mathbf q}{\xi,\xi^{\prime},s,s^{\prime}},
    \label{eq:WannierEquation}
\end{align}
with renormalized band gap $\tilde{E}_{\text{gap}}^{\xi,\xi^{\prime}}$, relative momentum $\mathbf q$, reduced mass $m_{\text{r},\xi,\xi^{\prime}}^{s,s^{\prime}} = \left( \frac{1}{m_{e}^{\xi^{\prime},s^{\prime}}} + \frac{1}{m_{h}^{\xi,s}} \right)^{-1}$ with effective masses of the electron/hole $m_{e/h}^{\xi,s}$, Coulomb potential $V_{\mathbf q}$ and excitonic energy $E_{\mu}^{\xi,\xi^{\prime},s,s^{\prime}}$ with excitonic quantum number $\mu$, 
we finally obtain:
\begin{align}
    \hat H_{\text{BR-X}} = &\, \sum_{\substack{\mu,\nu,\mathbf Q,\\\xi,\xi^{\prime},s,s^{\prime}}} \left(S_{\mu,\nu,\mathbf Q}^{h,\xi,\xi^{\prime},s,s^{\prime}} \Poldagtwo{\mu,\mathbf Q}{\xi,\xi^{\prime},\bar s,s^{\prime}}\Poloptwo{\nu,\mathbf Q}{\xi,\xi^{\prime},s,s^{\prime}} - S_{\mu,\nu,\mathbf Q}^{e,\xi,\xi^{\prime},s,s^{\prime}} \Poldagtwo{\mu,\mathbf Q}{s,\bar s^{\prime}}\Poloptwo{\nu,\mathbf Q}{\xi,\xi^{\prime},s,s^{\prime}} \right),
\end{align}
cf.~Eq.~\eqref{eq:RashbaHamiltonianIntrinsicExciton} in the main text, 
with matrix elements:
\begin{align}
S_{\mu,\nu,\mathbf Q}^{h,\xi,\xi^{\prime},s,s^{\prime}} 
    = &\, \sum_{\mathbf q,\mathbf q^{\prime}}\alpha_{\text{BR},v}^{\xi}\left( \mathrm i\left(\delta_{\bar s,\uparrow} - \delta_{\bar s,\downarrow}\right)q_x - q_y \right) E_{z,\mathbf q-\mathbf q^{\prime}}^{\text{loc}}\left(\ExWFstartwo{\mu,\mathbf q+\beta_{\xi,\xi^{\prime}}^{\bar s,s^{\prime}}\mathbf Q}{\xi,\xi^{\prime},\bar s,s^{\prime}}
    - \ExWFstartwo{\mu,\mathbf q^{\prime}+\beta_{\xi,\xi^{\prime}}^{\bar s,s^{\prime}}\mathbf Q}{\xi,\xi^{\prime},\bar s,s^{\prime}}
    \right)
    \ExWFtwo{\nu,\mathbf q^{\prime}+\beta_{\xi,\xi^{\prime}}^{s,s^{\prime}}\mathbf Q}{\xi,\xi^{\prime},s,s^{\prime}}, \label{eq:Sh_app}\\
    S_{\mu,\nu,\mathbf Q}^{e,\xi,\xi^{\prime},s,s^{\prime}} 
     = &\, - \sum_{\mathbf q,\mathbf q^{\prime}}\alpha_{\text{BR},c}^{\xi^{\prime}}\left( \mathrm i\left(\delta_{s^{\prime},\uparrow} - \delta_{s^{\prime},\downarrow}\right)q_x - q_y \right) E_{z,\mathbf q-\mathbf q^{\prime}}^{\text{loc}}\left(\ExWFstartwo{\mu,\mathbf q^{\prime}-\alpha_{\xi,\xi^{\prime}}^{s,\bar s^{\prime}}\mathbf Q}{\xi,\xi^{\prime},s,\bar s^{\prime}} - \ExWFstartwo{\mu,\mathbf q-\alpha_{\xi,\xi^{\prime}}^{s,\bar s^{\prime}}\mathbf Q}{\xi,\xi^{\prime},s,\bar s^{\prime}}\right)\ExWFtwo{\nu,\mathbf q-\alpha_{\xi,\xi^{\prime}}^{s,s^{\prime}}\mathbf Q}{\xi,\xi^{\prime},s,s^{\prime}},\label{eq:Se_app}
\end{align}
cf.~Eq.~\eqref{eq:RashbaMatrixElementIntrinsicHole} and Eq.~\eqref{eq:RashbaMatrixElementIntrinsicElectron} in the main text.

\section{Excitonic Spin Eigenvalue Equation}
\label{app:SpinEigenvalueEquation}
The excitonic spin eigenvalue equation reads \cite{mittenzwey2025ultrafast}:
\begin{align}
\begin{split}
    E_{\mu,\mathbf Q}^{\xi,\xi^{\prime},s_1,s_2}\mathcal C_{\mu,\mathbf Q,\mathcal S}^{\xi,\xi^{\prime},s_1,s_2} + S_{\mu,\nu,\mathbf Q}^{h,\xi,\xi^{\prime},\bar s_1, s_2}\delta_{\mu,\nu}\mathcal C_{\mu,\mathbf Q,\mathcal S}^{\xi,\xi^{\prime},\bar s_1,s_2} - S_{\mu,\nu,\mathbf Q}^{e,\xi,\xi^{\prime},s_1,\bar s_2}\delta_{\mu,\nu}\mathcal C_{\mu,\mathbf Q,\mathcal S}^{\xi,\xi^{\prime},s_1,\bar s_2} = E_{\mu,\mathbf Q}^{\xi,\xi^{\prime},\mathcal{S}}\mathcal C_{\mu,\mathbf Q,\mathcal S}^{\xi,\xi^{\prime},s_1,s_2},
    \end{split}
    \label{eq:SpinEigenvalueEquation}
\end{align}
where $E_{\mu,\mathbf Q}^{\mathcal{S},\xi,\xi^{\prime}}$ is the spin-diagonal exciton dispersion with respect to the excitonic spin-diagonal quantum number $\mathcal S$ and $\mathcal C_{\mu,\mathbf Q,\mathcal S}^{\xi,\xi^{\prime},s_1,s_2}$ is the spin-diagonal eigenfunction. Note, that we neglected possible interactions between distinct excitonic states $\mu \neq \nu$ in \eqref{eq:SpinEigenvalueEquation}, as we are only interested in the dynamics of the lowest-lying $\mu=1s$ exciton. However, such interactions, which have been shown to hybridize $s$-$p$ excitons \cite{cao2024emergent,cao2025tunable}, can be straightforwardly included. At zero electric field (or zero momentum), the excitonic spin-diagonal quantum number $\mathcal{S}$ corresponds to the uncoupled spins of the exciton, i.e.\ if $E_{z,\mathbf q}\rightarrow 0$ (or $\mathbf Q\rightarrow \mathbf 0$), then $\mathcal{S}\rightarrow \{(\uparrow,\uparrow),(\uparrow,\downarrow),(\downarrow,\uparrow),(\downarrow,\downarrow)\}$.

The excitonic transitions are transformed as follows:
\begin{align}
    \Pndagtwo{\mu,\mathbf Q}{\xi,\xi^{\prime},s,s^{\prime}} = \sum_{\mathcal S}\mathcal C_{\mu,\mathbf Q,\mathcal S}^{\xi,\xi^{\prime},s,s^{\prime}}\Pndagtwo{\mu,\mathbf Q}{\xi,\xi^{\prime},\mathcal S},
\end{align}
with:
\begin{align}
    \sum_{\mathcal S} \mathcal C^*\vphantom{\mathcal C}_{\mu,\mathbf Q,\mathcal S}^{\xi,\xi^{\prime},s,s^{\prime}} \mathcal C_{\mu,\mathbf Q,\mathcal S}^{\xi,\xi^{\prime},s^{\prime\prime},s^{\prime\prime\prime}} =  \delta_{s,s^{\prime\prime}}\delta_{s^{\prime},s^{\prime\prime\prime}},\quad\quad \sum_{s,s^{\prime}} \mathcal C^*\vphantom{\mathcal C}_{\mu,\mathbf Q,\mathcal S}^{\xi,\xi^{\prime},s,s^{\prime}} \mathcal C_{\mu,\mathbf Q,\mathcal S^{\prime}}^{\xi,\xi^{\prime},s,s^{\prime}} =  \delta_{\mathcal S,\mathcal S^{\prime}},
\end{align}
which enables a formulation of the excitonic theory in the spin-diagonal basis $\mathcal S$.

\section{Parameters}

\begin{table*}[]
    \centering
    \begin{tabular}{lll}
    Quantity & MoSe$_2$ & MoS$_2$\\
    \hline
        Bright-dark splitting $E_{1s,\mathbf Q=\mathbf 0}^{K,K,\uparrow,\uparrow}-E_{1s,\mathbf Q=\mathbf 0}^{K,K,\uparrow,\downarrow}$ (meV) & $-1.45$ \cite{robert2020measurement,lu2019magnetic} & 14 \cite{robert2020measurement}\\
        Effective spin-up electron mass $m_e^{K,\uparrow}/m_0$ \cite{kormanyos2015k} & 0.495 & 0.435\\
        Effective spin-down electron mass $m_e^{K,\uparrow}/m_0$ \cite{kormanyos2015k} & 0.57 & 0.465\\
        Effective spin-up hole mass $m_h^{K,\uparrow}/m_0$ \cite{kormanyos2015k} & 0.595 & 0.54\\
        Rashba coupling constant for electrons $\alpha_{\text{BR}}^{e,K}$ ($e$\,nm$^2$) \cite{yang2015hole} & 0.011 & 0.003 \\
        Rashba coupling constant for holes $\alpha_{\text{BR}}^{h,K}$ ($e$\,nm$^2$) \cite{yang2015hole} & 0.0422 & 0.041 \\
        Monolayer width $d$ (nm) \cite{kylanpaa2015binding} & 0.6527  & 0.618\\
        Static out-of-plane dielectric constant $\epsilon_{\bot,0}$ \cite{laturia2018dielectric} & 7.2 & 6.2  \\
Static bulk dielectric constant 
$\epsilon_{\text{bulk},0} = \sqrt{\epsilon_{\text{bulk},\parallel,0}\epsilon_{\text{bulk},\bot,0}}$ \cite{laturia2018dielectric} & 12.05  & 10.47\\
        Plasmon peak energy $\hbar\omega_{\text{pl}}$ (eV) \cite{kumar2012tunable} & 22.0  & 22.5 \\
    Thomas-Fermi parameter $\alpha_{\text{TF}}$ (fit to \textit{ab-initio} results \cite{andersen2015dielectric}) & 1.9 & 1.5 \\
    \hline
    \hline
    Calculated $1s$ excitonic binding energies\\
    with $l=0.3$\,nm \cite{florian2018dielectric,druppel2017diversity} (meV)\\
    \hline
    Spin-bright, SiO$_2$ substrate  & $-441$ & $-478$ \\
    Spin-dark, SiO$_2$ substrate  & $-455$ & $-486$ \\
    Spin-bright, Al$_2$O$_3$ substrate  & $-341$ & $-367$ \\
    Spin-dark, Al$_2$O$_3$ substrate  & $-355$ & $-375$  \\
    Spin-bright, h-BN encapsulated & $-325$ & $-349$ \\
    Spin-dark, h-BN encapsulated & $-339$ & $-357$
    \end{tabular}
    \caption{Parameters used in the simulations.}
    \label{tab:parameters}
\end{table*}

All parameters are known from first-principle calculations or taken from literature. No additional fitting parameters have been introduced.

In Tab.~\ref{tab:parameters}, we display the parameters, which are not related to phonon scattering, used in all numerical simulations.

For the numerical evaluation of the Wannier equation in Eq.~\eqref{eq:WannierEquation}, which provides the excitonic binding energies and wave functions $\ExWFtwo{\mu,\mathbf q}{\xi,\xi^{\prime},s,s^{\prime}}$ necessary to estimate the excitonic energy landscape and the Rashba-, cf.~Eq.~\eqref{eq:RashbaMatrixElementIntrinsicHole} and Eq.~\eqref{eq:RashbaMatrixElementIntrinsicElectron}, and exciton-phonon coupling elements, cf.~Ref.~\cite{mittenzwey2025ultrafast}, we use the analytical dielectric function from Ref.~\cite{trolle2017model}, which is fitted to \textit{ab initio} calculations from the Computational Materials Repository \cite{andersen2015dielectric}, and the screened Coulomb potential $V_{\mathbf q} = \frac{e^2}{\mathcal A}\int\mathrm dz\,\mathrm dz^{\prime}\,|\zeta(z)|^2G_{\mathbf q}^{}(z,z^{\prime})|\zeta(z^{\prime})|^2$ with the Green's function from Eq.~\eqref{eq:GreensFunction} in the isotropic limit with $\epsilon_s\rightarrow \epsilon_{s,\mathbf q}$, where $\epsilon_{s,\mathbf q}$ is the bulk dielectric function of the analytic model from Ref.~\cite{trolle2017model}.

All relevant parameters for the exciton-phonon interaction such as electron-phonon interaction potentials, phonon dispersions and phonon modes are taken from \textit{ab initio} calculations in Refs.~\cite{jin2014intrinsic,li2013intrinsic} and are not displayed in Tab.~\ref{tab:parameters} for better readability. For intravalley scattering, we take into account $A^{\prime}$, TO, LA and TA modes and for intervalley scattering we take into account $A^{\prime}$, LO, TO, LA and TA modes. The electron-phonon interaction potentials in deformation-potential approximation from Refs.~\cite{jin2014intrinsic,li2013intrinsic} for acoustic phonons are rescaled by a factor of $\frac{1}{\sqrt{2}}$ to properly consider the fact, that piezoelectric coupling, which is naturally included in the effective electron-phonon interaction potentials from Refs.~\cite{jin2014intrinsic,li2013intrinsic}, is ineffective for exciton-phonon interaction \cite{lengers2020theory}.

\FloatBarrier
\twocolumngrid

\bibliography{bibliography_full}

@article{mittenzwey2026coulombinteraction,
      title={Coulomb Interaction in Atomically Thin Semiconductors and Density-Independent Exciton-Scattering Processes}, 
      author={Henry Mittenzwey and Andreas Knorr and Thorsten Deilmann},
      journal={arXiv preprint},
      year={2026},
      doi={10.48550/arXiv.2602.13763}, 
}

@article{snoeken2025theory,
  title={Theory of In-Plane-Magnetic-Field-Dependent Excitonic Spectra in Atomically Thin Semiconductors},
  author={Snoeken, Michiel and Steeger, Paul and Schmidt, Robert and de Vasconcellos, Steffen Michaelis and Bratschitsch, Rudolf and Knorr, Andreas and Mittenzwey, Henry},
  journal={arXiv preprint},
  year={2025},
doi = {10.48550/arXiv.2511.02524}
}

@article{echeverry2016splitting,
  title = {Splitting between bright and dark excitons in transition metal dichalcogenide monolayers},
  author = {Echeverry, J. P. and Urbaszek, B. and Amand, T. and Marie, X. and Gerber, I. C.},
  journal = {Phys. Rev. B},
  volume = {93},
  issue = {12},
  pages = {121107},
  numpages = {5},
  year = {2016},
  month = {Mar},
  publisher = {American Physical Society},
  doi = {10.1103/PhysRevB.93.121107},
  url = {https://link.aps.org/doi/10.1103/PhysRevB.93.121107}
}

@article{kosmider2013large,
  title = {Large spin splitting in the conduction band of transition metal dichalcogenide monolayers},
  author = {Ko\ifmmode \acute{s}\else \'{s}\fi{}mider, K. and Gonz\'alez, J. W. and Fern\'andez-Rossier, J.},
  journal = {Phys. Rev. B},
  volume = {88},
  issue = {24},
  pages = {245436},
  numpages = {7},
  year = {2013},
  month = {Dec},
  publisher = {American Physical Society},
  doi = {10.1103/PhysRevB.88.245436},
  url = {https://link.aps.org/doi/10.1103/PhysRevB.88.245436}
}

@article{junior2022first,
doi = {10.1088/1367-2630/ac7e21},
url = {https://doi.org/10.1088/1367-2630/ac7e21},
year = {2022},
month = {aug},
publisher = {IOP Publishing},
volume = {24},
number = {8},
pages = {083004},
author = {Faria Junior, Paulo E and Zollner, Klaus and Woźniak, Tomasz and Kurpas, Marcin and Gmitra, Martin and Fabian, Jaroslav},
title = {First-principles insights into the spin-valley physics of strained transition metal dichalcogenides monolayers},
journal = {New J. Phys.},
}

@article{combescot2019spin,
  title = {Spin-orbit coupling: Atom versus semiconductor crystal},
  author = {Combescot, Monique and Shiau, Shiue-Yuan and Voliotis, Valia},
  journal = {Phys. Rev. B},
  volume = {99},
  issue = {24},
  pages = {245202},
  numpages = {13},
  year = {2019},
  month = {Jun},
  publisher = {American Physical Society},
  doi = {10.1103/PhysRevB.99.245202},
  url = {https://link.aps.org/doi/10.1103/PhysRevB.99.245202}
}

@article{steeger2023pressure,
author = {Steeger, Paul and Graalmann, Jan-Hauke and Schmidt, Robert and Kupenko, Ilya and Sanchez-Valle, Carmen and Marauhn, Philipp and Deilmann, Thorsten and de Vasconcellos, Steffen Michaelis and Rohlfing, Michael and Bratschitsch, Rudolf},
title = {Pressure Dependence of Intra- and Interlayer Excitons in {2H}-{Mo}{S}$_2$ Bilayers},
journal = {Nano Lett.},
volume = {23},
number = {19},
pages = {8947-8952},
year = {2023},
doi = {10.1021/acs.nanolett.3c02428}}

@article{pasenow2005excitonic,
  title = {Excitonic wave packet dynamics in semiconductor photonic-crystal structures},
  author = {Pasenow, B. and Reichelt, M. and Stroucken, T. and Meier, T. and Koch, S. W.},
  journal = {Phys. Rev. B},
  volume = {71},
  issue = {19},
  pages = {195321},
  numpages = {15},
  year = {2005},
  month = {May},
  publisher = {American Physical Society},
  doi = {10.1103/PhysRevB.71.195321},
  url = {https://link.aps.org/doi/10.1103/PhysRevB.71.195321}
}

@article{zimmermann2016poisson,
  title = {{P}oisson {G}reen's function method for increased computational efficiency in numerical calculations of {C}oulomb coupling elements},
  author = {Zimmermann, Anke and Kuhn, Sandra and Richter, Marten},
  journal = {Phys. Rev. B},
  volume = {93},
  issue = {3},
  pages = {035308},
  numpages = {9},
  year = {2016},
  month = {Jan},
  publisher = {American Physical Society},
  doi = {10.1103/PhysRevB.93.035308},
  url = {https://link.aps.org/doi/10.1103/PhysRevB.93.035308}
}

@article{boccuni2024unveiling,
author = {Boccuni, Alberto and Peluzo, Bárbara Maria Teixeira Costa and Bodo, Filippo and Ambrogio, Giacomo and Maul, Jefferson and Mitoli, Davide and Vignale, Giovanni and Pittalis, Stefano and Kraka, Elfi and Desmarais, Jacques K. and Erba, Alessandro},
title = {Unveiling the Role of Spin Currents on the Giant {R}ashba Splitting in Single-Layer {W}{Se}$_2$},
journal = {J. Phys. Chem. Lett.},
volume = {15},
number = {29},
pages = {7442-7448},
year = {2024},
doi = {10.1021/acs.jpclett.4c01607}
}

@book{chang1993many,
author = {Chang, T-N},
title = {Many-Body Theory of Atomic Structure and Photoionization},
publisher = {WORLD SCIENTIFIC},
year = {1993},
doi = {10.1142/1578},
address = {},
edition   = {},
}

@article{rajagopal1998spin,
  title = {Spin-orbit interactions in the many-body theory of magnetic electron systems},
  author = {Rajagopal, A. K. and Mochena, Mogus},
  journal = {Phys. Rev. B},
  volume = {57},
  issue = {18},
  pages = {11582--11591},
  numpages = {0},
  year = {1998},
  month = {May},
  publisher = {American Physical Society},
  doi = {10.1103/PhysRevB.57.11582},
  url = {https://link.aps.org/doi/10.1103/PhysRevB.57.11582}
}

@article{yi2000breit,
  title={{B}reit interaction, level spacing statistics and far-infrared absorption in small metal clusters},
  author={Yi, Lin and Sheng, Ping},
  journal={Solid State Commun.},
  volume={114},
  number={4},
  pages={177--192},
  year={2000},
  publisher={Elsevier},
doi = {10.1016/S0038-1098(00)00022-3}
}

@article{gindikin2022spin,
  title={Spin-dependent electron-electron interaction in {R}ashba materials},
  author={Gindikin, Yasha and Sablikov, Vladimir A.},
  journal={J. Exp. Theor. Phys.},
  volume={135},
  number={4},
  pages={531--539},
  year={2022},
  publisher={Springer},
doi = {10.1134/S1063776122100041}
}

@article{gindikin2018spin,
  title = {Spin-orbit-driven electron pairing in two dimensions},
  author = {Gindikin, Yasha and Sablikov, Vladimir A.},
  journal = {Phys. Rev. B},
  volume = {98},
  issue = {11},
  pages = {115137},
  numpages = {7},
  year = {2018},
  month = {Sep},
  publisher = {American Physical Society},
  doi = {10.1103/PhysRevB.98.115137},
  url = {https://link.aps.org/doi/10.1103/PhysRevB.98.115137}
}

@article{breit1929effect,
  title = {The Effect of Retardation on the Interaction of Two Electrons},
  author = {Breit, G.},
  journal = {Phys. Rev.},
  volume = {34},
  issue = {4},
  pages = {553--573},
  numpages = {0},
  year = {1929},
  month = {Aug},
  publisher = {American Physical Society},
  doi = {10.1103/PhysRev.34.553},
  url = {https://link.aps.org/doi/10.1103/PhysRev.34.553}
}

@article{gindikin2025electron,
  title = {Electron interactions in {R}ashba materials},
  author = {Gindikin, Yasha and Kamenev, Alex},
  journal = {Phys. Rev. B},
  volume = {111},
  issue = {3},
  pages = {035104},
  numpages = {18},
  year = {2025},
  month = {Jan},
  publisher = {American Physical Society},
  doi = {10.1103/PhysRevB.111.035104},
  url = {https://link.aps.org/doi/10.1103/PhysRevB.111.035104}
}

@article{liu2024ordered,
  title = {Ordered phases and superconductivity in two-dimensional electron systems subject to pair spin-orbit interaction},
  author = {Liu, Feng and Principi, Alessandro},
  journal = {Phys. Rev. B},
  volume = {109},
  issue = {7},
  pages = {075163},
  numpages = {9},
  year = {2024},
  month = {Feb},
  publisher = {American Physical Society},
  doi = {10.1103/PhysRevB.109.075163},
  url = {https://link.aps.org/doi/10.1103/PhysRevB.109.075163}
}

@article{dogadov2026diss,
author={Dogadov, Oleg
and Mittenzwey, Henry
and Bertolotti, Micol
and Olsen, Nicholas
and Deckert, Thomas
and Trovatello, Chiara
and Zhu, Xiaoyang
and Brida, Daniele
and Cerullo, Giulio
and Knorr, Andreas
and Dal Conte, Stefano},
title={Dissecting intervalley coupling mechanisms in monolayer transition metal dichalcogenides},
journal={npj 2D Mater. Appl.},
year={2026},
month={Jan},
day={31},
volume={10},
number={1},
pages={21},
issn={2397-7132},
doi={10.1038/s41699-025-00653-2},
url={https://doi.org/10.1038/s41699-025-00653-2}
}

@article{druppel2017diversity,
  title={Diversity of trion states and substrate effects in the optical properties of an {Mo}{S}$_2$ monolayer},
  author={Dr{\"u}ppel, Matthias and Deilmann, Thorsten and Kr{\"u}ger, Peter and Rohlfing, Michael},
  journal={Nat. Commun.},
  volume={8},
  number={1},
  pages={2117},
  year={2017},
  publisher={Nature Publishing Group UK London},
doi = {10.1038/s41467-017-02286-6}
}

@article{he2020valley,
  title={Valley phonons and exciton complexes in a monolayer semiconductor},
  author={He, Minhao and Rivera, Pasqual and Van Tuan, Dinh and Wilson, Nathan P. and Yang, Min and Taniguchi, Takashi and Watanabe, Kenji and Yan, Jiaqiang and Mandrus, David G. and Yu, Hongyi and Dery, Hanan and Yao, Wang and Xu, Xiaodong},
  journal={Nat. Commun.},
  volume={11},
  number={1},
  pages={618},
  year={2020},
  publisher={Nature Publishing Group UK London},
doi = {10.1038/s41467-020-14472-0}
}

@article{yang2022relaxation,
  title={Relaxation and darkening of excitonic complexes in electrostatically doped monolayer {W}{Se}$_2$: Roles of exciton-electron and trion-electron interactions},
  author={Yang, Min and Ren, Lei and Robert, Cedric and Van Tuan, Dinh and Lombez, Laurent and Urbaszek, Bernhard and Marie, Xavier and Dery, Hanan},
  journal={ Phys. Rev. B},
  volume={105},
  number={8},
  pages={085302},
  year={2022},
  publisher={APS},
doi = {10.1103/PhysRevB.105.085302}
}

@article{li2022intervalley,
  title={Intervalley electron-hole exchange interaction and impurity-assisted recombination of indirect excitons in {W}{S}$_2$ and {W}{Se}$_2$ monolayers},
  author={Li, Pengke and Robert, Cedric and Van Tuan, Dinh and Ren, Lei and Yang, Min and Marie, Xavier and Dery, Hanan},
  journal={ Phys. Rev. B},
  volume={106},
  number={8},
  pages={085414},
  year={2022},
  publisher={APS},
doi = {10.1103/PhysRevB.106.085414}
}

@article{wiser1963dielectric,
  title={Dielectric Constant with Local Field Effects Included},
  author={Wiser, Nathan},
  journal={Phys. Rev.},
  volume={129},
  number={1},
  pages={62},
  year={1963},
  publisher={APS},
doi = {10.1103/PhysRev.129.62}
}

@article{zhang2022ab,
  title={Ab initio calculations of spin-nonconserving exciton-phonon scattering in monolayer transition metal dichalcogenides},
  author={Zhang, Xiao-Wei and Cao, Ting},
  journal={J. Phys. Condens. Matter},
  volume={34},
  number={26},
  pages={264002},
  year={2022},
  publisher={IOP Publishing},
doi = {10.1088/1361-648X/ac6649}
}

@article{wang2018colloquium,
  title={Colloquium: Excitons in atomically thin transition metal dichalcogenides},
  author={Wang, Gang and Chernikov, Alexey and Glazov, Mikhail M. and Heinz, Tony F. and Marie, Xavier and Amand, Thierry and Urbaszek, Bernhard},
  journal={Rev. Mod. Phys.},
  volume={90},
  number={2},
  pages={021001},
  year={2018},
  publisher={APS},
doi = {10.1103/RevModPhys.90.021001}
}

@article{cao2025tunable,
  title = {Tunable resonant $s\ensuremath{-}p$ mixing of excitons in van der {W}aals heterostructures},
  author = {Cao, Jiayu David and Denisov, Konstantin S. and \ifmmode \check{Z}\else \v{Z}\fi{}uti\ifmmode \acute{c}\else \'{c}\fi{}, Igor},
  journal = {Phys. Rev. B},
  volume = {112},
  issue = {16},
  pages = {L161405},
  numpages = {7},
  year = {2025},
  month = {Oct},
  publisher = {American Physical Society},
  doi = {10.1103/kmsm-h1wc},
  url = {https://link.aps.org/doi/10.1103/kmsm-h1wc}
}

@article{mittenzwey2025ultrafast,
  title={Ultrafast Optical Control of {R}ashba Interactions in a {TMDC} Heterostructure},
  author={Mittenzwey, Henry and Kumar, Abhijeet M. and Dhingra, Raghav and Watanabe, Kenji and Taniguchi, Takashi and Gahl, Cornelius and Bolotin, Kirill I. and Selig, Malte and Knorr, Andreas},
  journal={ Phys. Rev. Lett.},
  volume={134},
  number={2},
  pages={026901},
  year={2025},
  publisher={APS},
doi = {10.1103/PhysRevLett.134.026901}
}

@article{chan2025exciton,
  title={Exciton thermalization dynamics in monolayer {Mo}{S}$_2$: A first-principles {B}oltzmann equation study},
  author={Chan, Yang-hao and Haber, Jonah B. and Naik, Mit H. and Louie, Steven G. and Neaton, Jeffrey B. and da Jornada, Felipe H. and Qiu, Diana Y.},
  journal={ Phys. Rev. B},
  volume={111},
  number={18},
  pages={184305},
  year={2025},
  publisher={APS},
doi = {10.1103/PhysRevB.111.184305}
}

@article{wallauer2021momentum,
  title={Momentum-resolved observation of exciton formation dynamics in monolayer {W}{S}$_2$},
  author={Wallauer, Robert and Perea-Causin, Raul and Münster, Lasse and Zajusch, Sarah and Brem, Samuel and Güdde, Jens and Tanimura, Katsumi and Lin, Kai-Qiang and Huber, Rupert and Malic, Ermin and Höfer, Ulrich},
  journal={Nano Lett.},
  volume={21},
  number={13},
  pages={5867--5873},
  year={2021},
  publisher={ACS Publications},
doi = {10.1021/acs.nanolett.1c01839}
}

@article{brem2020phonon,
  title={Phonon-Assisted Photoluminescence from Indirect Excitons in Monolayers of Transition-Metal Dichalcogenides},
  author={Brem, Samuel and Ekman, August and Christiansen, Dominik and Katsch, Florian and Selig, Malte and Robert, Cedric and Marie, Xavier and Urbaszek, Bernhard and Knorr, Andreas and Malic, Ermin},
  journal={Nano Lett.},
  volume={20},
  number={4},
  pages={2849--2856},
  year={2020},
  publisher={ACS Publications},
doi = {10.1021/acs.nanolett.0c00633}
}

@article{jin2014intrinsic,
  title={Intrinsic transport properties of electrons and holes in monolayer transition-metal dichalcogenides},
  author={Jin, Zhenghe and Li, Xiaodong and Mullen, Jeffrey T. and Kim, Ki Wook},
  journal={ Phys. Rev. B},
  volume={90},
  number={4},
  pages={045422},
  year={2014},
  publisher={APS},
doi = {10.1103/PhysRevB.90.045422}
}

@article{min2024chalcogen,
  author = {Min, Jinhong and Kim, Jae Hyung and Kang, Joohoon},
title = {Chalcogen Vacancy Engineering of Two-Dimensional Transition Metal Dichalcogenides for Electronic Applications},
journal = {ACS Appl. Nano Mater.},
volume = {7},
number = {23},
pages = {26377-26396},
year = {2024},
doi = {10.1021/acsanm.3c06263}
}

@article{ozcan2024point,
  title={Point-like vacancies in two-dimensional transition metal dichalcogenides},
  author={{\"O}zcan, Sibel and Gallardo, Aurelio and Biel, Blanca},
  journal={Electron. Struct.},
  volume={6},
  number={1},
  pages={015006},
  year={2024},
  publisher={IOP Publishing},
doi = {10.1088/2516-1075/ad2090}
}

@article{fontanella1974low,
  title={Low-frequency dielectric constants of $\alpha$-quartz, sapphire, {Mg}{F}$_2$, and {Mg}{O}},
  author={Fontanella, John and Andeen, Carl and Schuele, Donald},
  journal={J. Appl. Phys.},
  volume={45},
  number={7},
  pages={2852--2854},
  year={1974},
  publisher={American Institute of Physics},
doi = {10.1063/1.1663690}
}

@article{molas2019probing,
  title={Probing and Manipulating Valley Coherence of Dark Excitons in Monolayer {W}{Se}$_2$},
   author = {Molas, M. R. and Slobodeniuk, A. O. and Kazimierczuk, T. and Nogajewski, K. and Bartos, M. and Kapu\ifmmode \acute{s}\else \'{s}\fi{}ci\ifmmode \acute{n}\else \'{n}\fi{}ski, P. and Oreszczuk, K. and Watanabe, K. and Taniguchi, T. and Faugeras, C. and Kossacki, P. and Basko, D. M. and Potemski, M.},
  journal={ Phys. Rev. Lett.},
  volume={123},
  number={9},
  pages={096803},
  year={2019},
  publisher={APS},
doi = {10.1103/PhysRevLett.123.096803}
}

@article{mapara2022bright,
  title={Bright and Dark Exciton Coherent Coupling and Hybridization Enabled by External Magnetic Fields},
  author={Mapara, Varun and Barua, Arup and Turkowski, Volodymyr and Trinh, M. Tuan and Stevens, Christopher and Liu, Hengzhou and Nugera, Florence A. and Kapuruge, Nalaka and Gutierrez, Humberto Rodriguez and Liu, Fang and Zhu, Xiaoyang and Semenov, Dmitry and McGill, Stephen A. and Pradhan, Nihar and Hilton, David J. and Karaiskaj, Denis},
  journal={Nano Lett.},
  volume={22},
  number={4},
  pages={1680--1687},
  year={2022},
  publisher={ACS Publications},
doi = {10.1021/acs.nanolett.1c04667}
}

@article{guo2019exchange,
  title={Exchange-driven intravalley mixing of excitons in monolayer transition metal dichalcogenides},
  author={Guo, Liang and Wu, Meng and Cao, Ting and Monahan, Daniele M. and Lee, Yi-Hsien and Louie, Steven G. and Fleming, Graham R.},
  journal={Nat. Phys.},
  volume={15},
  number={3},
  pages={228--232},
  year={2019},
  publisher={Nature Publishing Group UK London},
doi = {10.1038/s41567-018-0362-y}
}

@article{xue2011scanning,
  title={Scanning tunnelling microscopy and spectroscopy of ultra-flat graphene on hexagonal boron nitride},
  author={Xue, Jiamin and Sanchez-Yamagishi, Javier and Bulmash, Danny and Jacquod, Philippe and Deshpande, Aparna and Watanabe, K. and Taniguchi, T. and Jarillo-Herrero, Pablo and LeRoy, Brian J.},
  journal={Nat. Mater.},
  volume={10},
  number={4},
  pages={282--285},
  year={2011},
  publisher={Nature Publishing Group UK London},
doi = {10.1038/nmat2968}
}

@article{lagarde2024efficient,
  title={Efficient electron spin relaxation by chiral phonons in {W}{Se}$_2$ monolayers},
  author = {Lagarde, D. and Glazov, M. and Jindal, V. and Mourzidis, K. and Gerber, Iann and Balocchi, A. and Lombez, L. and Renucci, P. and Taniguchi, T. and Watanabe, K. and Robert, C. and Marie, X.},
  journal={ Phys. Rev. B},
  volume={110},
  number={19},
  pages={195403},
  year={2024},
  publisher={APS},
doi = {10.1103/PhysRevB.110.195403}
}

@article{kumar2012tunable,
  title={Tunable dielectric response of transition metals dichalcogenides {M}{X}$_2$ ({M}={Mo}, {W}; {X}={S}, {Se}, {Te}): Effect of quantum confinement},
  author={Kumar, Ashok and Ahluwalia, P.~K.},
  journal={Phys. B: Condens. Matter},
  volume={407},
  number={24},
  pages={4627--4634},
  year={2012},
  publisher={Elsevier},
doi = {10.1016/j.physb.2012.08.034}
}

@article{andersen2015dielectric,
  title={Dielectric Genome of van der {W}aals Heterostructures},
  author={Andersen, Kirsten and Latini, Simone and Thygesen, Kristian S.},
  journal={Nano Lett.},
  volume={15},
  number={7},
  pages={4616--4621},
  year={2015},
  publisher={ACS Publications},
doi = {10.1021/acs.nanolett.5b01251}
}

@article{kylanpaa2015binding,
  title={Binding energies of exciton complexes in transition metal dichalcogenide monolayers and effect of dielectric environment},
  author={Kyl{\"a}np{\"a}{\"a}, Ilkka and Komsa, Hannu-Pekka},
  journal={ Phys. Rev. B},
  volume={92},
  number={20},
  pages={205418},
  year={2015},
  publisher={APS},
doi = {10.1103/PhysRevB.92.205418}
}

@article{yu2014valley,
  title={Valley depolarization due to intervalley and intravalley electron-hole exchange interactions in monolayer {Mo}{S}$_2$},
  author={Yu, Tao and Wu, M.~W.},
  journal={ Phys. Rev. B},
  volume={89},
  number={20},
  pages={205303},
  year={2014},
  publisher={APS},
doi = {10.1103/PhysRevB.89.205303}
}

@article{kwong2021effect,
  title={Effect of intravalley and intervalley electron-hole exchange on the nonlinear optical response of monolayer {Mo}{Se}$_2$},
  author={Kwong, Nai-Hang and Schaibley, John R. and Binder, Rolf},
  journal={Phys. Rev. B},
  volume={104},
  number={24},
  pages={245434},
  year={2021},
  publisher={APS},
doi = {10.1103/PhysRevB.104.245434}
}

@article{yang2015hole,
  title={Hole spin relaxation in bilayer {W}{Se}$_2$},
  author={Yang, F. and Wang, L. and Wu, M.~W.},
  journal={Phys. Rev. B},
  volume={92},
  number={15},
  pages={155414},
  year={2015},
  publisher={APS},
doi = {10.1103/PhysRevB.92.155414}
}

@article{molas2017brightening,
  title={Brightening of dark excitons in monolayers of semiconducting transition metal dichalcogenides},
  author={Molas, Maciej R. and Faugeras, Clement and Slobodeniuk, Autur O. and Nogajewski, Karol and Bartos, Miroslav and Basko, D.~M. and Potemski, Marek},
  journal={2D Mater.},
  volume={4},
  number={2},
  pages={021003},
  year={2017},
  publisher={IOP Publishing},
doi = {10.1088/2053-1583/aa5521}
}

@article{zhou2017probing,
  title={Probing dark excitons in atomically thin semiconductors via near-field coupling to surface plasmon polaritons},
  author={Zhou, You and Scuri, Giovanni and Wild, Dominik S. and High, Alexander A. and Dibos, Alan and Jauregui, Luis A. and Shu, Chi and De Greve, Kristiaan and Pistunova, Kateryna and Joe, Andrew Y. and Taniguchi, Takashi and Watanabe, Kenji and Kim, Philip and Lukin, Mikhail D. and Park, Hongkun},
  journal={Nat. Nanotechnol.},
  volume={12},
  number={9},
  pages={856--860},
  year={2017},
  publisher={Nature Publishing Group UK London},
doi = {10.1038/nnano.2017.106}
}

@article{wang2018intravalley,
  title={Intravalley Spin-Flip Relaxation Dynamics in Single-Layer {W}{S}$_2$},
  author = {Wang, Zilong and Molina-Sánchez, Alejandro and Altmann, Patrick and Sangalli, Davide and De Fazio, Domenico and Soavi, Giancarlo and Sassi, Ugo and Bottegoni, Federico and Ciccacci, Franco and Finazzi, Marco and Wirtz, Ludger and Ferrari, Andrea C. and Marini, Andrea and Cerullo, Giulio and Dal Conte, Stefano},
  journal={Nano Lett.},
  volume={18},
  number={11},
  pages={6882--6891},
  year={2018},
  publisher={ACS Publications},
doi = {10.1021/acs.nanolett.8b02774}
}

@article{molina2017ab,
  title={Ab Initio Calculations of Ultrashort Carrier Dynamics in Two-Dimensional Materials: Valley Depolarization in Single-Layer {W}{Se}$_2$},
  author={Molina-S{\'a}nchez, Alejandro and Sangalli, Davide and Wirtz, Ludger and Marini, Andrea},
  journal={Nano Lett.},
  volume={17},
  number={8},
  pages={4549--4555},
  year={2017},
  publisher={ACS Publications},
doi = {10.1021/acs.nanolett.7b00175}
}

@article{deilmann2020ab,
  title = {Ab Initio Studies of Exciton $g$ Factors: Monolayer Transition Metal Dichalcogenides in Magnetic Fields},
  author = {Deilmann, Thorsten and Kr\"uger, Peter and Rohlfing, Michael},
  journal={ Phys. Rev. Lett.},
  volume={124},
  number={22},
  pages={226402},
  year={2020},
  publisher={APS},
doi = {10.1103/PhysRevLett.124.226402}
}

@article{wang2017plane,
  title={In-Plane Propagation of Light in Transition Metal Dichalcogenide Monolayers: Optical Selection Rules},
  author = {Wang, G. and Robert, C. and Glazov, M. M. and Cadiz, F. and Courtade, E. and Amand, T. and Lagarde, D. and Taniguchi, T. and Watanabe, K. and Urbaszek, B. and Marie, X.},
  journal={Phys. Rev. Lett.},
  volume={119},
  number={4},
  pages={047401},
  year={2017},
  publisher={APS},
doi = {10.1103/PhysRevLett.119.047401}
}

@article{scharf2017magnetic,
  title = {Magnetic Proximity Effects in Transition-Metal Dichalcogenides: Converting Excitons},
  author = {Scharf, Benedikt and Xu, Gaofeng and Matos-Abiague, Alex and \ifmmode \check{Z}\else \v{Z}\fi{}uti\ifmmode \acute{c}\else \'{c}\fi{}, Igor},
  journal = {Phys. Rev. Lett.},
  volume = {119},
  issue = {12},
  pages = {127403},
  numpages = {6},
  year = {2017},
  month = {Sep},
  publisher = {American Physical Society},
  doi = {10.1103/PhysRevLett.119.127403},
  url = {https://link.aps.org/doi/10.1103/PhysRevLett.119.127403}
}

@article{cao2024emergent,
  title={Emergent bright excitons with {R}ashba spin-orbit coupling in atomic monolayers},
  author={Cao, Jiayu David and Xu, Gaofeng and Scharf, Benedikt and Denisov, Konstantin and {\v{Z}}uti{\'c}, Igor},
  journal={Phys. Rev. B},
  volume={109},
  number={8},
  pages={085407},
  year={2024},
  publisher={APS},
doi = {10.1103/PhysRevB.109.085407}
}

@article{berghauser2018inverted,
  title={Inverted valley polarization in optically excited transition metal dichalcogenides},
  author={Bergh{\"a}user, Gunnar and Bernal-Villamil, Ivan and Schmidt, Robert and Schneider, Robert and Niehues, Iris and Erhart, Paul and Michaelis de Vasconcellos, Steffen and Bratschitsch, Rudolf and Knorr, Andreas and Malic, Ermin},
  journal={Nat. Commun.},
  volume={9},
  number={1},
  pages={971},
  year={2018},
  publisher={Nature Publishing Group UK London},
doi = {10.1038/s41467-018-03354-1}
}

@article{steinhoff2021microscopic,
  title={Microscopic theory of exciton-exciton annihilation in two-dimensional semiconductors},
  author={Steinhoff, Alexander and Jahnke, Frank and Florian, Matthias},
  journal={Phys. Rev. B},
  volume={104},
  number={15},
  pages={155416},
  year={2021},
  publisher={APS},
doi = {10.1103/PhysRevB.104.155416}
}

@article{erkensten2021dark,
  title={Dark exciton-exciton annihilation in monolayer {W}{Se}$_2$},
   author = {Erkensten, Daniel and Brem, Samuel and Wagner, Koloman and Gillen, Roland and Perea-Caus\'{\i}n, Ra\"ul and Ziegler, Jonas D. and Taniguchi, Takashi and Watanabe, Kenji and Maultzsch, Janina and Chernikov, Alexey and Malic, Ermin},
  journal={Phys. Rev. B},
  volume={104},
  number={24},
  pages={L241406},
  year={2021},
  publisher={APS},
doi = {10.1103/PhysRevB.104.L241406}
}

@article{song2013transport,
  title={Transport Theory of Monolayer Transition-Metal Dichalcogenides through Symmetry},
  author={Song, Yang and Dery, Hanan},
  journal={Phys. Rev. Lett.},
  volume={111},
  number={2},
  pages={026601},
  year={2013},
  publisher={APS},
doi = {10.1103/PhysRevLett.111.026601}
}

@article{robert2020measurement,
  title={Measurement of the spin-forbidden dark excitons in {Mo}{S}$_2$ and {Mo}{Se}$_2$ monolayers},
  author={Robert, C. and Han, B. and Kapuscinski, P. and Delhomme, A. and Faugeras, C. and Amand, T. and Molas, M. R. and Bartos, M. and Watanabe, K. and Taniguchi, T. and Urbaszek, B. and Potemski, M. and Marie, X.},
  journal={Nat. Commun.},
  volume={11},
  number={1},
  pages={4037},
  year={2020},
  publisher={Nature Publishing Group UK London},
doi = {10.1038/s41467-020-17608-4}
}

@article{yang2020exciton,
  title={Exciton valley depolarization in monolayer transition-metal dichalcogenides},
  author={Yang, Min and Robert, Cedric and Lu, Zhengguang and Van Tuan, Dinh and Smirnov, Dmitry and Marie, Xavier and Dery, Hanan},
  journal={Phys. Rev. B},
  volume={101},
  number={11},
  pages={115307},
  year={2020},
  publisher={APS},
doi = {10.1103/PhysRevB.101.115307}
}

@article{ochoa2013spin,
  title={Spin-orbit-mediated spin relaxation in monolayer {Mo}{S}$_2$},
  author={Ochoa, H{\'e}ctor and Rold{\'a}n, Rafael},
  journal={Phys. Rev. B},
  volume={87},
  number={24},
  pages={245421},
  year={2013},
  publisher={APS},
doi = {10.1103/PhysRevB.87.245421}
}

@article{kovalchuk2025revealing,
  title={Revealing hidden interlayer excitons in {2D} bilayers via hybrid molecular gating},
  author={
      Kovalchuk, Sviatoslav and Greben, Kyrylo and Kumar, Abhijeet M. and Pessel, Simon and Soyka, Jan and Cao, Qing and Watanabe, Kenji and Taniguchi, Takashi and Christiansen, Dominik and Selig, Malte and Knorr, Andreas and Eigler, Siegfried and Bolotin, Kirill I.},
  journal={Nat. Commun.},
  volume={16},
  number={1},
  pages={9893},
  year={2025},
  publisher={Nature Publishing Group UK London},
doi = {10.1038/s41467-025-65431-6}
}

@book{jackson1999classical,
  title={Classical Electrodynamics},
  author={Jackson, John David},
  year={1999},
  publisher={Wiley, New York},
doi = {}
}

@article{florian2018dielectric,
  title={The Dielectric Impact of Layer Distances on Exciton and Trion Binding Energies in van der {W}aals Heterostructures},
  author={Florian, Matthias and Hartmann, Malte and Steinhoff, Alexander and Klein, Julian and Holleitner, Alexander W. and Finley, Jonathan J. and Wehling, Tim O. and Kaniber, Michael and Gies, Christopher},
  journal={Nano Lett.},
  volume={18},
  number={4},
  pages={2725--2732},
  year={2018},
  publisher={ACS Publications},
doi = {10.1021/acs.nanolett.8b00840}
}

@article{rohlfing2000electron,
  title={Electron-hole excitations and optical spectra from first principles},
  author={Rohlfing, Michael and Louie, Steven G.},
  journal={Phys. Rev. B},
  volume={62},
  number={8},
  pages={4927},
  year={2000},
  publisher={APS},
doi = {10.1103/PhysRevB.62.4927}
}

@article{trolle2017model,
  title={Model dielectric function for {2D} semiconductors including substrate screening},
  author={Trolle, Mads L. and Pedersen, Thomas G. and V{\'e}niard, Valerie},
  journal={Sci. Rep.},
  volume={7},
  number={1},
  pages={39844},
  year={2017},
  publisher={Nature Publishing Group UK London},
doi = {10.1038/srep39844}
}

@book{haug2009quantum,
author = {Haug, Hartmut and Koch, Stephan W.},
title = {Quantum Theory of the Optical and Electronic Properties of Semiconductors},
publisher = {World Scientific},
year = {2009},
doi = {10.1142/7184},
edition   = {5th}
}

@article{katzer2023exciton,
  title = {Exciton-phonon scattering: Competition between the bosonic and fermionic nature of bound electron-hole pairs},
  author = {Katzer, Manuel and Selig, Malte and Sigl, Lukas and Troue, Mirco and Figueiredo, Johannes and Kiemle, Jonas and Sigger, Florian and Wurstbauer, Ursula and Holleitner, Alexander W. and Knorr, Andreas},
  journal = {Phys. Rev. B},
  volume = {108},
  issue = {12},
  pages = {L121102},
  numpages = {6},
  year = {2023},
  month = {Sep},
  publisher = {American Physical Society},
  doi = {10.1103/PhysRevB.108.L121102},
  url = {https://link.aps.org/doi/10.1103/PhysRevB.108.L121102}
}

@article{li2013intrinsic,
  title={Intrinsic electrical transport properties of monolayer silicene and {Mo}{S}$_2$ from first principles},
   author = {Li, Xiaodong and Mullen, Jeffrey T. and Jin, Zhenghe and Borysenko, Kostyantyn M. and Buongiorno Nardelli, M. and Kim, Ki Wook},
  journal={Phys. Rev. B},
  volume={87},
  number={11},
  pages={115418},
  year={2013},
  publisher={APS},
doi = {10.1103/PhysRevB.87.115418}
}

@article{vasconcelos2018dark,
  title = {Dark exciton brightening and its engaged valley dynamics in monolayer {WSe}$_{2}$},
  author = {Vasconcelos, Railson and Bragan\ifmmode \mbox{\c{c}}\else \c{c}\fi{}a, Helena and Qu, Fanyao and Fu, Jiyong},
  journal = {Phys. Rev. B},
  volume = {98},
  issue = {19},
  pages = {195302},
  numpages = {14},
  year = {2018},
  month = {Nov},
  publisher = {American Physical Society},
  doi = {10.1103/PhysRevB.98.195302},
  url = {https://link.aps.org/doi/10.1103/PhysRevB.98.195302}
}

@article{feierabend2020brightening,
doi = {10.1088/2053-1583/abb876},
url = {https://doi.org/10.1088/2053-1583/abb876},
year = {2020},
publisher = {IOP Publishing},
volume = {8},
number = {1},
pages = {015013},
author = {Feierabend, Maja and Brem, Samuel and Ekman, August and Malic, Ermin},
title = {Brightening of spin- and momentum-dark excitons in transition metal dichalcogenides},
journal = {2D Mater.}
}

@article{lu2019magnetic,
  title={Magnetic field mixing and splitting of bright and dark excitons in monolayer {Mo}{Se}$_2$},
  author = {Lu, Zhengguang and Rhodes, Daniel and Li, Zhipeng and Van Tuan, Dinh and Jiang, Yuxuan and Ludwig, Jonathan and Jiang, Zhigang and Lian, Zhen and Shi, Su-Fei and Hone, James and Dery, Hanan and Smirnov, Dmitry},
  journal={2D Mater.},
  volume={7},
  number={1},
  pages={015017},
  year={2019},
  publisher={IOP Publishing},
doi = {10.1088/2053-1583/ab5614}
}

@article{zhang2017magnetic,
  title={Magnetic brightening and control of dark excitons in monolayer {W}{Se}$_2$},
  author={Zhang, Xiao-Xiao and Cao, Ting and Lu, Zhengguang and Lin, Yu-Chuan and Zhang, Fan and Wang, Ying and Li, Zhiqiang and Hone, James C. and Robinson, Joshua A. and Smirnov, Dmitry and Louie, Steven G. and Heinz, Tony F.},
  journal={Nat. Nanotechnol.},
  volume={12},
  number={9},
  pages={883--888},
  year={2017},
  publisher={Nature Publishing Group UK London},
doi = {10.1038/nnano.2017.105}
}

@article{dery2015polarization,
  title={Polarization analysis of excitons in monolayer and bilayer transition-metal dichalcogenides},
  author={Dery, Hanan and Song, Yang},
  journal={Phys. Rev. B},
  volume={92},
  number={12},
  pages={125431},
  year={2015},
  publisher={APS},
doi = {10.1103/PhysRevB.92.125431}
}

@article{maialle1993exciton,
  title={Exciton spin dynamics in quantum wells},
  author = {Maialle, M. Z. and de Andrada e Silva, E. A. and Sham, L. J.},
  journal={Phys. Rev. B},
  volume={47},
  number={23},
  pages={15776},
  year={1993},
  publisher={APS},
doi = {10.1103/PhysRevB.47.15776}
}

@article{latini2015excitons,
  title={Excitons in van der {W}aals heterostructures: The important role of dielectric screening},
  author={Latini, Simone and Olsen, Thomas and Thygesen, Kristian Sommer},
  journal={Phys. Rev. B},
  volume={92},
  number={24},
  pages={245123},
  year={2015},
  publisher={APS},
doi = {10.1103/PhysRevB.92.245123}
}

@article{qiu2015nonanalyticity,
  title={Nonanalyticity, Valley Quantum Phases, and Lightlike Exciton Dispersion in Monolayer Transition Metal Dichalcogenides: Theory and First-Principles Calculations},
  author={Qiu, Diana Y. and Cao, Ting and Louie, Steven G.},
  journal={ Phys. Rev. Lett.},
  volume={115},
  number={17},
  pages={176801},
  year={2015},
  publisher={APS},
doi = {10.1103/PhysRevLett.115.176801}
}

@article{laturia2018dielectric,
  title={Dielectric properties of hexagonal boron nitride and transition metal dichalcogenides: from monolayer to bulk},
  author={Laturia, Akash and Van de Put, Maarten L. and Vandenberghe, William G.},
  journal={npj 2D Mater. Appl.},
  volume={2},
  number={1},
  pages={6},
  year={2018},
  publisher={Nature Publishing Group UK London},
doi = {10.1038/s41699-018-0050-x}
}

@article{article:THEORY_DispersionBychkovRashba1984,
  title={Properties of a {2D} electron gas with lifted spectral degeneracy},
  author={Bychkov, Yu. A. and Rashba, {\'E}.~I.},
  journal={P'isma Zh. Eksp. Teor. Fiz.},
  volume={39},
  number={2},
  pages={66--69},
  year={1984},
doi = {},
url = {http://jetpletters.ru/ps/0/article_19121.shtml}
}

@book{book:Spin_orbit_coupling_Winkler2003,
  title={Spin-Orbit Coupling Effects in Two-Dimensional Electron and Hole Systems},
  author={Winkler, Roland},
  volume={191},
  year={2003},
  publisher={Springer Berlin, Heidelberg},
doi = {10.1007/b13586}
}

@article{article:THEORY_Bychkov_Rashba_Coupling_Kormanyos2014,
  title = {Spin-Orbit Coupling, Quantum Dots, and Qubits in Monolayer Transition Metal Dichalcogenides},
  author = {Korm\'anyos, Andor and Z\'olyomi, Viktor and Drummond, Neil D. and Burkard, Guido},
  journal = {Phys. Rev. X},
  volume = {4},
  issue = {1},
  pages = {011034},
  numpages = {16},
  year = {2014},
  month = {Mar},
  publisher = {American Physical Society},
  doi = {10.1103/PhysRevX.4.011034},
  url = {https://link.aps.org/doi/10.1103/PhysRevX.4.011034}
}

@article{lengers2020theory,
  title = {Theory of the absorption line shape in monolayers of transition metal dichalcogenides},
  author = {Lengers, F. and Kuhn, T. and Reiter, D.~E.},
  journal = {Phys. Rev. B},
  volume = {101},
  issue = {15},
  pages = {155304},
  numpages = {12},
  year = {2020},
  month = {Apr},
  publisher = {American Physical Society},
  doi = {10.1103/PhysRevB.101.155304},
  url = {https://link.aps.org/doi/10.1103/PhysRevB.101.155304}
}

@article{selig2020suppression,
  title={Suppression of intervalley exchange coupling in the presence of momentum-dark states in transition metal dichalcogenides},
  author={Selig, Malte and Katsch, Florian and Brem, Samuel and Mkrtchian, Garnik F. and Malic, Ermin and Knorr, Andreas},
  journal={Phys. Rev. Res.},
  volume={2},
  number={2},
  pages={023322},
  year={2020},
  publisher={APS},
doi = {10.1103/PhysRevResearch.2.023322}
}

@book{book:graphene_carbon_nanotubes_malic2013,
  title={Graphene and Carbon Nanotubes: Ultrafast Relaxation Dynamics and Optics},
  author={Malic, Ermin and Knorr, Andreas},
  year={2013},
  publisher={John Wiley \& Sons, Ltd},
doi = {10.1002/9783527658749}
}

@article{selig2018dark,
	doi = {10.1088/2053-1583/aabea3},
	url = {https://doi.org/10.1088/2053-1583/aabea3},
	year = 2018,
	month = {may},
	publisher = {{IOP} Publishing},
	volume = {5},
	number = {3},
	pages = {035017},
	author = {Malte Selig and Gunnar Berghäuser and Marten Richter and Rudolf Bratschitsch and Andreas Knorr and Ermin Malic},
	title = {Dark and bright exciton formation, thermalization, and photoluminescence in monolayer transition metal dichalcogenides},
	journal = {2D Mater.}
}

@article{article:THEORY_spin_valley_selective_excitation_Yao2012,
  title = {Coupled Spin and Valley Physics in Monolayers of {Mo}{S}$_2$ and Other Group-{VI} Dichalcogenides},
  author = {Xiao, Di and Liu, Gui-Bin and Feng, Wanxiang and Xu, Xiaodong and Yao, Wang},
  journal = {Phys. Rev. Lett.},
  volume = {108},
  issue = {19},
  pages = {196802},
  numpages = {5},
  year = {2012},
  month = {May},
  publisher = {American Physical Society},
  doi = {10.1103/PhysRevLett.108.196802},
  url = {https://link.aps.org/doi/10.1103/PhysRevLett.108.196802}
}

@article{ivanov1993self,
  title = {Self-consistent theory of the biexciton optical nonlinearity},
  author = {Ivanov, A. L. and Haug, H.},
  journal = {Phys. Rev. B},
  volume = {48},
  issue = {3},
  pages = {1490--1504},
  numpages = {0},
  year = {1993},
  month = {Jul},
  publisher = {American Physical Society},
  doi = {10.1103/PhysRevB.48.1490},
  url = {https://link.aps.org/doi/10.1103/PhysRevB.48.1490}
}

@article{katsch2018theory,
author = {Katsch, Florian and Selig, Malte and Carmele, Alexander and Knorr, Andreas},
title = {Theory of Exciton–Exciton Interactions in Monolayer Transition Metal Dichalcogenides},
journal = {Phys. Status Solidi (B)},
volume = {255},
number = {12},
pages = {1800185},
year={2018},
doi = {10.1002/pssb.201800185},
url = {https://onlinelibrary.wiley.com/doi/abs/10.1002/pssb.201800185}
}

@article{selig2019ultrafast,
  title = {Ultrafast dynamics in monolayer transition metal dichalcogenides: Interplay of dark excitons, phonons, and intervalley exchange},
  author = {Selig, Malte and Katsch, Florian and Schmidt, Robert and Michaelis de Vasconcellos, Steffen and Bratschitsch, Rudolf and Malic, Ermin and Knorr, Andreas},
  journal = {Phys. Rev. Res.},
  volume = {1},
  issue = {2},
  pages = {022007},
  numpages = {6},
  year = {2019},
  month = {Sep},
  publisher = {American Physical Society},
  doi = {10.1103/PhysRevResearch.1.022007},
  url = {https://link.aps.org/doi/10.1103/PhysRevResearch.1.022007}
}

@article{kormanyos2015k,
	doi = {10.1088/2053-1583/2/2/022001},
	url = {https://doi.org/10.1088/2053-1583/2/2/022001},
	year = 2015,
	month = {apr},
	publisher = {{IOP} Publishing},
	volume = {2},
	number = {2},
	pages = {022001},
	author = {Andor Korm{\'{a}}nyos and Guido Burkard and Martin Gmitra and Jaroslav Fabian and Viktor Z{\'{o}}lyomi and Neil D. Drummond and Vladimir Fal'ko},
	title = {k$\cdot$p theory for two-dimensional transition metal dichalcogenide semiconductors},
	journal = {2D Mater.}
}

@article{thranhardt2000quantum,
  title = {Quantum theory of phonon-assisted exciton formation and luminescence in semiconductor quantum wells},
  author = {Thr\"anhardt, A. and Kuckenburg, S. and Knorr, A. and Meier, T. and Koch, S. W.},
  journal = {Phys. Rev. B},
  volume = {62},
  issue = {4},
  pages = {2706--2720},
  numpages = {0},
  year = {2000},
  publisher = {American Physical Society},
  doi = {10.1103/PhysRevB.62.2706},
  url = {https://link.aps.org/doi/10.1103/PhysRevB.62.2706}
}
\end{document}